\documentclass[universe,review,accept,moreauthors,pdftex,10pt,a4paper]{mdpi}
\firstpage{1} 
\articlenumber{9}
\doinum{10.3390/universe7090330}
\pubvolume{7}
\pubyear{2021}
\copyrightyear{2021}
\externaleditor{}
\history{}
\pdfoutput=1
 \usepackage{amssymb}
 \usepackage{amsmath}
  \usepackage{amsfonts}
   \usepackage{slashed}
  \usepackage{epsfig}
   \usepackage{soul}
   \usepackage{bm} 
 \theoremstyle{mdpi}
 \newcounter{thm}
 \newcounter{ex}
 \newcounter{re}
 \theoremstyle{mdpidefinition}

\Title{Different faces of confinement}

\Author{Roman Pasechnik$^{1}$ and Michal~\v{S}umbera$^{2}$}
\address{$^{1}$ \quad
Department of Astronomy and Theoretical Physics, Lund
University, SE-223 62 Lund, Sweden \\
$^{2}$ \quad
Nuclear Physics Institute CAS, 25068 \v{R}e\v{z}, Czech Republic\\}

\abstract{In this review, we provide a short outlook of some of the currently most popular pictures and promising approaches to non-perturbative physics and confinement in gauge theories. A qualitative and by no means exhaustive discussion presented here covers such key topics as the phases of QCD matter, the order parameters for confinement, the central vortex and monopole pictures of the QCD vacuum structure, fundamental properties of the string tension, confinement realisations in gauge-Higgs and Yang-Mills theories, magnetic order/disorder phase transition among others.}

\keyword{magnetic disorder; confinement models; lattice QCD; center vortices; magnetic monopoles; quark condensate} 

\PACS{12.38.Aw,12.38.Gc}

\begin{document}
\section{Introduction}
\label{Sect:intro}

Quantum Chromodynamics (QCD) based upon $SU(3)_c$ gauge theory of color represents a real-world example of a fundamental Yang-Mills (YM) theory applied to description of strong interactions and is an organic part of the Standard Model (SM) of particle physics. This theory is extremely successful in predicting various measurable phenomena at particle colliders. The class of phenomena that originate from (or driven by) strong interactions is extremely wide and covers such areas as nuclear physics, hadron physics, physics of quark-gluon plasma, high-temperature and high-density QCD, high-energy particle production and hadronisation. Depending on characteristic length scales, QCD behaves very differently. At short space-time separations, e.g. once we zoom into distances much shorter than the proton radius, QCD appears as a weakly-coupled theory enabling a precise Perturbation Theory (PT) analysis. Much of its success has been achieved in this {\it asymptotic freedom} or ultraviolet (UV) regime where the quark-gluon interaction strength recedes. Thus success highlights the QCD theory as the correct theory of strong interactions at the fundamental level precisely matching all the existing observations up to very high momentum transfers reached by the Large Hadron Collider (LHC) so far. However, on the opposite side of length-scales in the infrared (IR) limit, QCD enters entirely different, strongly-coupled domain, rendering the PT inapplicable and creating substantial problems for making reliable predictions at intermediate and low momentum transfers, i.e. at large distances. While it is conventionally believed that QCD should remain the correct theory of strong interactions also at large distances, in the so-called {\it confined regime}, deriving reliable predictions remains a big theoretical challenge. For one of the broadest and comprehensive overviews of many phenomenological and theoretical aspects of QCD and QCD-like gauge theories spanning from IR to UV, from dilute to dense regimes, see Ref.~\cite{Brambilla:2014jmp}.

The {\it problem of confinement} concerns the strongly-coupled sector of QCD composed of interacting colored partons (quarks and gluons). In virtue of color confinement, the colored particles appear to be always trapped (confined) inside colorless composites. The latter emerge as asymptotic states rendering the long-distance regime of hadron physics described by Effective Field Theory (EFT) approaches such as the chiral PT, as well as a variety of non-perturbative techniques realised in numerical simulations on the lattice. Much of the discussion in the current review is devoted to highlighting main ideas and possible existing ways to address the confinement problem that is known as the main unsolved problem in the SM framework. Despite the major efforts of the research community and tremendous progress made over last few decades, it does not appear to be fully and consistently resolved yet. There are several important subtleties in formulation of this problem to be discussed in what follows. One of the standard ways of formulating the problem is that there is no complete understanding of why these fundamental degrees of freedom (DoFs) of QCD (or, generically, of any strongly-coupled YM theory) do not emerge in the physical spectrum of asymptotic states and how the composite hadrons are dynamically produced starting from the fundamental DoFs in the initial state. In a phenomenological sense, there is a fundamental mismatch between the underlined DoFs of QCD in its short- and long-distance regimes manifest in experimental measurements, and there is not a single consistent theoretical framework that goes beyond the framework of PT and treats both weakly- and strongly-coupled regimes on the same footing.

For practical purposes, various phenomenological approaches have been proposed that characterise the long-distance effects of QCD absorbing them into universal elements of a given scattering process, such as non-perturbative matrix elements, fragmentation functions or parton distributions. As a commonly adopted picture, a color-electric flux tube (also known as a color string) is stretched among the partons produced in a high-energy collision. A string-like picture emerges in the limit of large number of colors already in $D=1+1$ dimensions as has been advocated by t'Hooft back in early 70'es -- see e.g. Ref.~\cite{tHooft:1974pnl}. As produced partons move away from each other at large enough distances, those flux tubes fragment into composite particles such as mesons and baryons where initial (anti)quarks and gluons get necessarily combined with newly emerged ones from the vacuum into color-neutral configurations. In a nutshell, the basic problem concerns a first-principle derivation of long-distance hadron spectrum and dynamics from an underlined strongly-coupled gauge theory. More specifically, a successful model of confinement is expected to provide a first-principle dynamical description of the string formation, its basic characteristics and string-breaking effects, also connecting those unambiguously to dynamics of the fundamental DoFs of the underlined gauge theory and deducing the phase structure of the theory at various densities and temperatures. While there are no compelling solutions yet available, there are several distinct approaches to confinement treatment being actively developed in the literature. Not only a large variety of treatments of confinement has hit the literature in past decades, but also a proper definition of confinement, what we actually mean by this word, posses a notorious difficulty as was thoroughly discussed in Refs.~\cite{Greensite:2011zz,Greensite:2016pfc}. In this review, we will try to summarise some of the existing attractive treatments of confinement and ideas and why confinement occurs in the way it does in a conceptual and qualitative manner, without pretending to provide an exhaustive overview of all relevant details and corresponding references.

The review is organised as follows. In Sect.~\ref{Sect:QCDphases} we discuss the basic ingredients of the QCD phase diagram at different temperatures and values of the baryon chemical potential. In Sect.~\ref{Sect:Latt} we provide a brief description of magnetic order/disorder phases and introduce the basic notions of the lattice gauge theory that would be used in follow-up discussions. In Sect.~\ref{Sect:Wilson} we overview basic concepts and ideas that lead to different asymptotic behaviors of Wilson loop VEV as an order parameter for confining phase. Such distinct properties of QCD scattering amplitudes as the Regge trajectories and the associated picture of a color string have been outlined in Sect.~\ref{Sect:Regge}. In Sect.~\ref{Sect:Higgs-conf}, we provide a detailed outlook on the complementarity between the Higgs and confining phases and describe such a common feature for both phases as color confinement. In Sect.~\ref{Sect:hadronisation}, a brief description of the string hadronisation picture realised in the Lund model is given. Sect.~\ref{Sect:conf-criteria} elaborates on why confinement criteria based upon gauge symmetry remnants (un)breaking may be spoiled by gauge-fixing artefacts, highlighting the need for a gauge-invariant description of confinement. Sect.~\ref{Sect:center} introduces the basics of the center symmetry based confinement criterion and its implications. Sect.~\ref{Sect:PL} gives a brief outlook on another order parameter of confinement, the Polyakov loop, particularly suitable for confinement description at finite temperatures. In Sect.~\ref{Sect:tHL}, yet another important order parameter of confinement probing the vortex structure of the QCD vacuum, the t'Hooft loop, has been introduced and the basic features of the center vortices have been described. Sect.~\ref{Sect:tension} elaborates on the most important characteristics of the string tension as the probes for a confining phase. The foundations and implications of the center vortex mechanism of confinement, with its basic tests performed in the literature, have been discussed in Sect.~\ref{Sect:vortex}. Sect.~\ref{Sect:condensates} connects the chiral symmetry breaking and the topological charge to the existence of vortex configurations. In Sect.~\ref{Sect:GZ-scenario}, we briefly describe the Gribov-Zwanziger scenario of confinement relating it to the non-perturbative behavior of propagators and describing how a color string could emerge in this scenario by considering constituent gluons in the gluon chain model. An renown dual superconductivity picture of confinement and the fundamental role of magnetic monopoles have been briefly described in Sect.~\ref{Sect:monopoles}. A novel generalisation of the confinement criterion applicable in gauge theories with matter in the fundamental representation has been briefly discussed in Sect.~\ref{Sect:separation}. Sect.~\ref{Sect:Hconf-transitions} highlights an important recent development in understanding the confining property of the gauge-field vacuum and Higgs-confinement transitions via a novel non-local order parameter. A summary and concluding remarks are given in Sect.~\ref{Sect:Summary}.

\section{Phase structure of QCD matter}
\label{Sect:QCDphases}

Following the discovery of asymptotic freedom in QCD \cite{Gross:1973id,Politzer:1973fx}, it has been realised that phase transitions in the hot and dense QCD matter between the hadronic (confined) and quark-gluon (deconfined) phases are crucial for understanding the cosmological evolution as well as the state of matter and dynamics of neutron stars \cite{Collins:1974ky,Cabibbo:1975ig, Shuryak:1977ut, Shuryak:1978ij,Freedman:1976ub, Polyakov:1978vu, Kapusta:1979fh, Witten:1984rs}. Besides, the idea of experimental measurements through heavy-ion collisions has been offered as a tantalising opportunity for explorations of this interesting physics. In those early times, a hypothetical state of QCD matter at characteristic temperatures of around 100 MeV has been envisaged as existing in two possible states of ``hadronic plasma'' \cite{Shuryak:1977ut} and ``quark-gluon plasma'' (QGP) \cite{Shuryak:1978ij}, with an energy density of order 1 GeV$/$fm$^3$. Later on, it has been understood that the QCD phase diagram has a much richer structure, particularly, at high baryon number densities, with a lot of important implications for understanding, for instance, neutron star physics as well as heavy-ion collisions at particle colliders.

Strongly-interacting QGP was first discovered at RHIC collider in 2005 \cite{BRAHMS:2004adc,PHOBOS:2004zne,STAR:2005gfr,PHENIX:2004vcz} and later has been confirmed at much higher energies at the CERN LHC (for a detailed review, see e.g. Refs.~\cite{Braun-Munzinger:2015hba, Pasechnik:2016wkt} and references therein). In the QGP phase, as the name suggests, the strong interactions between constituents of the plasma, ``dressed'' light quarks and gluons being its collective excitations, is driven by their $SU(3)_{\text c}$ color charges. For a comprehensive review of early developments and key ideas in analysis of strongly-coupled QCD phenomena and QGP in particular, see e.g. Ref.~\cite{Rafelski:2003zz}, while an overview of more recent theoretical and experimental studies can be found in Refs.~\cite{Shuryak:2014zxa, Braun-Munzinger:2015hba, Pasechnik:2016wkt}.
\begin{figure}[hbt]
\includegraphics[trim= 0 0 0 0, clip, width=.48\textwidth, height=.44\textwidth]{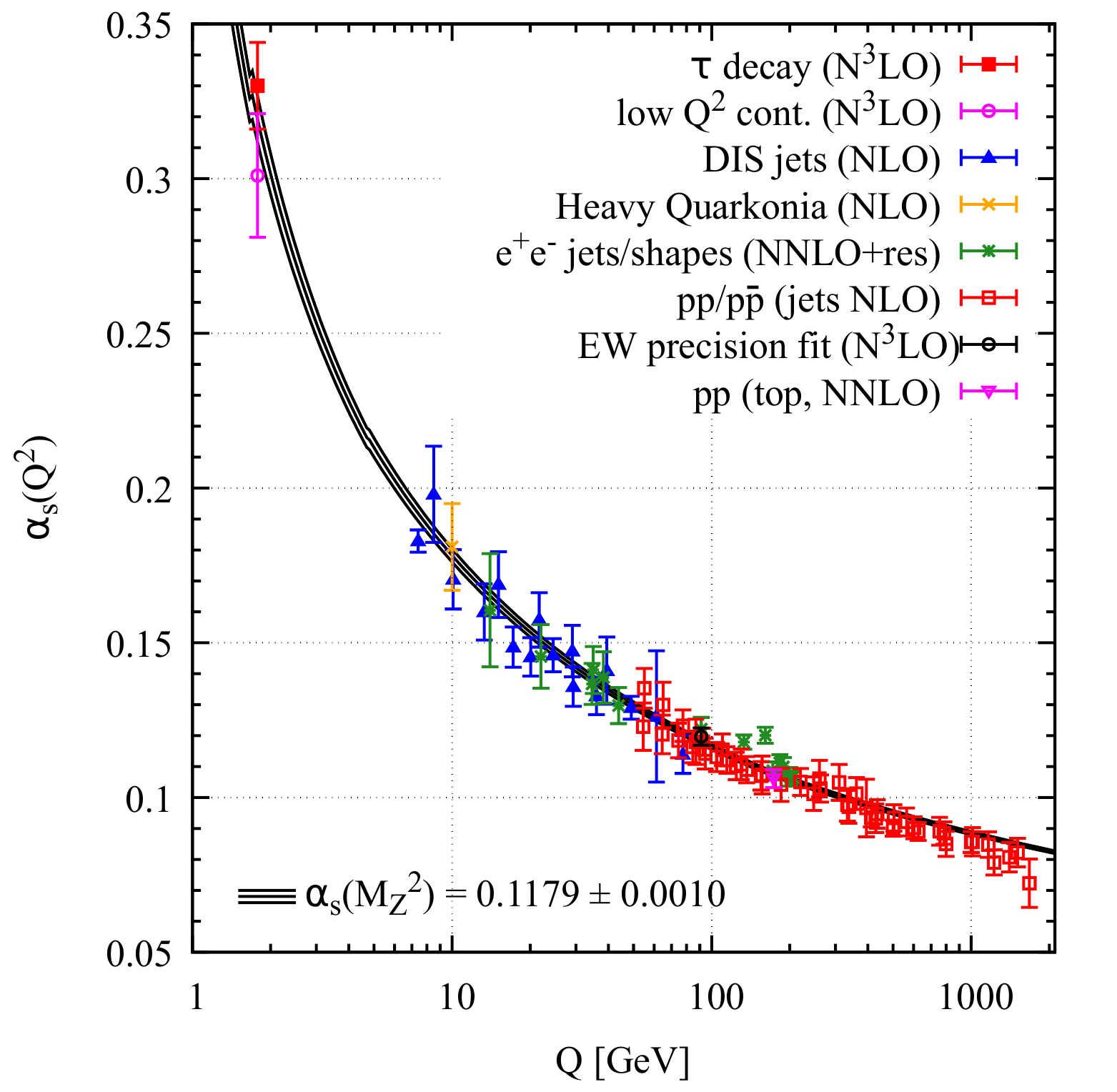}
\includegraphics[trim= 0 0 0 0, clip, width=.51\textwidth, ]{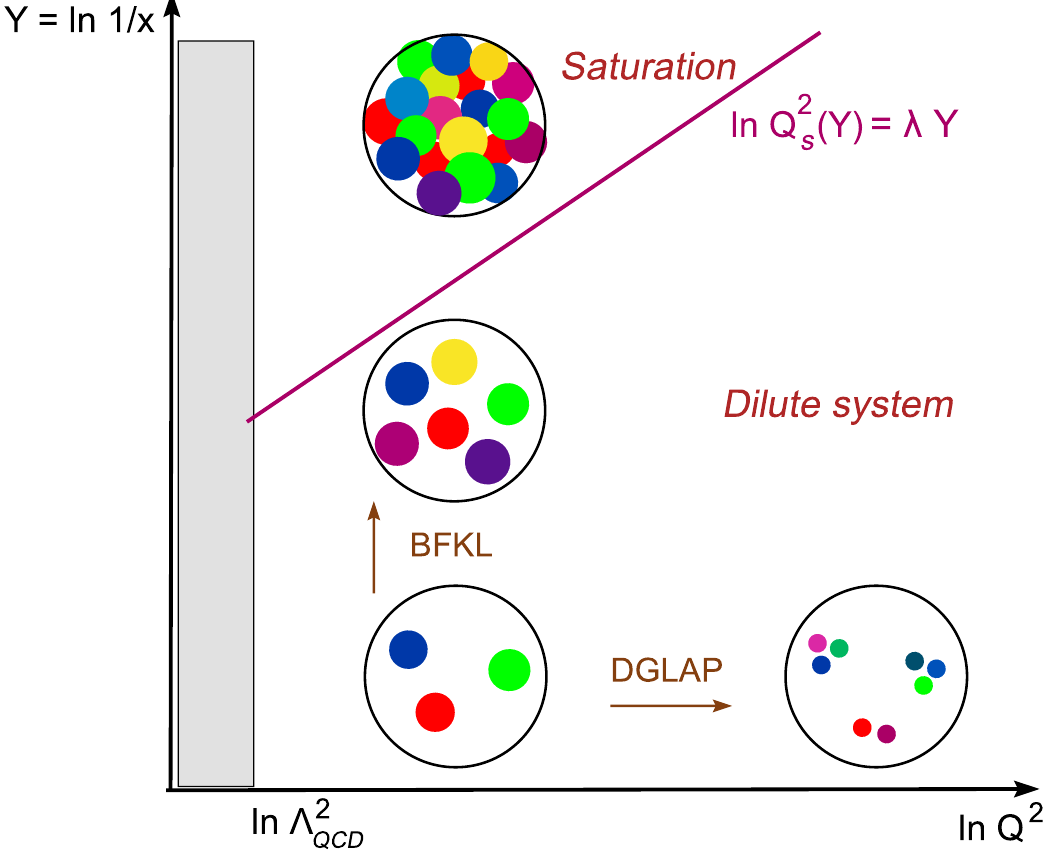}
\caption{On the left panel, the QCD interaction strength $\alpha_{\rm s}$ as a function of the momentum transfer $Q$ at next-to-leading order of the PT. The figure is taken from Ref.~\cite{ParticleDataGroup:2020ssz}. On the left panel, the QCD evolution of the characteristic parton (quark and gluon) density and length-scale with respect to rapidity $Y = \ln(1/x)$ and $\ln Q^2$. The figure is taken from Ref.~\cite{Gelis:2010nm}.
}
\label{fig:alphas}
\end{figure}

In a weakly-interacting QCD gas at very high $T$, the microscopic quark-gluons interactions are relatively weak and should obey the predictions of asymptotic freedom. The leading-order perturbative QCD coupling that determines the strength of QCD interactions at asymptotically short distances,
\begin{equation} 
\alpha_{\rm s}(Q) \simeq  \frac{2\pi}{b_0 \ln(Q /\Lambda_{\rm QCD})}\,, \qquad 
b_0\!=\!11\!-\!\frac{2}{3}N_f \,,
\label{eq:alpha_S}
\end{equation}
is given in terms of the QCD energy scale $\Lambda_{\rm QCD}\approx {\cal O}(1\,{\rm GeV})$, momentum transfer $Q\gg \Lambda_{\rm QCD}$ and the active quark flavors' number $N_f$. In a perturbative domain of QCD, when going towards shorter distances $l \ll \Lambda_{\rm QCD}^{-1}$, the color charge is being diluted compared to the ``soft'' and non-perturbative domain of QCD at larger distances $l \sim \Lambda_{\rm QCD}^{-1}$ where the charge is being built-up effectively, due to the phenomenon called the color charge ``anti-screening'' \cite{Gross:1973id,Politzer:1973fx}. This is quite an opposite effect to what happens in QED. This behavior of the coupling is demonstrated in Fig.~\ref{fig:alphas} (left panel), together with experimentally measured values. As soon as $\alpha_{\rm s}(Q)$ hits large values entering the strongly-coupled (confined) regime at lower $T$, the PT ceases to work such that effective and non-perturbative methods are applied, being however often vastly disconnected from the microscopic QCD theory. One could perform a consistent matching of the fundamental QCD to the effective Lagrangian of chiral PT at the ``soft'' scale $Q \simeq 4\pi f_{\pi} \simeq 1$ GeV where both descriptions are expected to be valid and overlap. Such a matching provides a clue about the IR behavior of $\alpha_{\rm S}(Q)$ that tends to get ``frozen" at the value of $\langle\alpha_{\rm S} \rangle_{\rm IR} \simeq 0.56$ \cite{Fujii:1999xn}.

Besides the weakly-coupled short wavelength modes of partonic DoFs with $Q=2\pi T$ dominating thermodynamic evolution at very high $T$, the QGP features also long wavelength (non-perturbative) modes, with length scales of $l > T^{-1}$. The latter modes dominate the evolution at not-so-high $T$ forming a liquid effectively turning QGP into ideal fluid \cite{Lacey:2006bc, Heinz:2013th, Shuryak:2014zxa}. The latter fundamental property of QGP has been discovered first at RHIC \cite{BRAHMS:2004adc,PHOBOS:2004zne,STAR:2005gfr,PHENIX:2004vcz} and then confirmed at the LHC. Other effects of such strongly-interacting QGP are manifested through a collective flow phenomenon \cite{Heinz:2013th} as well as in an effective suppression of high-energy partons transiting through a hot and dense deconfined medium \cite{PHENIX:2001hpc,STAR:2002svs} (for a review, see Ref.~\cite{Pasechnik:2016wkt} and references therein).

Taking the ratio of interaction-to-kinetic energy of the QGP constituents and assuming equal contributions from chromo-electric and chromo-magnetic interactions, one introduces the so-called plasma parameter \cite{Thoma:2005aw} 
\begin{equation}
\Gamma \simeq 2\frac{C_{q,g} \alpha_{\rm S}}{a T} \,, \qquad 
C_q=\frac{N_c^2-1}{2 N_c}=\frac{4}{3} \,, 
\qquad C_g=N_c=3\,,
\label{eq:Thoma}
\end{equation}
expressed in terms of the fundamental (quark) and adjoint (gluon) Casimir invariants of $SU(3)_c$, $C_q$ and $C_g$, respectively, and $T$-dependent average distance between the partons $a$ satisfying $aT \sim d_F^{-1/3}$, where
\begin{eqnarray}
d_F\equiv 2\times 8 + \frac{3}{4}\left(3 \times N_f  \times 2 \times 2\right)\,.
\end{eqnarray}
The latter evolves in $T$ only through $N_f(T)$. Weakly-interacting (ideal) plasmas have a very low $\Gamma<10^{-3}$, while a strongly-interacting plasma typically has a much larger $\Gamma \gtrsim 1$. Taking a nearly-ideal (weakly-coupled) massless QCD gas, for instance, one obtains $\Gamma \sim \alpha_{\rm S} d_F^{1/3}$ serving as a lower estimate for the plasma parameter as it ignores the partonic interactions in the ideal gas approximation. In a realistic case of QGP created in heavy-ion collisions at RHIC, one finds $T\approx200$ MeV and $\alpha_{\rm S}=0.3-0.5$ with only two relevant active flavours, $N_F=2$, leading to a value of $\Gamma \simeq 1.5 - 6$, indeed being deeply inside the strongly-coupled plasma regime.

The QCD evolution of partonic matter in terms of basic kinematic parameters of resolved partons in the medium is illustrated in Fig.~\ref{fig:alphas} (right panel). For instance, developing the partonic cascades in typical momentum transfer $Q$ one resolves the partons with a transverse area $1/Q^2$, such that at larger $Q$ and $T\sim Q$ one observes a dilution of the parton density controlled by the DGLAP evolution equations (see for instance Refs.~\cite{Ioffe:2010zz, Campbell:2017hsr}). One may also observe how the parton density evolves with energy or, more conveniently, with a fraction of light cone momentum taken by a given radiated parton out of a parent particle, $x=k^+/P^+$. One may visualise the partonic cascade off the initial particle effectively as Brownian-like motion in the transverse plane that can be considered as the Gribov diffusion process in the evolution ``time'' $Y=\ln (1/x)$. The latter parameter is simply a rapidity difference between the radiated and parent partons, while the diffusion constant is $D \sim \alpha_{\rm S}$. Such an evolution is controlled by BFKL equations (for more details, see e.g. Refs.~\cite{Ioffe:2010zz, Campbell:2017hsr} and references therein). 

The partonic cascade is dominated essentially by soft gluons at high energies or at very small fractions $x\ll 1$, and they are of the same size at a fixed scale $Q$. As soon as the parton scattering cross section $\sim\alpha_{\rm S}/Q^2$ multiplied by the probability to find a parton at a given $Q$ with a fraction $x$, $x G_A(x,Q^2)$, becomes of the order of the geometrical cross-section of an area $A$ occupied by the gluons, $\sim \pi R_A^2$, the gluons start to overlap effectively. Due to a repulsive interaction between gluons, however, their occupation number saturates at $f_g \sim 1/\alpha_{\rm S}$. In particular, this occurs for gluons with transverse momenta below a certain emergent scale $Q_s(x)$, $k_{\perp} \leq Q_s(x)$, known as a saturation or ``close packing'' scale \cite{Gribov:1983ivg} (see also Refs.~\cite{Kharzeev:2002np,Berges:2020fwq}),
\begin{equation}
Q_s^2(x) = \frac{\alpha_{\rm S}(Q_s)}{2(N_c^2-1)} 
\frac{x G_A(x,Q_s^2)}{\pi R_A^2} \,,
\label{eq:Q_s}
\end{equation} 
thus, representing a fixed point in the parton $x$-evolution. Such a {\it saturation} phenomenon is rather generic as an analogical scaling of the density $\sim \alpha^{-1}$ characterises various Bose-Einstein condensation phenomena, in particular, those in the Higgs mechanism and in superconductivity \cite{McLerran:2008es}. Such a highly coherent gluonic state of matter has properties of a classical field \cite{Kharzeev:2002np} and is known in the literature as the {\it Color Glass Condensate} (CGC) \cite{McLerran:1993ni, Gelis:2010nm, Berges:2020fwq}, or {\it glasma} \cite{Kovner:1995ja}. 

Indeed, in the path integral formulation of the $SU(N)$ gauge theory, for instance, one sums over all gauge-field configurations weighted with $\exp(-iS_g/\hbar)$, where the action can be written as
\begin{eqnarray}
\label{eq:rescaleNotation}
&& S_g=-\frac{1}{4g_s^2}\int {\mathcal F^{\mu\nu,a}} 
{\mathcal F_{\mu\nu}^a} d^4x \,, \\
&& A_\mu^a \rightarrow {\mathcal A}_\mu^a  \equiv g_s A_\mu^a \,, \quad
F_{\mu\nu}^a \rightarrow g_s F_{\mu\nu}^a \equiv {\mathcal F}_{\mu\nu}^a =
\partial_\mu {\mathcal A}_\nu^a - \partial_\nu {\mathcal A}_\mu^a +
f^{abc}{\mathcal A}_\mu^b {\mathcal A}_\nu^c \,, \nonumber
\end{eqnarray}
such that $g_s^2$ multiplies $\hbar$ in the exponent. Here, $f^{abc}$, $(a,b,c)\in \{ 1,\ldots,N^2-1\}$ are the ${\rm SU}(N)$ structure constants. The path integral would be dominated by the classical configurations for $\hbar \rightarrow 0$ (classical limit) which is therefore equivalent to taking the weak coupling limit of the theory $g_s^2 \rightarrow 0$ where the action is large, $S_g \gg \hbar$, and so is the number of quanta in these configurations, $f_g \sim S_g/\hbar$ \cite{Kharzeev:2002np}. There are certain reasons to believe that such classical-field configurations should describe the state of cold nuclear matter in initial stages of ultra-relativistic heavy-ion collisions \cite{McLerran:1993ni, Gelis:2010nm}.

Needless to mention, strongly-interacting QCD exhibits a variety of emergent collective effects and phenomena other then those of QGP that are very difficult to understand and to predict starting from the first-principle microscopic theory of QCD. Observable predictions of the hot/dense QCD theory depend on the equation of state (EoS) of compressed nuclear matter but the latter has not been fully understood yet. This situation is analogical to emergent phenomena in atomic and condensed-matter physics driven by the QED interaction theory at the microscopic level. Notably enough, besides the hadronic and QGP phases, QCD matter features also other distinct phases predicted in various approaches \cite{Braun-Munzinger:2008szb, Fukushima:2010bq}. 
\begin{figure}[hbt] 
\begin{center}
\includegraphics[height=25em]{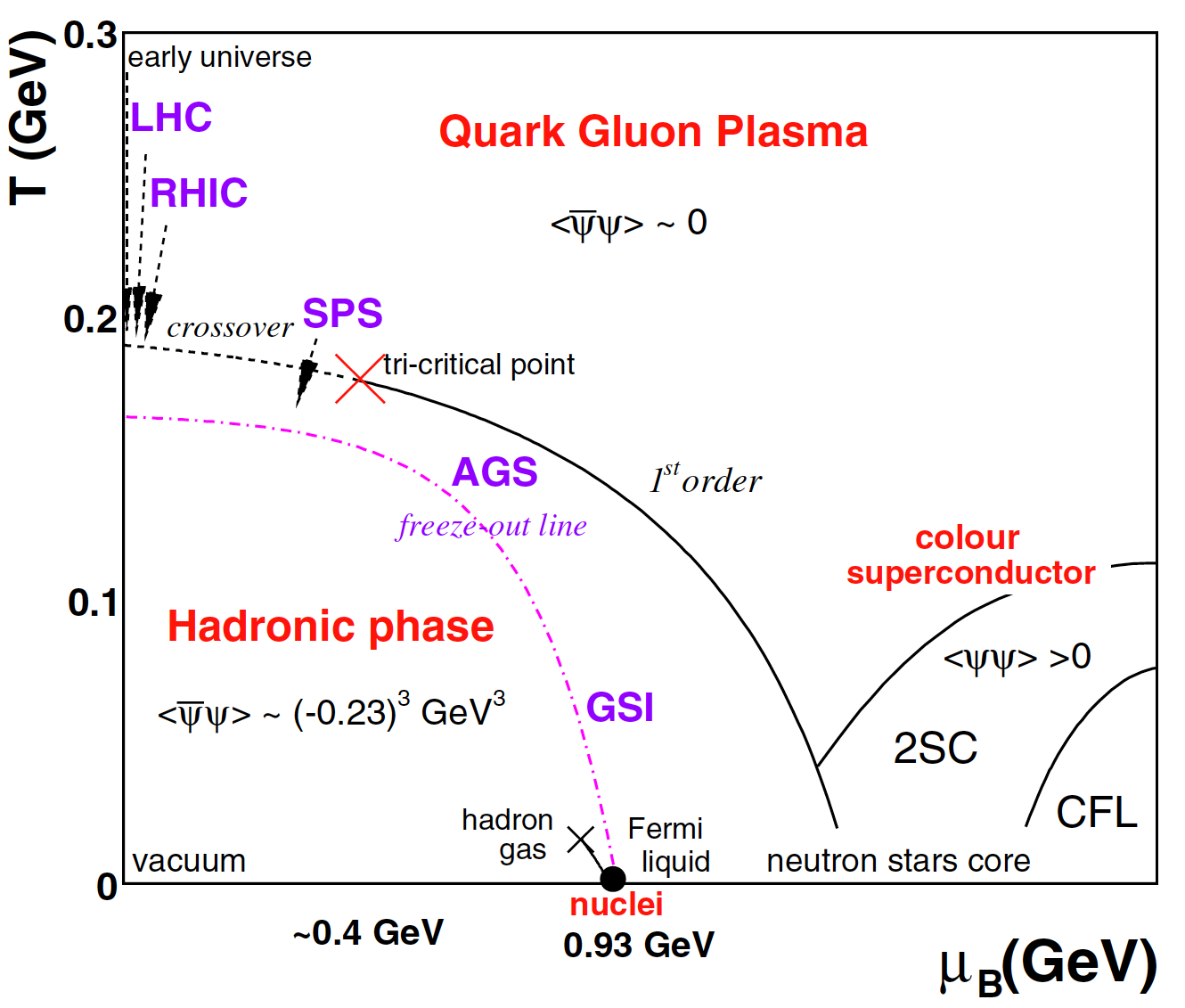}
\caption{An illustration of typical phases of QCD matter that are expected to emerge at various values of the temperature $T$ and the baryonic chemical potential $\mu_B$ associated with $U(1)_{\rm B}$ breaking (see Ref.~\cite{Pasechnik:2016wkt} and references therein). Accelerators operating at different center-of-mass (c.m.) energies are depicted here.}
\label{fig:besqcd}
\end{center}
\end{figure}

Among important examples of various realisations of confining non-abelian gauge-field dynamics in cosmology are the relaxation phenomena in real-time cosmological evolution of the QCD vacuum \cite{Addazi:2018ctp} and a possibility of phase transitions in a ``dark'' strongly-coupled $SU(N)$ gauge sectors \cite{Huang:2020mso}, both potentially testable via detection of stochastic primordial gravitational-wave spectra in future measurements. The homogeneous gluon condensates in the effective $SU(N)$ theory (such QCD gluodynamics) have also been found to play an important role in generation of the observable cosmological constant \cite{Pasechnik:2016twe,Pasechnik:2013poa,Pasechnik:2013sga,Pasechnik:2016wkt,Addazi:2018ctp,Addazi:2019mlo}. For a recent review of implications of the quantum YM vacuum for the Dark Energy problem, see Ref.~\cite{Pasechnik:2016sbh}.

Systematic explorations of QCD matter at high densities and temperatures including the search for critical end point (CEP) in the middle of the phase diagram at $\mu_{\rm B}$ $\sim$ 0.4 GeV shown in Fig.~\ref{fig:besqcd} has started only about ten years ago. The CEP is located at the end of the first-order phase transition boundary between the hadronic phase and QGP, where a second-order phase transition is predicted to occur. One expects a number of new phenomena in a vicinity of that point \cite{Stephanov:1998dy, Gupta:2011wh, STAR:2017sal, Bzdak:2019pkr} that have been searched for by the RHIC Beam Energy Scan program. 

Currently, a number of different studies of QCD phases in various parts of the $(T,\mu_{\rm B})$-diagram are being deployed, both experimentally and theoretically, and a high complexity has started to emerge. Particularly intense are explorations of low $\mu_{\rm B} \!\simeq\!0$ \cite{Bellwied:2015rza, Ding:2016qdj, Bazavov:2017dus, Philipsen:2019rjq} and high $\mu_{\rm B}\!\simeq 100-600$ MeV \cite{Fukushima:2010bq, Gupta:2011wh, Bzdak:2019pkr, Philipsen:2019rjq} domains, with possible transitions in between, also indicated in Fig.~\ref{fig:besqcd}. Other CEPs may also be expected to emerge such as those for chiral (crossover at low-$T$, not shown in the figure) and nuclear liquid-gas (in the nuclear matter ground-state at nearly-zero $T$ and $\mu_{\rm B}$ = 0.93 GeV) transitions.

More specifically, looking at the QCD phase diagram in Fig.~\ref{fig:besqcd} along the direction of increasing baryon chemical potential $\mu_B$, we notice that at energies close to the binding energy of bulk nuclear matter the so-called {\it cold nuclear matter} phase is found. Interactions between nucleons (quark bound states) may lead to pairing and di-baryon condensation that spontaneously break the U$(1)_{\rm B}$ baryon number symmetry (see e.g. Refs.~\cite{Dean:2002zx,Gandolfi:2015jma}). Physically, this also means that the system is in a superfluid (confining) phase. This system has a close analogy, for instance, with liquid helium where one also finds the Bose-Einstein condensation and Goldstone modes, both associated with the superfluidity property. The same physics emerges in ordinary nuclear matter based upon nuclear many-body theory which is applicable at not too large densities.

There is still a substantial lack of knowledge on a transition between the cold nuclear matter and high-density QCD phases, particularly relevant for physics of neutron stars. Since QCD is asymptotically free, one can go to very high $\mu_B$ in the quark-matter phase and employ weak-coupling techniques \cite{Cherman:2020hbe}. In this {\it quark-matter} phase of dense QCD, such calculations predict a nearly Fermi-liquid with residual interactions that lead to pairing among quarks in a gauge-dependent way. This is described by means of a gauge-dependent di-quark condensate $\langle qq \rangle$ playing a role of an order parameter in the dynamical Higgs mechanism such that we deal with a Higgs phase. Indeed, such a di-quark condensate emerges due to long-range attractive forces between the quarks through a Cooper-like pairs' condensation \cite{Barrois:1977xd, Bailin:1983bm}. Such a high-density (baryon) superfluid phase where the $SU(3)_c$ gluon field is fully ``Higgsed'' is known in the literature as a {\it color superconductor}\footnote{Real superconductors have observable phenomena such as persistent currents distinguishing the superconducting phase from a normal one. The name ``color superconductor'' come out rather misleading in a sense that there are no observable persistent color-charge currents associated to this phase \cite{Cherman:2020hbe}} (CSC) (for a comprehensive review on key aspects of dense QCD, see Refs.~\cite{Alford:2007xm,Baym:2017whm}). Formation of such Cooper pairs of quarks can be seen in QCD with three massless $u,d,s$ flavours at large baryon number densities featuring the following color and flavor symmetries' reduction \cite{Alford:1998mk, Fukushima:2010bq}
\begin{eqnarray}
SU(3)_{\text c}\times 
SU(3)_{\text R} \times SU(3)_{\text L} \times 
3U(1)_{\text B} \rightarrow SU(3)_{\rm c+L+R} \times \mathbb{Z}(2)
\end{eqnarray}
down to a diagonal subgroup $SU(3)_{\rm c+L+R}$. The corresponding symmetry transformations involve a simultaneous ``rotation'' of color and flavor group representations known as the color-flavor locking (CFL). Such a CFL phase is known not to be topologically ordered \cite{Cherman:2018jir}. Then, in the CFL quark-matter phase one could also find an order parameter for $U(1)_{\rm B}$ symmetry breaking (down to $\mathbb{Z}_2$) in analogy to the di-baryon condensate in the nuclear-matter phase -- it can be viewed as a cubic power of the di-quark condensate thus being associated with a superfluid flow.

In fact, both quark matter and nuclear matter phases were found to be relevant for the EoS of neutron stars (see e.g. Refs.~\cite{Alford:2002rj,Steiner:2002gx,Baym:2017whm}), and the signatures of possible phase transitions might show up in mass-radii relations for neutron stars and gravitational-wave spectra from neutron star collisions. As at high temperatures no baryon number symmetry breaking occurs, one supposedly crosses the line where U$(1)_{\rm B}$ gets restored when the system heats up. As we noticed above, at low temperatures, both low- and high-density phases have the same order parameter w.r.t. U$(1)_{\rm B}$ breaking, and one of the fundamental open questions is whether a boundary between the quark-matter (Higgs) and nuclear-matter (confinement) phases actually exists. Following Refs.~\cite{Schafer:1998ef,Schafer:1999pb,Schafer:1999jg}, one could consider a simplified picture of pure QCD and include three massless flavors in a maximally symmetric realisation, such that there is no distinction in symmetry realisations between the hadronic phase and asymptotically high-density phase. The latter means there may be no phase transition that is consistent with identical global symmetry realisations in both regimes, t'Hooft anomalies' matching and with smoothly connecting low-lying excitations (see e.g. Refs.~\cite{Schafer:1998ef,Alford:2018mqj,Wan:2019oax}). Such an assumption has become a working one for many phenomenological studies modelling the EoS for neutron star physics (see e.g.~Ref.~\cite{Alford:2019oge} and references therein). Below, following the recent results of Ref.~\cite{Cherman:2020hbe} one may conclude, however, that the Sch\"afer-Wilczek conjecture about quark-hadron continuity at large $\mu_B$ may be largely oversimplified. The reality may be even more complex than what emerges in existing theoretical approaches. The basic problem is that there are no well-justified theoretical methods available for treatment of the strong-coupling regime of QCD, with a non-zero chemical potential, where lattice simulations may not be very reliable.

Finally, yet another QCD phase which is believed to be located somewhere between the chirally restored and confined phases is known as quarkyonic matter \cite{McLerran:2007qj} that may also have some relevance for neutron star physics \cite{McLerran:2018hbz}. In the limit of large number of color charges $N_c$, the gluons' contribution scales as $\sim N_c^2$ compared to that of quarks $\sim N_c$ such that this phase is assumed to have energy densities well beyond $\Lambda_{\rm QCD}^4$. Since gluons are bounded in glueballs, one ends up with $N_c$ DoFs in this phase.

Let us now turn to a discussion of methods of the lattice gauge theory that became the main tool for explorations of non-perturbative physics in gauge theories and, in particular, QCD in the strongly-coupled regime and the associated dynamics of confinement, at least, at not too large chemical potentials.

\section{Ising model and lattice gauge theory}
\label{Sect:Latt}

To what extent one can expect to derive precision results for low-energy observables from the first-principle QCD theory? A default answer to this question is that we should not expect that, at least, analytically. The collective phenomena that are manifest in the strongly-coupled regime of a gauge theory are so complex that none of the existing analytic approaches captures all the relevant dynamics and yields satisfactory results. At the same time, a theory may remain to be correct even if methods of extracting observable information from it are not perfect or suitable. Often though, we start with a simplified model that hopefully captures the same physics as a realistic one but where we have a better control, and then we abstract the lessons that we learn from such a model back to more complicated theories such as QCD.

Luckily, a precise and reliable analysis is possible, but only numerically. The best available framework so far is the lattice gauge theory providing a first-principle numerical approach for strongly-coupled theories like QCD. In fact, this framework is often considered as a ``numerical experiment'' and may be regarded as a black-box whose results need to fit a certain theoretical picture of real underlined physical phenomena and objects providing means to understand those phenomena qualitatively. Whether or not the lattice results fit a particular picture of confinement is an ongoing long-standing debate in the literature. For relatively recent detailed reviews on non-perturbative physics and the confinement problem, see e.g. Refs.~\cite{Shifman:2010jp,Ogilvie:2010vx,Greensite:2011zz,Greensite:2016pfc,Reinhardt:2018roz} and references therein. Here and below, we follow the notation adopted in Ref.~\cite{Greensite:2011zz} unless noted otherwise, acknowledging that the latter reference represents one of the most complete, pedagogical and sophisticated reviews available in the literature on what the confinement problem actually is from various perspectives and approaches.

In order to built a consistent picture of confinement, we need to elaborate on such important notions as ordered and disordered systems. One of the simplest examples of lattice field theory follows the basic principles of statistical mechanics where the most relevant properties of these systems are readily seen in the Ising model of ferromagnetism. For illustration, consider a simple system -- a square ($D=2$), cubic ($D=3$) or hypercubic ($D>3$) array (or {\it lattice}) of atoms, each with two spin states -- in external magnetic field $h$. This system is described by the Hamiltonian,
\begin{equation}
    H=-J\sum_x \sum_{\mu=1}^D s(x)s(x+\mu) - h\sum_x s(x) \,, \quad J>0 \,,
\end{equation}
where $s(x) = +1$ and $-1$ would correspond to an atom at a point $x$ with spin up and down, respectively, and we denote here the total number of spins as $N$. Probability for a specific configuration of spins, $\{s(x)\}$, at a given temperature $T$ can be written as
\begin{equation}
    P_{\{s(x)\}}=\frac{1}{Z} \exp\left[ -\frac{H}{kT} \right]\,, \qquad
    Z = \sum_{\{s(x)\}} \exp[-H/kT] \,.
\end{equation}
In the case of zero external field, $h=0$, the system apparently possesses a {\it global} $\mathbb{Z}_2$ symmetry w.r.t. transformations
\begin{equation}
    s(x) \to s'(x)=\xi s(x)\,, \qquad \xi = \pm 1 \,,
\end{equation}
such that the mean magnetisation (average spin)
\begin{equation}
    \langle s \rangle = \sum_{\{s(x)\}} \frac{P_{\{s(x)\}}}{N} \sum_y s(y)
\end{equation}
vanishes. This is a system in a so-called {\it disordered} state. 

Assume that the spins in the initial state are aligned. The exact $\mathbb{Z}_2$ symmetry means that at any given temperature any finite system would end up in a disordered state provided that one waits for long enough for that to occur. This leads to non-existence of permanent magnets as any alignment of the spins would be destroyed by thermal fluctuations. However, for large $N$, i.e. for macroscopic magnets, the time between sizable fluctuations that could flip a lot of spins would grow exponentially and eventually exceeds the lifetime of the Universe. For non-zero $h$, however, the $\mathbb{Z}_2$ symmetry appears to be explicitly broken enabling $\langle s \rangle \not=0$ at any temperature. In this case, the system appears to be in an {\it ordered} state where a large amount of spins point in the same direction.

Now, consider magnetisation of a large system in the limit of vanishing $h$. One could show that, in general, this quantity is non-vanishing
\begin{equation}
    \lim_{h\to 0} \lim_{N\to \infty} \langle s \rangle \not=0 \,,
\end{equation}
yielding the so-called {\it spontaneous symmetry breaking} (SSB) of the global $\mathbb{Z}_2$ symmetry which occurs particularly at low temperatures (ordered state). A global symmetry is said to be broken spontaneously when the Hamiltonian and the corresponding equations of motion are symmetric but the solutions for physical observables (such as the magnetisation introduced above) are not. At high $T$ above a certain critical temperature (Curie temperature), the averaged spin vanishes, and the spin system appears again in a symmetric (disordered) state. Considering the vacuum expectation value (VEV) of a product of two spins we notice, $G(r)\equiv \langle s(0)s(r) \rangle \sim \exp(-r/l)$, i.e. it falls off exponentially with the distance between atoms $r$ in a disordered state, where $l$ is the correlation length. There is a phase transition between the ordered and disordered phases of the system at the Curie temperature for any $D>1$, while for $D=1$ the system is in a disordered phase at any $T$. The existence of such phase transitions associated with a global symmetry breaking is a generic property of many different systems and is also manifest in strongly-coupled gauge theories as will be discussed below.

Let us further promote the global $\mathbb{Z}_2$ symmetry to a local one whose transformation parameter depends on position of the associated DoFs, $\xi(x)=\pm 1$, and can be chosen independently at each site ({\it gauge transformations}). For this purpose, let us consider the links of the lattice $s_\mu(x)$ along each dimension $\mu=1\dots D$ as dynamical DoFs subjected to the gauge transformation
\begin{equation}
    s_\mu(x) \to \xi(x) s_\mu(x) \xi(x+\hat \mu) \,,
\end{equation}
and write down the Hamiltonian of the gauge-invariant Ising model
\begin{equation}
    H=-J\sum_x \sum_{\mu = 1}^{D-1} \sum_{\nu > \mu}^D 
    s_\mu(x) s_\nu(x+\hat \mu) s_\mu(x+\hat \nu) s_\nu(x) \,.
    \label{Hamil}
\end{equation}
Thereby we arrive at the simplest example of the $\mathbb{Z}_2$ lattice gauge theory. In order to describe such systems, one considers observables that are invariant under gauge transformations. A particularly important class of observables can be obtained by taking the VEV of the so-called {\it Wilson loop} -- a product of links on the lattice around a given closed contour $C$ \cite{Wegner:1971app},
\begin{equation}
    W(C)=\left\langle \Pi_{(x,\mu)\subset C} s_{\mu}(x) \right\rangle \,.
\end{equation}
The Hamiltonian (\ref{Hamil}) is given by the simplest Wilson loop given by a plaquette, the minimal closed loop on the lattice.

In analogy to the gauged Ising model, in a generic lattice gauge theory described by a certain (discrete or continuous) gauge group $G$, one starts with the Euclidean action where the link variables are the elements of the gauge group. For instance, in the case of a non-abelian group $G\equiv$ $SU(2)$ the group elements in discretized spacetime are
\begin{equation}
    U_\mu(x) = e^{iag A_\mu(x)} \,, \qquad A_\mu(x)=\frac{1}{2} \sigma^a A^a_\mu(x) \,,
\end{equation}
in terms of the lattice spacing $a$, the gauge coupling $g$, the Pauli spin matrices $\sigma^a$, $a=1,2,3$, and the $SU(2)$ gauge field $A^a_\mu(x)$. By convention, the link variable $U_\mu(x)$ is associated with a line running from site $x$ on the lattice to a neighbor site $x+\hat\mu$ in the positive direction $\mu$. The probability distribution of lattice configurations of the gauge field is found in full analogy to that of the Ising model, namely,
\begin{equation}
    P_{\{s(x)\}}=\frac{1}{Z} \exp(-S[U])\,,
\end{equation}
where the Euclidean action, also known as the Wilson action,
\begin{equation}
    S[U]=-\frac{\beta}{2} \sum_{x,\mu < \nu} {\rm Tr}[ U_\mu(x)U_\nu(x+\hat\mu)U^\dagger_\mu(x+\hat\nu)U^\dagger_\nu(x)]
\end{equation}
is invariant under local gauge transformations
\begin{equation}
    U_\mu(x) \to G(x) U_\mu(x) G^\dagger(x+\hat\mu) \,, \qquad G(x) \subset {\rm SU}(2) \,.
\end{equation}
We used the fact that the trace of any $SU(2)$ group element is real. A straightforward extension to $SU(N)$ gauge theory leads to
\begin{equation}
    S[U]=-\frac{\beta}{2N} \sum_{x,\mu < \nu} 
    \Big\{ {\rm Tr}[U_\mu(x)U_\nu(x+\hat\mu)U^\dagger_\mu(x+\hat\nu)U^\dagger_\nu(x)] + {\rm c.c.} \Big\} \,,
\label{pure-SUN}
\end{equation}
with suitably generalised group elements $U_\mu(x)$. 

Expanding the latter in powers of $A^a_\mu(x)$, taking $\beta=2N/g^2$ and turning to the continuum limit of vanishing lattice spacing $a\to 0$, one arrives at the standard expressions for the action and gauge transformations in Euclidean spacetime
\begin{eqnarray}
    && S=\frac{1}{2} \int d^4x {\rm Tr}[F_{\mu\nu}F_{\mu\nu}]\,, \qquad 
    F_{\mu\nu} = \partial_\mu A_\nu - \partial_\nu A_\mu - ig[A_\mu,A_\nu] \,, \\
    && A_\mu(x) \to G(x) A_\mu(x) G^\dagger(x) - \frac{i}{g} G(x)\partial_\mu G^\dagger(x) \,,
\end{eqnarray}
in terms of the field strength tensor $F_{\mu\nu}$ and a gauge group element $G(x)$. Here, the repeated indices are summed over as usual. 

The formulation of the lattice gauge theory in Euclidean spacetime has quickly become the cornerstone and the main reference for numerical analysis of basic characteristics of the corresponding quantum field theory (QFT) in Minkowski spacetime (such as its low lying spectrum and the static potential). This is due to a single most important fact that the Euclidean formulation of the field theory is conveniently considered as a statistical (not quantum) system whose analysis can be performed using the power of the lattice Monte Carlo methods. For a detailed description of these methods, see e.g. Ref.~\cite{DeGrand:2006zz}. 

The Euclidean formulation is particularly designed for studies of QFT at finite temperatures in equilibrium and works in Euclidean space with periodic time direction for bosonic fields while fermion fields fulfill antiperiodic boundary conditions in the time direction (for a recent review, see e.g. Refs.~\cite{Ghiglieri:2020dpq, Lundberg:2020mwu}). A finite $T$ theory is then constructed from its zero-temperature counterpart by replacing bosonic and fermionic four-momenta $k^{\mu}$ in Euclidean integrals by $2\pi n T$ and $(2 n + 1)\pi T$, respectively, and then switching from $k^{\mu}$ integration to summation over $n$. In a hot medium, an average momentum transfer is the given in terms of temperature, $Q=2\pi T$. The study of thermodynamics and phase transitions is performed in the Hamiltonian formalism starting from the thermal partition function, and the ``time'' is Euclidean in the path integral formalism from the beginning, at any temperature. The order of deconfinement phase transition in Euclidean $SU(3)$ lattice gauge theory has been studied in this approach by Monte Carlo methods in Ref.~\cite{Celik:1983wz}. 

In the continuum limit, in order to obtain the Minkowski action of the corresponding QFT starting from the thermal theory action in Euclidean spacetime one conventionally adopts the Wick rotation $t\to -it$ and $A_0\to iA_0$ relying on analyticity property of the vector-potential. Then, an assumption that a numerical simulation successfully set up in Euclidean spacetime yields relevant results to the corresponding QFT in Minkowski spacetime would be justified only for smooth transitions between short-distance to long-distance physics enabling analytic (in physical time and in $A_0$) continuations of amplitudes from Minkowski to Euclidean spacetime, and backwards. Indeed, such an assumption is violated in the most general case as stated by the so-called Maiani-Testa no-go theorem \cite{Maiani:1990ca} related to the ``failure'' of the Wick rotation mentioned above. Indeed, when going out from thermodynamics approaching the study of bound states, the Wick rotation is applicable only to compute static characteristics of the QCD medium such as vacuum condensates as well as masses of stable particles which are the minority of the QCD spectrum. Resonances such as the majority of mesons, charmed and stranged baryons, tetraquarks, pentaquarks, and hadron molecules are accessible in Euclidean space only indirectly and only under restrictive assumptions. For more details on the associated problems in treatment of two-particle systems, see Ref.~\cite{Luscher:1990ux}, while a review on the status of three-particle systems can be found e.g. in Ref.~\cite{Hansen:2019nir}. 

A manifestation of non-analytic structures (domain walls) in the YM vacuum in physical time has also been discussed recently in the context of the non-stationary background of expanding Universe in Ref.~\cite{Addazi:2019mlo}. Such structures were found as attractor cosmological solutions at sufficiently large physical times asymptotically matching the YM dynamics on the Minkowski background. In the essence of Maiani-Testa theorem, such non-analytic (domain-wall) solutions found in (nearly) Minkowski background would in general not match the corresponding lattice simulations in Euclidean spacetime, so their implications for confinement are yet unclear and should be studied separately. As long as such solutions are concerned, one may conjecture that the Euclidean YM field theory predictions match those in Minkowski spacetime only in regions sufficiently far away from the non-analytic phase boundaries. This conjecture however requires further in-depth studies of implications of these novel solutions for confinement dynamics.

Another crucial limitation of Monte-Carlo lattice simulations concerns the thermal gauge theory with non-vanishing chemical potential. Indeed, the action becomes complex if the temperature $T$ and the chemical potential $\mu$ are both non-zero, meaning that standard Monte-Carlo methods fail in this case (for a thorough review on this issue, see e.g.~Ref.~\cite{Aarts:2015tyj}). In particular, due to the sign problem the lattice simulations of QCD at $\mu_B > 0$  exhibit difficulties in reproducing the quark-gluon plasma as observed in heavy-ion collisions, even under an assumption of thermal equilibrium. The situation becomes even worse when considering the nuclear matter in neutron stars or collapsing black holes at very large density in the curved spacetime. The way to proceed is to expand the pressure in $\mu_B/T$ and calculate the physical observables as Taylor expansions in this quantity, see e.g.~Ref.~\cite{Bollweg:2020yum}. In practice, this requires calculating operators of high order, which are noisy and require very large statistics \cite{Bazavov:2019lgz}. Recently an alternative summation scheme for the equation of state of QCD at finite real chemical potential was proposed in \cite{Borsanyi:2021sxv}, designed to overcome those shortcomings. 
Using simulations at zero and imaginary chemical potentials the extracted LO and NLO parameters describing the chemical potential dependence of the baryon density were extrapolated to large real chemical potentials. Proposed expansion scheme converges faster than the Taylor series at finite density, thus, leading to an unprecedented coverage up to $\mu_B/T\le 3.5$ and to more precise results for the thermodynamic observables.

\section{Asymptotic behavior of large Wilson loop VEVs}
\label{Sect:Wilson}

Different phases of a gauge theory are classified based on the behavior of Wilson loop VEVs at large Euclidean times compared to spacial separations, i.e. $T_{\rm E}\gg R$. Computing those in Euclidean spacetime provides a direct access to the interaction energy between the static field sources in Minkowski QFT when the mass of the sources (and hence the fundamental energy scale of a confining gauge theory) is taken to infinity. Introducing a massive scalar field (a ``scalar quark'') in an arbitrary representation $r$ to the gauge theory on $D$-dimensional lattice, the corresponding action
\begin{equation}
    S = -\frac{\beta}{N} \sum_p {\rm ReTr}[U(p)] - \gamma \sum_{x,\mu} (\phi^\dagger(x) U^{(r)}_\mu(x) \phi^\dagger(x+\hat\mu) + {\rm c.c.}) + \sum_x(m^2+2D)\phi^\dagger(x) \phi(x)
    \label{action}
\end{equation}
is invariant under gauge transformation of the scalar field: $\phi(x)\to G(x) \phi(x)$, where the link variable is $U^{(r)}_\mu(x)$, and the gauge-field holonomy is $U(p)$ for a given plaquette $p$. 

Consider an operator that creates a particle-antiparticle pair in a color-singlet state at a given time $T_{\rm E}$ and separation $R$,
\begin{equation}
    {\cal C}(T_{\rm E}) = \phi^\dagger(0,T_{\rm E}) 
    \Big[ \Pi_{n=0}^{R-1} U_i^{(r)}(n\hat i,T_{\rm E}) \Big]
    \phi(R\hat i,T_{\rm E}) \,,
\end{equation}
that also creates a color-electric flux tube (or string) stretched between the charges. In the limit of heavy static color-charged sources, $m\gg 1$ in lattice units, the second term in Eq.~(\ref{action}) may be considered as a small perturbation, so the string-breaking effect can be neglected to a first approximation. Indeed, as matter fields are very heavy in this limit, it would take an infinite energy to pull them out of the vacuum and to place them on mass shell in order for them to bind to the sources and hence to screen their charge. This means that one would stretch the flux tube to an infinite length before it can ever break apart, which is of course an unrealistic but still useful picture to test the confinement property of the quantum vacuum.

Thus, by integrating out $\phi$ in the functional integral one finds for the VEV
\begin{equation}
    \langle {\cal C}(T_{\rm E})^\dagger {\cal C}(0) \rangle \sim W_r(R,T_{\rm E}) \,,
\end{equation}
to the leading order in $1/m^2$ expansion, where
\begin{equation}
    W_r(R,T_{\rm E}) = \langle {\rm Tr}[U^{(r)}U^{(r)}\dots U^{(r)}]_C \rangle \equiv \langle \chi_r[U(R,T_{\rm E})] \rangle
\end{equation}
is the VEV of the Wilson loop written in terms of the time-like holonomy $U(R,T_{\rm E})$ of the pure gauge theory. Here, the link variables run counterclockwise on a time-like rectangular contour $C=R\times T_{\rm E}$, the group character is $\chi_r$ and the sum runs over states with two static charges. In the continuum limit, the corresponding holonomy is given by the path-ordered exponential
\begin{eqnarray}
    U(C)= P \exp\left[ ig \oint_C dx^\mu A_\mu(x) \right] \,.
\end{eqnarray}
So the Wilson loop (holonomy) operator in this case represents a rectangular time-like loop describing the creation, propagation and, finally, destruction of two static quark and antiquark placed at certain fixed spacial points. The time-like links in a given Wilson loop, hence, can be considered as the worldlines of static heavy charges.

On the other hand, in the operator formalism one deduces that \cite{Greensite:2011zz}
\begin{equation}
    \langle {\cal C}(T_{\rm E})^\dagger {\cal C}(0) \rangle \propto \sum_n |c_n|^2 e^{\Delta E_n T_{\rm E}} \sim e^{-\Delta E_{\rm min}T_{\rm E}} \,, \qquad T_{\rm E}\to \infty \,,
\end{equation}
where $\Delta E_n$ is the energy of the $n$th excited state above the vacuum, and in the last part of this relation only the dominant contribution (at large $T_{\rm E}$) from the minimum-energy eigenstate has been taken into account. In this case, $\Delta E_{\rm min}=V_r(R)$ corresponds to the energy difference between two static charges, being in other words the interaction (static) potential between them $V_r(R)$. Hence, the VEV of the rectangular Wilson loop
\begin{equation}
    W_r(R,T_{\rm E}) \sim e^{-V_r(R)T_{\rm E}}
\end{equation}
is characterised by the potential $V(R)$, which can be inverted as
\begin{eqnarray}
    V_r(R) = \lim_{T_{\rm E}\to \infty} 
    \log \Big[ \frac{W_r(R,T_{\rm E}+1)}{W_r(R,T_{\rm E})} \Big] \,.
\end{eqnarray}

Now consider, for instance, a planar non self-intersecting Wilson loop in $U(1)$ gauge theory, and using the Stokes law, it can be written as
\begin{equation}
    U(C)=\exp \Big[ ie\oint_C dx^k A_k(x) \Big] = \exp \Big[ ie\int_C dS_C F_{ij}(x) \Big] \,,
\end{equation}
where the areal integration represents the magnetic flux and proceeds through the minimal area of the large Wilson loop. Thus, due to additive nature of the flux, such a planar Wilson loop can be arbitrarily split into a product of smaller loops whose areas add up to the one of the large loop
\begin{equation}
    U(C)=\Pi_{i=1}^n U(C_i) \,.
\end{equation}
Here, the orientations of the smaller loops are chosen in such a way that neighboring contours run in opposite directions to each other. In the case of {\it magnetic disorder}, the magnetic fluxes through smaller loops $C_i$ (e.g. plaquette variables, in the case of smallest loops) are completely uncorrelated, such that the VEV factorises as
\begin{eqnarray}
    W_r(C) \equiv \langle U(C)\rangle = \Pi_{i=1}^n \langle U(C_i)\rangle = \exp[-\sigma_r A(C)] \,, \qquad \sigma_r = - \frac{\ln \langle U(C_i)\rangle}{A'} \,,
\end{eqnarray}
where $A$ and $A'$ are the larger and smaller Wilson loop areas, respectively.
\begin{figure}[hbt]
\begin{center}
\includegraphics[height=15em]{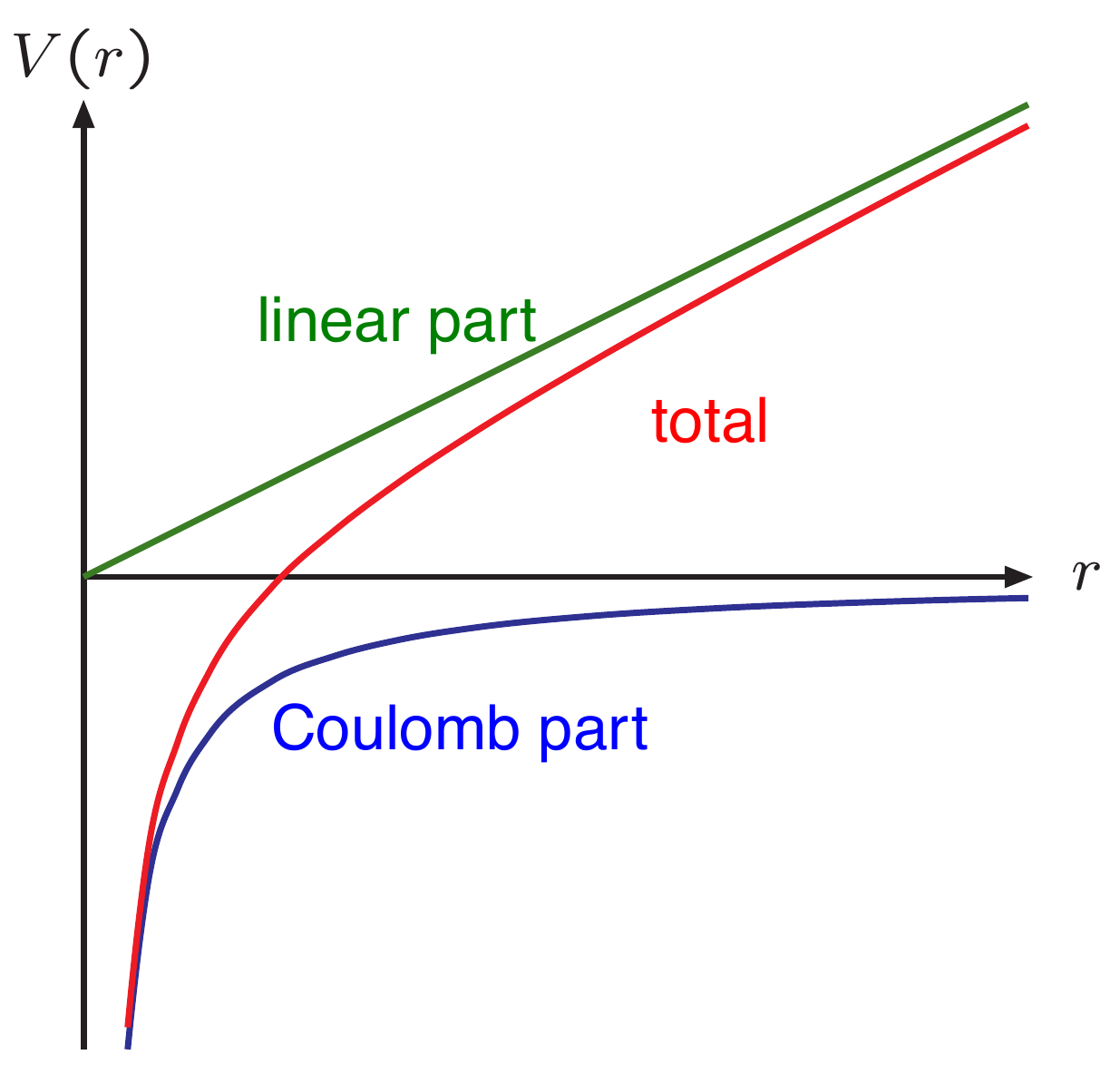}
\caption{An illustration of the total static quark potential as a function of interquark separation.
}
\label{fig:Vr}
\end{center}
\end{figure}

Assuming the absence of light matter fields that could, in principle, screen the color charge of the massive sources, and considering a rectangular Wilson loop with $C=R \times T_{\rm E}$, the magnetically disordered state is characterised by the linear growth of the interaction potential with distance $R$ between the static charges asymptotically,
\begin{equation}
    V_r(R)=\sigma_r R + 2V_0 \,,
\end{equation}
hence, represents a potential of a linear string. Here, $V_0$ is interpreted as a self-energy contribution, and $\sigma_r$ has the meaning of the string tension in a given group representation $r$ that does not depend on the subloop area $A'$. For an illustration of the total potential interpolating small-$R$ (Coulomb) and large-$R$ (confining) regimes, see Fig.~\ref{fig:Vr}. The area-law for the Wilson loop VEV $\sim \exp[-\sigma_r\, RT_{\rm E} - 2V_0T_{\rm E}]$ is then reproduced for $T_{\rm E}\gg R$ as expected, or for a generic contour enclosing a large minimal area $A(C)$,
\begin{equation}
    W_r(C) \sim \exp[-\sigma_r\,A(C) - V_0\,P(C)] \,,
\end{equation}
including also a dependence on the perimeter of the contour $P(C)$. Note, the gluon propagator is singular in the UV regime in the continuum limit which generically induces a singular term that is interpreted as a divergent self-energy $V_0$ of the charged particles and antiparticles propagating in the loop. The latter produces a perimeter-law contribution to the large Wilson-loop VEV in the above expression. So, usually a kind of smearing of the loop via a superposition of nearby loops is required to regularise the Wilson loop in the continuum limit (see e.g. Ref.~\cite{Narayanan:2006rf}), while on the lattice such a short-distance regularisation is always implicit.

It is straightforward to show that for any gauge group and $D=2$, only magnetically disordered phase is realised, reproducing the area-law falloff due to the absence of a Bianchi constraint on the components of the field strength tensor \cite{Halpern:1978ik}. It is, however, a much harder problem to prove the area-law falloff of large Wilson loop VEVs in a generic YM theory with a non-trivial center symmetry which represents the basic confinement problem (for more details, see below). A remarkable property of a Wilson loop is that it characterises vacuum fluctuations of the gauge field, i.e. without the presence of any external sources,
\begin{equation}
    W_r(C)=\langle \Psi_0 | \chi_r[U(C)] | \Psi_0 \rangle \,,
\end{equation}
with a spacelike loop $C$, in terms of the ground-state $\Psi_0$ of the Hamiltonian of the pure gauge theory. As the spacelike and timelike loops are related by a Lorentz transformation, one deduces that the potential energy of interaction between static charges is directly connected to the gauge-field vacuum fluctuations in the absence of color-charged sources.
    
In $D>2$ lattice, the Bianchi constraint emerges that correlates the field strength values at neighbor sites so that those no longer fluctuate independently from one point to another \cite{Batrouni:1984rb}. The absence of those correlations among the smallest Wilson loops, the plaquette variables, is the single most important requirement that provides the area-law relation for Wilson loops of arbitrary sizes. For $D>2$ such correlations disappear and the area-law is established in the strong-coupling limit only, i.e. in the leading order in $\beta\ll 1$. In the weakly-coupled regime $\beta\gg 1$ in $D=3+1$ electrodynamics, this property does not hold and one recovers the massless phase instead, with the potential \cite{Intriligator:1995er}
\begin{equation}
    V(R)=-\frac{g^2(R)}{R} + 2V_0 \,,
\end{equation}
corresponding to a perimeter-law falloff of the Wilson loop VEV,
\begin{equation}
    W(C) \sim \exp[- V_0\,P(C)] \,,
\end{equation}
where $P(C)=2T_{\rm E}$ for a rectangular loop $C$ (with $T_{\rm E}\gg R$), while the coupling $g(R)$ is a slow function of $R$ that approaches a constant in the Coulomb phase. In non-abelian theories the magnetically disorder phase has been established for sufficiently large Wilson loops using the non-abelian Stokes law (see e.g.~Refs.~\cite{Arefeva:1979dp,Fishbane:1980eq,Diakonov:1989fc,Karp:1999vq,Hirayama:1999ar,Diakonov:2000kw,Kondo:2000pp,Kondo:1999tj}), and also employing a finite-range behavior of field strength correlators \cite{DiGiacomo:2000irz,Kuzmenko:2004hk}. Let us now briefly discuss one of the most distinctive features of long-range dynamics of QCD associated with Regge trajectories.

\section{Regge trajectories and QCD strings}
\label{Sect:Regge}

We have seen that the magnetic disorder phase manifests itself through linear dependence of the static potential, and this behavior is inherent to that of a string. What is the nature of such a ``color string'', and how is it formed? Which phenomenological implications such strings may have?

In hadronic scattering processes, the $t$-channel exchanges of QCD resonances are considered to be important at high energies. As suggested by quantum mechanics, a given scattering amplitude can be represented as a series expansion in partial waves,
\begin{equation}
    A(k,\cos\theta)=\sum_{l=0}^\infty (2l+1)a_l(k) P_l(\cos\theta) \,,
\end{equation}
in terms of the Legendre polynomials of the first kind and of order $l$, $P_l(\cos\theta)$, the scattering angle $\theta$ and the partial wave amplitudes $a_l$. For a $2\to 2$ process and particles of equal mass, for instance,
\begin{equation}
    \cos\theta = 1 + \frac{2s}{t-4m^2} \,.
\end{equation}
Considering an exchange of a single resonance only, with spin $l_0$ and at large $s\to \infty$, the amplitude behaves as $A(s,t)\propto s^{l_0}$, such that by means of the optical theorem, the corresponding total cross section, $\sigma_{\rm tot}\propto s^{l_0-1}$. This result does not work very well against the experimental data for an integer value of $l_0$. The way out is to adopt that there are several resonances being exchanged in the $t$-channel that should all be taken into account. This is consistently done in the formalism of the Regge theory operating with an analytical continuation of partial amplitudes $a_l$ to the complex angular momentum plane (for a thorough discussion of Regge theory principles and applications, see e.g. Ref.~\cite{Collins:1977jy}). The poles in this plane are traced out by straight lines known as Regge trajectories, $l=\alpha(t)$, and are associated with particles. The squared mass of an exchanged resonance with spin $l$ corresponds to those $t$ at which $l$ is an integer. As a result of the Regge theory, the asymptotic energy dependence of the scattering amplitude reads
\begin{eqnarray}
    A(s,t) \to \beta(t)s^{\alpha(t)} \,, \qquad s\to \infty \,.
\end{eqnarray}
As a striking feature of QCD that has not been observed e.g. in the electroweak (EW) theory, the Regge trajectories appear to be almost linear functions,
\begin{equation}
    \alpha(t) = \alpha(0) + \alpha' t \,,
\end{equation}
and one of the big questions is which dynamics could provide such a simple behavior confirmed experimentally. Namely, hadrons of a given flavour quantum number appear to lie at almost parallel Regge trajectories. 

It is clear that such a behaviour must be specific to confining dynamics of QCD. Apparently, the potential that binds the quark and anti-quark together into a meson and rises with the interquark separation linearly should be responsible for such a behaviour. One adopts the physical picture of a string stretched between $q$ and $\bar q$ as a narrow color-electric flux tube, which carries the energy $E=\sigma r$, so that one can neglect the quark masses. Considering for simplicity the leading Regge trajectory that maximises $l$ at a given $t$, the flux tube of length $r$ rotates about its center such that its end points move with the speed of light, and
\begin{equation}
    \sqrt{t}=\int^{r/2}_{-r/2} dx \frac{\sigma}{\sqrt{1-v_{\perp}^2}} =\frac{\pi r \sigma}{2} \,,
\end{equation}
in terms of the string tension $\sigma$, and the transverse velocity $v_{\perp}=2x/r$. Analogously, the angular momentum of such a system
\begin{equation}
    l=\int^{r/2}_{-r/2} dx \frac{\sigma v_{\perp} x}
    {\sqrt{1-v_{\perp}^2}} = \frac{\pi r^2 \sigma}{8} \,,
\end{equation}
providing us finally with the Regge slope $l/t=1/2\pi\sigma\equiv \alpha'={\rm const}$. The latter can be extracted by fitting to the experimental data $\alpha'\simeq 0.9$ GeV$^{-2}$ yielding the string tension value of $\sigma \simeq 0.18$ GeV$^2$ = 0.91 GeV$/$fm.

The fundamental question is how non-local string-like objects emerge from the local microscopic parton (quark and gluon) dynamics in QCD. For some peculiar reasons, the gluon field between a static quark and anti-quark gets ``squeezed'' into a narrow cylindrical domain whose transverse area is nearly independent on the interquark distance -- the main effect of the magnetic disorder phase. In a color electric flux tube picture, the energy stored in such a QCD string is proportional to the string tension $\sigma$ that can be found in terms of the color electric field $E^a_i\equiv F_{0k}^a$ as an integral over the transverse area of the flux tube as \cite{Greensite:2011zz}
\begin{eqnarray}
    \sigma = \frac12 \int d^2 y_\perp\, (E^a_i(y))^2 \,.
\end{eqnarray}
Such a string then wildly fluctuates in transverse directions, and the energy of such fluctuations tends to grow with the distance between the static sources. At some critical distance, the strong fluctuations destabilise the flux tube making the longer strings less energetically favourable than the shorter ones. So, instead of indefinitely (and linearly) rising energy stored in a flux tube with its length, one encounters a string breaking effect realised due to the presence of quarks in QCD or, in a general YM theory, matter fields in fundamental representation of the gauge group. Let us elaborate on this point in some more details in what follows.

\section{Color confinement and Higgs-confinement complementarity}
\label{Sect:Higgs-conf}

A traditional and rather generic question one may ask here is what we actually mean by confinement in a gauge theory with and without matter fields that transform in the fundamental representation of the gauge group. As was discussed above, in pure non-abelian gauge theories without dynamical matter fields, the existing attempts to prove confinement consist in demonstrating the area-law dependence of $W(C)$, or equivalently, in showing linear dependence of the static quark potential at large separations\footnote{One should make a side remark here: considering static charges in fundamental representation, with a non-zero coupling to the gauge field in the action, automatically implies that the theory is {\it not} a pure non-abelian gauge theory. Obviously, a pure gauge theory feature neither ``static'' nor ``dynamic'' quark fields, and moreover as such the latter are not distinguished by the action unless the static ones are made very much heavier than the dynamic ones. So any statements about the linear static potential in pure non-abelian gauge theories should be taken with reservations and does make sense only when taking a limit of heavy (static) matter fields that can be effectively integrated out in the corresponding path integral of the theory. However, the latter procedure, formally, eliminates such heavy charges from asymptotic states of the resulting EFT entirely making it impossible to use them as probes for vacuum dynamics and hence confinement in pure gauge theories. So, in practice, one does not eliminate them from the asymptotic states of a gauge theory but rather retains them as heavy sources but with a finite mass.}. As we will elaborate in more formal details below, confinement in a pure YM theory is associated with an unbroken center symmetry. Thus, the non-perturbative vacuum of QCD or, in general, a non-abelian gauge theory in the range of length-scales where the static potential satisfies a linearly-rising behavior is considered to be in a {\it confined phase}.

In the presence of dynamical quarks in the theory, there would actually be no a linear static potential between heavy test quarks at asymptotically large $R$. Indeed, if one attempts to pull them apart one eventually observes pair creation (out of the vacuum), thus, ending up with formation of mesons at very large distances. In this picture, such a dynamical quark-antiquark pair creation occurs at the ends of the two shorter strings at the breaking point of the larger one such that the color charge of the static charges gets effectively screened off. Such a {\it string breaking} or fragmentation phenomenon in QCD causes flattening out of the static quark potential at large distances in consistency with the Regge trajectories of QCD and with vast phenomenology of particle physics processes with hadronic final states. Such a picture has become the cornerstone of the hadronisation modelling when long strings loose their stability and decay into shorter strings yielding the spree of hadrons measurable by experiments at long distances. As we will discuss more later on, no exact center symmetry can be found in such a theory since it generically gets broken by the presence of matter in fundamental gauge-group representation. There are reasons to expect a finite range in intermediate distances where the potential could be seen as approximately linear and hence string-like. So even as confinement is unquestionably useful way of thinking about long-range physics of QCD, it is by far a more complex phenomenon than an assumption about an asymptotically linear static potential associated with unbroken center symmetry.

The phenomenological reality is that coloured quarks and anti-quarks at long distances are always bind together into composite states -- mesons and baryons -- and do not exist as isolated color charges. This is realised in an effective string-based hadronisation picture that is proven to work very well phenomenologically in a variety of high-energy scattering processes with hadron final states (see below). The corresponding dynamics has been studied in lattice gauge theory simulations in the strong-coupling regime when matter fields are present in the action \cite{Philipsen:1998de,Duncan:2000kr,Bernard:2001tz}. The resulting hadrons are automatically color-neutral and are the true asymptotic states of QCD, not the colored quarks and gluons. Hence, sometimes QCD confinement is naively identified with {\it color confinement} (known also as $C$-confinement) due to the color charge being effectively screened away at large distances by dynamical matter fields such that the colored partons may only propagate at short distances. However, one must be a little more careful with such an identification. If color confinement were the only property of the confining phase, than typical Higgs theories (such as the weak interactions' theory in the SM) should also be considered as confining \cite{Greensite:2011zz} although they do not feature such phenomena as flux tube formation and Regge trajectories \cite{Frohlich:1981yi,Fradkin:1978dv}. This is why ``true confinement'' appears to be a more complex phenomenon and, in addition to $C$-confinement, it should also be connected to other distinct properties of the quantum ground state such as magnetic disorder associated with an unbroken global symmetry \cite{Greensite:2011zz}. It does appear indeed rather obvious that $C$-confinement always accompanies the magnetic disorder phase while the opposite may not necessarily be always true \cite{Greensite:2017ajx}.

Indeed, consider an even simpler $SU(2)$-invariant gauge-Higgs theory \cite{Lang:1981qg}, with Yukawa-type interaction term that can be straightforwardly deduced from Eq.~(\ref{action}). Here, the confinement regime is reproduced for small $\beta,\gamma\ll 1$ characterised by the linear rise of the static potential followed by its flattening at large separations due to string breaking. So, this regime is very similar to the long-range dynamics of real QCD. However, at large values of $\beta,\gamma\gg 1$ one enters the Higgs regime characterised by the presence of massive vector bosons, analogues to those in the EW theory. This is the so-called {\it massive phase} characterised by a Yukawa-type potential for $T_{\rm E}\gg R$
\begin{equation}
    V(R)=-g^2\frac{e^{-mR}}{R} + 2V_0 \,,
\end{equation}
corresponding to a perimeter-law for a generic large planar loop $C$, $W(C)\sim \exp[-V_0 P(C)]$ with $R\gg 1/m$. In fact, in both confinement and Higgs (massive) regimes the color field is not detectable far from its source. Indeed, while in the confinement regime there are only color-singlets in the physical spectrum of this theory, in the Higgs regime the gauge forces are the short-range ones, such that one charge screening mechanism transforms into another as the couplings change. This is due to the fact the gauge-invariant operators in $SU(2)$ theory that create color-singlet states in the confinement domain are also responsible for creation of massive vector bosons in the Higgs domain (for a first discussion on role of the EW theory operators for generation of particle spectra, see e.g. Ref.~\cite{Frohlich:1981yi}), and those states evolve into each other with varying the model parameters. Whether this happens continuously or via a first-order phase transition is a subject of ongoing research in the literature, to be discussed below.

Referring to the EW theory as a particularly important example one should be also very careful about what one actually means by the Higgs phase and the associated Higgs mechanism. Conventionally, the Higgs phase is described in terms of a Mexican-hat shape potential emerging due to formation of classical scalar fields' (Higgs) condensates in a weakly coupled regime and, as a cause, leading to spontaneous breaking of a given symmetry. While the gauge symmetry is manifest at the Lagrangian level, due to its spontaneous breakdown by means of the Higgs condensate, it is not a symmetry of solutions of the corresponding equations of motion. Note, however, that it is meaningless to talk about spontaneous breaking of a gauge symmetry without specifying a certain gauge-fixing condition. Indeed, the Higgs vacuum VEV depends on the gauge choice that we make in practical calculations and can be fixed to any value by an appropriate choice of the gauge while the actual physical observables and physical states must be gauge-invariant and do not depend on this choice. The gauge symmetry SSB phase cannot be regarded as a true physical system provided that the gauge symmetries are redundancies of description and cannot actually break spontaneously. The latter is the statement of the so-called Elitzur's theorem \cite{Elitzur:1975im}. Indeed, according to this theorem, a local gauge symmetry, in variance to less powerful global symmetries, can not break spontaneously such that VEVs of any gauge-noninvariant observables must be zero.

In general, in a gauge theory with fundamental-representation matter fields such as a gauge-Higgs theory, for instance, one typically does not expect to physically identify a {\it local} order parameter which would distinguish between the Higgs and confinement phases as qualitative descriptions of the corresponding field configurations. If there is no gauge-invariant way to distinguish between these regimes than it would be justified to attribute them to a single phase as mentioned earlier. A discussion of this issue known as {\it Higgs-confinement complementarity} goes back to as early as late 70'es and early 80'es. In Refs.~\cite{Osterwalder:1977pc,Fradkin:1978dv,Banks:1979fi}, by varying parameters in relatively simple lattice gauge-Higgs theories with a global symmetry, analyticity over a set of observables has been rigorously proven when going from a confining regime in the phase diagram to a regime characteristic for the Higgs phase. Although at certain large values of $\beta$ such a phase boundary emerges (see e.g. Ref.~\cite{Bonati:2009pf}), one can find an analyticity line continuously connecting any two points in the parameter space except $\gamma=0$\footnote{At $\gamma=0$ the theory is in the magnetic disorder phase which cannot be continuously evolved from other regions in parameter space with $\gamma\not=0$.}. In other words, in those models where this is true there would indeed be no thermodynamical phase transitions (or phase boundaries) along this path that separate the two regimes suggesting a possible existence of a single, massive phase all along the phase diagram (see Ref.~\cite{Greensite:2011zz} for a more elaborate discussion). Can this statement be applied only for some specific models, or it is always true?

This important result, first obtained in specific models, was then conjectured by some of the authors into a kind of ``folk theorem'' (known also as the Fradkin-Shenker-Banks-Rabinovici theorem) stating that the corresponding conclusion is expected to be always correct. Namely, if there is no local order parameter distinguishing different symmetry realisations one should probably expect continuity of phases. There are many examples where such a continuity has indeed been confirmed in simulations such as in transition from low- to high-temperature QCD when turning from physics of dilute gas of hadronic resonances to the physics of quark-gluon plasma (at low $\mu_B$). Indeed, in Euclidean description of real QCD there are certain reasons to believe that there is no thermodynamic phase transition that separates these two regimes. However, as will be discussed below, the analyticity conjecture may not actually be always true. As was argued in Ref.~\cite{Cherman:2020hbe} considering a discontinuity in a {\it non-local} order parameter, the Fradkin-Shenker-Banks-Rabinovici theorem does not apply to models where a global symmetry is broken in the same way in both the Higgs and confinement regimes, i.e. where the Higgs fields are charged under global symmetries.  

In fact, already in the string-breaking picture of hadronisation, by construction, the gluon vector-potential cannot retain its analyticity and is inherently discontinuous in the effective string-length (or string-time) scale as the string breaks apart and no gluon field is expected to retain between the daughter strings. Whether or not the observables still remain analytic upon such a string breaking is one of the big questions for confinement models.
One interesting example of the analyticity breakdown is associated with the notion of ``dense QCD'' or QCD at large baryon chemical potentials in the phase with broken $U(1)_{\rm B}$. We will elaborate on this aspect in the end of this review.

\section{String hadronisation and the Lund model}
\label{Sect:hadronisation}

\begin{figure}[hbt]
\begin{center}
\includegraphics[height=15em]{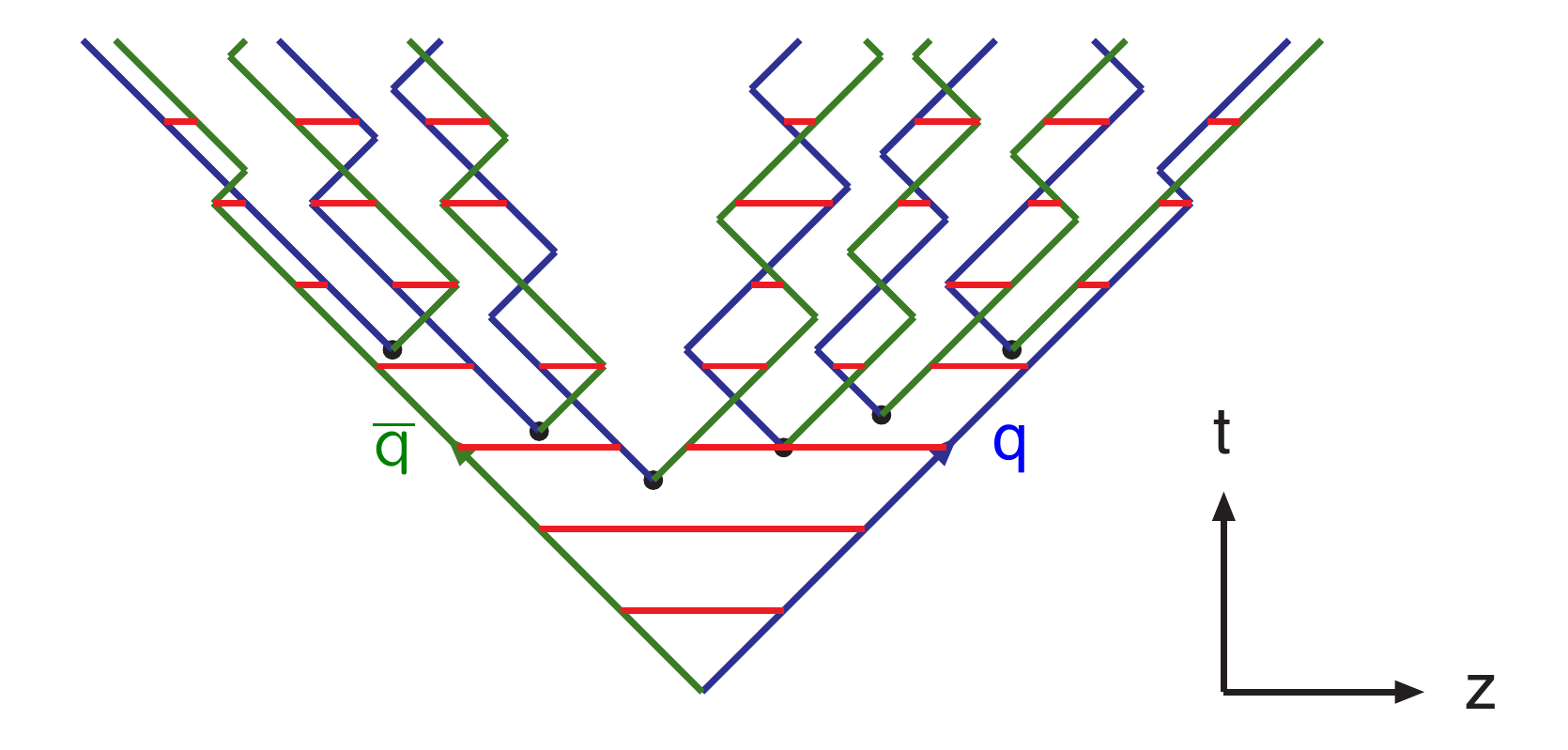}
\caption{An illustration of the string hadronisation picture in the Lund model.
}
\label{fig:yoyo}
\end{center}
\end{figure}

One of the existing successful realisations of the string hadronisation picture is the so-called the Lund string fragmentation model \cite{Andersson:1983ia} implemented in Monte-Carlo event generators widely-used in phenomenology of particle physics such as Pythia \cite{Sjostrand:2006za,Sjostrand:2014zea}. It realises the basic picture of linear confinement described above, where a flux tube is stretched between the color-charged endpoints of the back-to-back $q\bar q$ system that is characterised by the string tension $\sigma\simeq 1$ GeV$/$fm and the transverse size close to that of the proton, $r_p\simeq 0.7$ fm. 
In the simplest formulation of the hadronisation model, the quarks at the endpoints are assumed to be massless and to have zero transverse momenta. As the energy transfers between the endpoint quarks and the flux tube, they move along the light cone experiencing the ``yo-yo''-type oscillations. As the quarks move apart and pair-creation of dynamical $q\bar q$ pairs is enabled, there is non-zeroth probability for the initial ``quark-string-antiquark'' system to break up into smaller strings. For a simple illustration of this phenomenon, see Fig.~\ref{fig:yoyo}. Ordering the newly produced pairs as $q_i\bar q_i$, with $i=1,\dots,n-1$, into a chain along the string, depending on the initial energy of $q$ and $\bar q$ one eventually ends up with production of a set of $n$ mesons, $\{q\bar q_1,q_1\bar q_2,\dots,q_{n-2}\bar q_{n-1},q_{n-1}\bar q\}$ moving along $x$ axis of the initial string. The $q_i\bar q_i$ production vertices with coordinates $(t_i,x_i)$ have a spacelike separation, with no unique time-ordering, satisfying the constraint that the produced $i$th meson must be on its mass shell, i.e. $\sigma^2[(x_i-x_{i-1})^2-(t_i-t_{i-1})^2=m_i^2]$. 

In a more elaborate formulation, quarks have mass $m_q$ while the color string wildly fluctuates not only in longitudinal but also in transverse directions, and the amplitude of those fluctuations tend to grow with the string length and may eventually destabilise the system causing the string to break-up. The transverse momenta $p_{\perp}$ of the (anti)quarks are then naturally incorporated by giving $q$ and $\bar q$ opposite kicks in the transverse plane, with the mean square $\langle p_\perp^2 \rangle=\sigma/\pi\equiv \kappa^2\simeq (0.25\, {\rm GeV})^2$, such that the produced meson receives $\langle p_{\perp{\rm had}}^2 \rangle=2\kappa^2$. The virtual (anti)quarks tunnel over a distance $m_\perp/\sigma$, with $m_\perp=\sqrt{m_q^2+p_{\perp}^2}$ the transverse quark mass, before they become on-shell, and the tunneling probability of the produced pair provides an extra Gaussian suppression factor $\exp(-\pi m_\perp^2/\sigma)$.

In the framework of Lund model, a consistent selection of the produced DoFs is performed according to the probability distribution \cite{Andersson:1983ia},
\begin{eqnarray}
 f(z)\sim \frac{(1-z)^a}{z}\, e^{-bm_\perp^2/z} \,,
\end{eqnarray}
implying an equilibrium distribution of the production vertices on the string
\begin{eqnarray}
 P(\Gamma) \sim \Gamma^a\, e^{-b\Gamma} \,,
\end{eqnarray}
where $\Gamma=\sigma^2(t^2-x^2)$, $a,b$ are free parameters, and $z$ is the light-cone momentum fraction carried away by a produced meson. The remaining $(1-z)$-part of the momentum is kept by the string and is then redistributed among other mesons in its subsequent fragmentation. Even though the hadron masses do not enter this approach directly, a good description of the produced particle spectra can be reached with only a few free parameters.

More complicated $q\bar q gg\dots$ topologies can be introduced considering a gluon as a state with separate color and anticolor indices, well justified in the large-$N_c$ limit \cite{tHooft:1973alw}. The string gets then stretched between $q$ and $\bar q$ as usual while each of the gluons attach at intermediate points along the string respecting the color flow that goes in and out of each gluon. Notably, the fragmentation procedure of such a string does not require any extra free parameters \cite{Sjostrand:1984ic}. The fact that there is no string that connects $q$ and $\bar q$ directly in this case leads to asymmetries in the produced particle spectra in consistency with experimental observations \cite{Andersson:1980vk}. At last, baryon production can be conceptually tackled by enabling a diquark–antidiquark breaking e.g. via sequential $q\bar q$ production stages (for more details on this mechanism, see e.g. Refs.~\cite{Andersson:1981ce,Andersson:1984af}).

\section{Gauge symmetry remnants and confinement criteria}
\label{Sect:conf-criteria}

Due to the Elitzur’s theorem \cite{Elitzur:1975im} described above the phases of a gauge theory cannot be distinguished by means of the breaking of any local gauge symmetry. Thus, there must be an additional, global symmetry whose breaking enables us to identify those phases, at least, when a {\it local} order parameter is concerned. In the Ising model, the role of such a global symmetry is played by the $\mathbb{Z}_2$ symmetry as we have noticed earlier. Fixing a covariant gauge, in general, does not eliminate the gauge freedom entirely, but leaves certain remnant (both dependent and independent on spacetime coordinates) symmetries that can in principle get spontaneously broken since the Elitzur’s theorem does not apply to those. 

One of the examples of a possible confinement criterion known as the Kugo–Ojima condition \cite{Kugo:1979gm,Kugo:1995km} states that the full residual gauge symmetry in Landau gauge $\partial^\mu A^a_\mu = 0$ must remain unbroken in order to ensure that the expectation value of color charge operator $\langle \psi | Q_a | \psi \rangle$ vanishes in any physical state $\psi$. The spacetime-dependent (but global) part of such a full residual gauge symmetry w.r.t. gauge transformations $A_\mu \to G A_\mu G^\dagger$ in Landau gauge is known to take the following form \cite{Hata:1981nd,Hata:1983cs}
\begin{eqnarray}
    G(x)=\exp\Big( \frac{i}{2} \Xi^a(x) \sigma^a  \Big) \,,
    \label{spacetime-dependent}
\end{eqnarray}
where
\begin{eqnarray}
    \Xi^a(x) = \epsilon^a_\mu x^\mu - g \frac{1}{\partial^2}(A_\mu \times \epsilon^\mu)^a 
    + {\cal O}(g^2) \,,
\end{eqnarray}
in terms of a finite number of arbitrary parameters $\epsilon^a_\mu$, and the $SU(2)$ gauge coupling constant $g$. Besides, for confinement to hold yet another, spacetime-independent, part of the full residual gauge symmetry is required to be unbroken in addition to that in Eq.~(\ref{spacetime-dependent}). An analogical criterion of confinement has also been formulated in Coulomb gauge \cite{Marinari:1992kh,Greensite:2004ke}.
\begin{figure}[hbt]
\begin{center}
\includegraphics[height=25em]{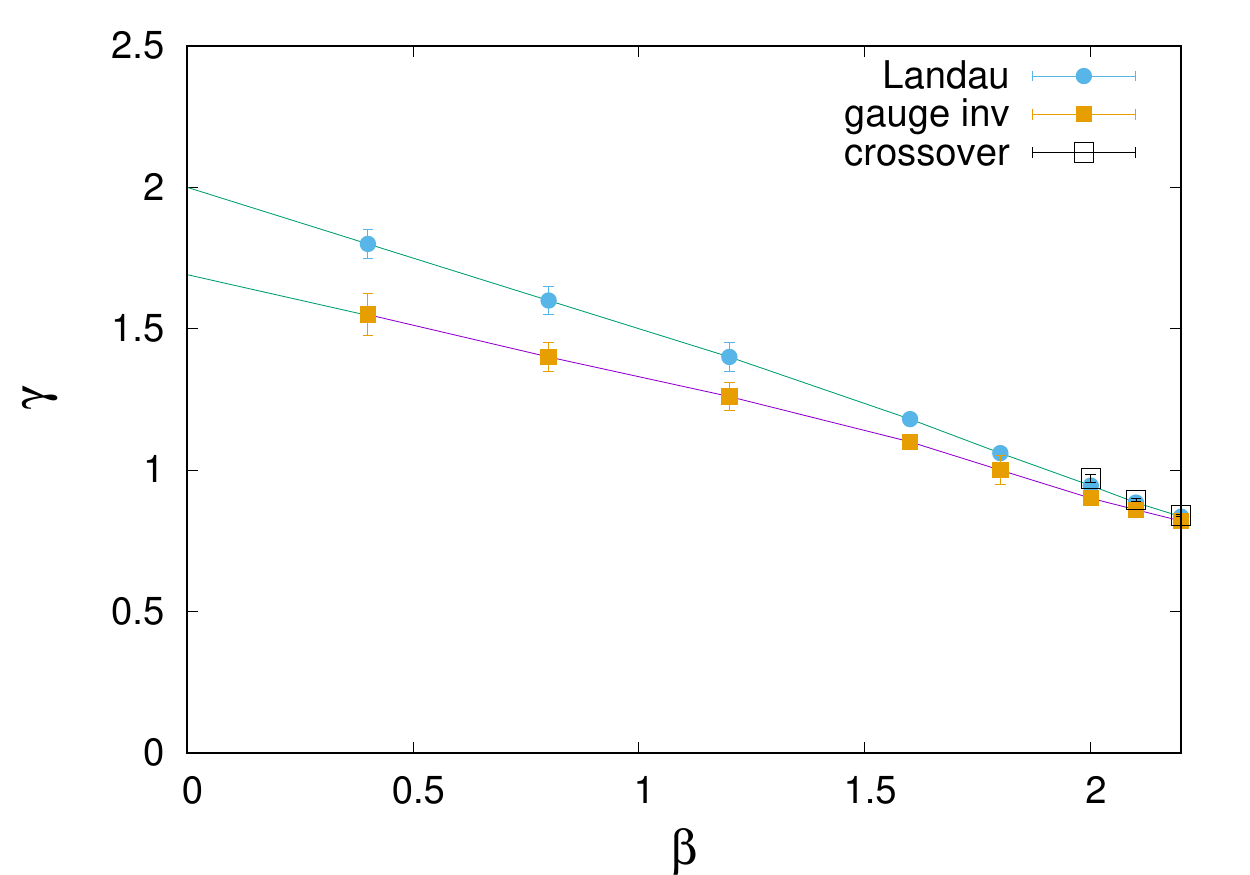}
\caption{Phase boundaries of global gauge symmetry breaking obtained in Landau gauges and in the gauge-invariant approach in the $SU(2)$ gauge-Higgs theory, along with the sharp crossover line at $\beta>2$. The figure is taken from Ref.~\cite{Greensite:2018ebg}.
}
\label{fig:remn}
\end{center}
\end{figure}

Thus, according to the Kugo–Ojima and Coulomb confinement criteria the phase boundary between the confining and de-confining regimes of a gauge theory is associated with the boundary between the unbroken and broken full ($x$-dependent and independent) remnants of the gauge symmetry in Landau and Coulomb gauges, respectively. However, a problem highlighted by lattice simulations and demonstrated in Fig.~\ref{fig:remn} is that these criteria predict transitions between confinement and deconfinement phases where actually no such transitions appear in the exact numerical analysis \cite{Caudy:2007sf}.

Different remnant symmetries emergent in different gauges break at different values of the couplings, so the resulting phase boundary is in fact gauge-dependent and might indeed emerge even when there is no actual change in the physical state of the system \cite{Caudy:2007sf}. In order to distinguish a confining state from a non-confining one, one should instead come up with a gauge-invariant criterion whose violation would indicate a true boundary between the magnetic order and disorder states that is the same in any gauge. As was discussed earlier, there is no such criterion in a gauge-Higgs theory. This might indicate that there is no such gauge-invariant separation of phases that can be attributed to a spontaneous breaking of a given symmetry and the system is in the massive phase characterised by the string-breaking effects and a perimeter-law behaviour of Wilson loop VEVs \cite{Greensite:2011zz}.

There are compelling reasons to believe that the same picture is realised in QCD at very large separations, supported also by lattice simulations. In the gauge-Higgs theory, only in the limit of Higgs decoupling, $\gamma=0$, the state of magnetic disorder emerges indicated by the area-law falloff of large Wilson loops at arbitrary large spacetime separations. The same occurs in the infinite quark mass limit in QCD such that it takes an infinite amount of energy in order to put an infinitely heavy quark-antiquark pair on its mass-shell from the vacuum such that the area-law persists to arbitrarily large string lengths.

\section{Center symmetry}
\label{Sect:center}

So, when one talks about the true (gauge-invariant) separation of phases, one implies a strong first-order (non-analytic) phase transition between the magnetic order (massive) and disorder states that exists at a well-defined (unique!) combination of model parameters in any gauge. Such a non-analytic behaviour is associated with a spontaneous breaking of a certain symmetry and, to comply with the Elitzur’s theorem \cite{Elitzur:1975im}, such a symmetry must be global. This type of a symmetry exists and is called the {\it center symmetry} -- a specific subgroup of a given gauge symmetry group which is defined as a subset of the gauge group elements that commutes with {\it all} the elements of the gauge group. For instance, the center of the $SU(N)$ gauge symmetry group is its $\mathbb{Z}_N$ subgroup $\{ \exp{2\pi i n/N}\, \hat{1}_N \}$, with $n=0,\dots, N-1$. 

Each of an infinite number of $SU(N)$ representations can be separated into $N$ possible subsets or {\it N-alities} depending on the corresponding representation of $\mathbb{Z}_N$ (there are only $N$ of those). Hence, each $SU(N)$ representation is characterised by N-ality $k$ that is found as the number of boxes in the associated Young tableau mod $N$. In other words, N-ality reflects how a given representation transforms under the center symmetry subgroup of the gauge group. For instance, if for a matrix representation $M[g]$ of an $SU(N)$ group element $g$, $M[zg]=z^k M[g]$ for a center $\mathbb{Z}_N$ element $z$, one says that $g$ belongs to a representation of N-ality $k$ (for a more detailed pedagogical discussion, see e.g. Ref.~\cite{Greensite:2011zz}). 
In the lattice formulation, one could show that the action (\ref{pure-SUN}) of a pure gauge theory is invariant under the timelike link transformation
\begin{equation}
U_0(\vec x, t_0) \to zU_0(\vec x, t_0) \,, \qquad z\subset \mathbb{Z}_N \,,
\label{center-transform}
\end{equation}
on a fixed time slice $t=t_0$. This transformation is a particular case of the singular gauge transformation defined on a time-periodic lattice with a period $L_t$ as
\begin{eqnarray}
U_0(\vec x, t) \to G(\vec x, t) U_0(\vec x, t_0) G^\dagger(\vec x, t+1) \,,
\label{singular}
\end{eqnarray}
where $G(\vec x, t)$ is a periodic function up to a center symmetry transformation, i.e.
\begin{eqnarray}
    G(\vec x, L_t+1) = z^* G(\vec x, 1) \,,
\end{eqnarray}
that also leaves Wilson loops invariant on the lattice. Such a transformation corresponds to an ``almost'' gauge transformation in the continuum limit,
\begin{eqnarray}
    A_\mu(x) \to G(x) A_\mu(x) G^\dagger(x) - \frac{i}{g}G(x)\partial_\mu G^\dagger(x) \,,
    \label{singular-continuum}
\end{eqnarray}
where the second term is dropped for $t=L_t$ and for $\mu=0$ when it turns into delta-function.

Matter fields in the fundamental representation of the gauge group $SU(N)$, or any other fields with N-ality $k\not=0$, break the center symmetry $\mathbb{Z}_N$ explicitly if they are not decoupled from the theory -- like the Higgs field for a non-zero coupling $\gamma$ in the example discussed above or the quark sector of real QCD (with $k=1$), with finite quark masses. Such a breaking, which is also a necessary ingredient of the string hadronisation model (see above), causes the static potential to flatten out instead of growing linearly at asymptotically large distances as the matter fields are, in fact, responsible for the string breaking phenomenon. Gluons or other particles in the adjoint representation having N-ality $k=0$ do not break the center symmetry so they cannot screen the color charge of a static source if the latter has a non-zero N-ality. A well-known exception is the $G_2$ gauge symmetry which has a trivial center subgroup, with a single unit element only, such that the gluons can bind to any source producing a color-singlet state. 

An important criterion of confinement is thus associated with unbroken center symmetry in a pure YM theory implying an asymptotically and infinitely rising static quark potential and signalling the area-law falloff of large Wilson line VEVs and hence the presence of the magnetic disorder state. The center symmetry can also be spontaneously broken by thermal effects i.e. at high temperatures, in pure YM theories causing the same effect of flattening out the static potential asymptotically as that of the matter fields. Other possible sources of the center symmetry breaking should also be considered, in order to reconstruct a full picture of phases in the underlined gauge theory.

\section{Polyakov loop}
\label{Sect:PL}

Consider a finite (in space) lattice which is periodic in time. Such a lattice is used, in particular, in quantum statistical mechanics at finite temperatures $T$ where the partition function reads
\begin{equation}
    Z=\sum_n \langle n | \exp(-\beta_T H) | n \rangle \,, \qquad \beta_T = \frac{1}{T} \,.
\end{equation}
In the continuum limit of a field theory and in Euclidean time $T_{\rm E}$, the latter generalises to a path integral
\begin{eqnarray}
    Z=\int D\phi(x, 0 \leq T_{\rm E} < \beta_T) e^{-S} \,,
\end{eqnarray}
where the periodic boundary condition in time $\phi(\vec x,0)=\phi(\vec x,\beta_T)$ is imposed through an implicit delta-function. Upon lattice regularisation, the temperature is related to the lattice period in time $L_t$ as $T=1/(L_ta)$, with $a$ being the lattice spacing as usual, and hence $\beta_T = L_ta$ is the total time extension of the lattice.

While neither the gauge-field action nor Wilson loops are affected by the $\mathbb{Z}_N$ center symmetry transformation (\ref{center-transform}), the trace of the following holonomy winding in time around the periodic-time lattice, known as the Polyakov loop \cite{Polyakov:1975rs},
\begin{eqnarray}
    P(\vec x) = {\rm Tr}\Pi_{n=1}^{L_t}U_0(\vec x, n) \,,
\end{eqnarray}
is $\mathbb{Z}_N$ non-invariant, i.e. it transforms as $P\to z P$. In the continuum limit, one can represent the Polyakov loop holonomy as follows
\begin{eqnarray}
{\cal P}(\vec x) = P \exp\Big( i \int dt A_0(\vec x, T_{\rm E}) \Big) = 
S\,{\rm diag}\big[ e^{2\pi i \mu_1}, e^{2\pi i \mu_2}, \dots, 
e^{2\pi i \mu_N} \big]\, S^{-1} \,, \qquad \sum_j \mu_j = 0 \,,
\label{Polyakov-holonomy}
\end{eqnarray}
in terms of an $SU(N)$ matrix $S(x)$ that diagonalises ${\cal P}(\vec x)$. 

One can show that in the case of ${\cal P}(\vec x)$ being a center element, all $\mu_j$ are equal and such a holonomy determines the finite-temperature classical instanton solutions known from Refs.~\cite{Harrington:1978ua,Harrington:1978ve}. In fact, in center-projected configurations that will be discussed below the Polyakov loop holonomies ${\cal P}(\vec x)$ are the only center elements. In general, the Polyakov loop is non-trivially charged under $\mathbb{Z}_N$ meaning that its expectation value plays a role of an order parameter for spontaneous breaking of the center symmetry. Hence, the Polyakov loop is yet another important characteristics of the confined (magnetic disorder) phase of the gauge theory, and vacuum fluctuations of the gauge field are responsible for formation of this phase and in some ways are associated with the center symmetry. 

Indeed, a difference between free energies of two states, one containing a single isolated (heavy) static charge $q$ and the other one defined in a pure gauge theory is as follows
\begin{eqnarray}
    e^{-\beta_T F_q} = \frac{Z_q}{Z} \propto \langle P(\vec x) \rangle \,,
\end{eqnarray}
which is obtained by integrating out the massive quark field (in the $m\to \infty$ limit) in the path integral for $Z_q$ over the period of the lattice $0 \leq T_{\rm E} < \beta_T$. Indeed, $F_q$ for a single quark $q$ would be infinite if $\langle P(\vec x) \rangle \to 0$, i.e. in the case of unbroken center symmetry. At high temperatures (small $\beta_T$), the center symmetry is in general spontaneously broken such that the isolated charges are described by finite-energy states (deconfining phase). A magnetic disorder-to-order phase transition associated with thermal breaking of the center symmetry is expected to occur at a critical temperature \cite{McLerran:1981pb}.

\section{t'Hooft loop and center vortices}
\label{Sect:tHL}

The singular gauge transformation in the continuum limit (\ref{singular-continuum}), unlike the ordinary center symmetry transformation, leaves the action non-invariant. As a result of such a transform, a singular loop of magnetic flux, the so-called {\it thin center vortex}, is being created. For instance, as was mentioned earlier the holonomy for a closed spacelike loop $C$ in $U(1)$ gauge theory
\begin{equation}
    U(C)=e^{ie\Phi_B}
\end{equation}
is given in terms of the magnetic flux $\Phi_B$ through the loop. For a loop winding around a solenoid oriented along the $z$-axis, it is possible that $\Phi_B\not = 0$ even for a zeroth magnetic field along the closed loop which can be obtained as a result of a singular gauge transformation applied to $A_\mu = 0$ with a discontinuous $G(x)$. If in cylindrical coordinates $\{r,\theta,z,t\}$, the corresponding transformation function $G$ has a discontinuity in $\theta$ for $r>0$, then
\begin{eqnarray}
    U(C) \to e^{\pm ie\Phi_B} U(C) \,,
    \label{Wilson-transform}
\end{eqnarray}
where $\exp(\pm ie\Phi_B)$ is an element of $U(1)$ group, the sign $\pm$ depends on the orientation of the loop $C$, such that a singular line of magnetic flux (thin vortex) is produced along the $z$-axis. Instead of $z$-axis, one could introduce yet another closed contour $C'$ topologically linked to $C$ such that the singular gauge transformation operator $G$ that creates a magnetic flux along $C'$ would satisfy
\begin{eqnarray}
    G(\vec x(1)) = e^{\pm ie\Phi_B} G(\vec x(0))
    \label{U1-G}
\end{eqnarray}
on the contour $C'$ determined by the parametric equation $\vec x = \vec x(\xi)$, with $\xi=[0,\dots, 1]$, such that $\vec x(1) = \vec x(0)$ belong to a surface bounded by $C'$. Upon such a transformation, a Wilson loop $C$ linked to $C'$ appears to transform as in Eq.~(\ref{Wilson-transform}). The {\it winding number} is defined as the number of times a loop goes around a fixed point in $D=2$, while in $D=3$ such a topological invariant generalises to the so-called {\it linking number} that determines the number of times two loops can wind around each other. This can be generalised further on for $D$ dimensions where a loop $C$ links to a $D-2$ hypersurface $C'$ on which a $(D-2)$-dimensional thin vortex is created by the corresponding singular gauge transformation that is discontinuous in the $D-1$ (Dirac) region bounded by the $D-2$ hypersurface.

Switching over to $SU(N)$ YM theory, the $U(1)$ group element that multiplies a transformation operator in Eq.~(\ref{U1-G}) should be replaced by a center-group $\mathbb{Z}_N$ element
\begin{eqnarray}
    G(\vec x(1)) = z G(\vec x(0)) \,, \qquad U(C) \to (z^*)^l U(C) \,,
\end{eqnarray}
in order for such a transform to create a thin vortex (and hence to affect the action) on the $(D-2)$-dimensional hypersurface only, and not on the Dirac $D-1$ region that it envelops. Above, the spacelike Wilson loop $C$ is topologically linked to the $(D-2)$-dimensional thin vortex, with the corresponding linking number $l$. Upon quantisation of the non-abelian magnetic flux, its quanta are known in the literature as the {\it thin center vortices}, while a regularisation of the singular color-magnetic field by smearing it out in the transverse directions to the $(D-2)$ hypersurface leads to a vortex with finite thickness, or a {\it thick center vortex}. For a more detailed description of the vortex configurations and properties, see e.g.~Ref.~\cite{Greensite:2011zz} and references therein.

Consider an operator $B(C)$ that creates a thin center vortex at a fixed time $t_0$ along a given loop $C$ in a $D=3+1$ gauge theory \cite{tHooft:1977nqb}. If $C$ and another closed loop $C'$ are topologically linked (with $l=1$) in a three-dimensional surface, then 
\begin{eqnarray}
    B(C)U(C')=zU(C')B(C)\,, \qquad z\subset \mathbb{Z}_N \,,
\end{eqnarray}
is valid. In this case, the operator $B(C)$ is known as the {\it t'Hooft loop}. As was demonstrated in Ref.~\cite{tHooft:1977nqb}, the VEV of a Wilson loop $W(C)\equiv \langle U(C) \rangle$ and a t'Hooft loop $\langle B(C) \rangle$ may satisfy either a perimeter-law or an area-law falloffs, but not simultaneously. Indeed, the confined (magnetic disorder) phase corresponding to an unbroken center symmetry is realised when
\begin{eqnarray}
    W(C) \sim e^{-a A(C)} \qquad \Longleftrightarrow \qquad \langle B(C) \rangle \sim e^{-b P(C)}\,, \qquad a,b>0\,, 
\end{eqnarray}
while the opposite case,
\begin{eqnarray}
    W(C) \sim e^{-a' P(C)} \qquad \Longleftrightarrow \qquad \langle B(C) \rangle \sim e^{-b' A(C)}\,, \qquad a',b'>0\,, 
\end{eqnarray}
implies a spontaneously broken center symmetry (magnetically-ordered phase). Indeed, the Wilson and t'Hooft loop operators can be considered dual to each other as the first one creates a closed loop of color-electric flux, while the second one creates a closed loop of color-magnetic flux (thin center vortex), at a fixed time $t$ in both cases.

One could introduce a vortex on a finite lattice in $D=4$ by replacing $U(p') \to \xi U(p')$ for a given plaquette $p'$ in the $SU(N)$ gauge-field action \cite{tHooft:1979rtg}
\begin{eqnarray}
    && S=-\frac{\beta}{2N}\Big[\sum_{p\not=p'}({\rm Tr}[U(p)]+{\rm c.c.}) + ({\rm Tr}[\xi U(p')]+{\rm c.c.}) \Big] \,, 
    \qquad \xi\subset \mathbb{Z}_N \,, \\
    && U(p') = U_1(x_0,y_0,z,t)U_2(x_0+1,y_0,z,t)U_1^\dagger(x_0,y_0+1,z,t)U_2^\dagger(x_0,y_0,z,t)\,,
\end{eqnarray}
which can be viewed as a change in periodic boundary conditions, also referred to as twisted boundary conditions. Such a change creates a thick center vortex on the lattice parallel to $(z-t)$-plane, satisfying the ordinary periodic boundary conditions in $z,t$ coordinates.

In the simplest case of $SU(2)$ gauge symmetry, the (magnetic) free energy of a $\mathbb{Z}_2$-center vortex $F_m$ can be found as
\begin{eqnarray}
    e^{-F_m}=\frac{Z_{-}}{Z_{+}}
\end{eqnarray}
in terms of the partition functions with ordinary and twisted boundary conditions, $Z_{+}$ and $Z_{-}$, respectively, while the free energy of closed color-electric flux $F_e$ is
\begin{eqnarray}
    e^{-F_e}=1-e^{-F_m} \,.
\end{eqnarray}
It was shown in Ref.~\cite{Tomboulis:1985ah} that the VEV of a rectangular Wilson loop $C$ with area $A(C)$ is bounded from above as
\begin{eqnarray}
    W(C) \leq [\exp(-F_e) ]^{A(C)/(L_xL_y)} \,.
    \label{W_C-limit}
\end{eqnarray}
A sufficient condition for the existence of a magnetic-disorder phase, and hence confinement, in terms of the behavior of the magnetic {\it vortex free energy} then reads
\begin{equation}
   F_m \sim L_z L_t e^{-\kappa L_x L_y} \,,
\end{equation}
i.e. it falls off exponentially at large $L_x L_y$ area, such $\exp(-F_e) \simeq F_m \ll 1$. Indeed, the latter limit, together with Eq.~(\ref{W_C-limit}), implies an area-law upper bound for a large Wilson loop and, hence, the asymptotic string tension. In Ref.~\cite{Cornwall:1981zr} it has been pointed out that quark confinement emerges from a vortex condensate supported by the mass gap.

\section{Fundamental properties of the string tension}
\label{Sect:tension}

One of the fundamental characteristics of confinement is an non-vanishing asymptotic string tension or, equivalently, the asymptotic linearity of the static potential \cite{Greensite:2011zz,Greensite:2016pfc}. As was proven in Ref.~\cite{Bachas:1985xs}, the potential is always convex and is saturated by a straight line from above. At not too large distances, the string tension for a quark in a given representation $r$ of the gauge group interacting with an antiquark can be approximated as
\begin{eqnarray}
    \sigma_r = \frac{C_r}{C_F}\sigma_F \,.
\end{eqnarray}
This is the property known as the {\it Casimir scaling} which is strictly valid in the large-$N$ limit. Here, $\sigma_F$ is the string tension for the defining (fundamental) representation. Such a scaling can be proven in a two-dimensional theory and then to a good precision can be found also in 4D by means of the dimensional reduction \cite{Ambjorn:1984mb}, supported also by numerical simulations \cite{Bali:2000un}. For a more recent analysis of the Casimir scaling in $D=2+1$ $SU(N)$ theory in the vortex picture, see Ref.~\cite{Junior:2019fty}. Asymptotically at very large distances, the Casimir scaling does not hold (apart from $N=2$ and large-$N$ cases), and can be effective at intermediate distances only.

The dimensional reduction is a specific (approximate) property of quantum state of the theory $\Psi_0[A]$ emergent at large length scales. According to this property a calculation of the VEV of a large Wilson loop $W(R,T)$ in fundamental representation in a $D=4$ gauge theory can be sequentially reduced to that in $D=3$ theory \cite{Greensite:1979yn, Greensite:1979ha} and then down to $D=2$ case \cite{Halpern:1978ik}. In this case,
\begin{eqnarray}
 W(R,T)=\langle {\rm Tr}[U(C)] \rangle_{D=4} \equiv \langle \Psi_0 | {\rm Tr}[U(C)] | \Psi_0 \rangle \simeq \langle {\rm Tr}[U(C)] \rangle_{D=3} \simeq \langle {\rm Tr}[U(C)] \rangle_{D=2}=e^{-\sigma A(C)} \,,
\end{eqnarray}
where in last relation corresponds to the fact that in $D=2$ the Wilson loop VEVs obey an area-law falloff. For this property to hold in the strong coupling limit, the vacuum functional should take the same form in $D=2,3,4$ at large length-scales:
\begin{eqnarray}
 \Psi_0[A] \propto \exp\Big[ -\frac{1}{4g_{\rm eff}^2} \int d^3x \, {\rm Tr}[F_{ij}^2] \Big] \,.
 \label{reduction}
\end{eqnarray}
Note, this form can not be correct at short distances in PT, so it should be regarded as approximate and generically valid in the non-perturbative regime only. It is also not correct for Wilson loops in the adjoint representation which follow a perimeter-law due to the color screening effect. An elaborate form for the vacuum functional that matches both the dimensional reduction form and the correct free-field limit has been proposed in Refs.~\cite{Leigh:2006vg} predicting the glueball mass spectrum in $D=2+1$ in consistency with the lattice calculations. For other proposals, see e.g. Refs.~\cite{Karabali:1998yq,Karabali:2009rg,Reinhardt:2004mm,Feuchter:2004mk,Greensite:2007ij}.

Another fundamental property of the string tension, presumably closely related to confinement, is the observation that the string tension depends only on N-ality of the gauge group representation. For static quarks in the adjoint representation, for instance, gluons screen their charges at large distances causing the string to break at separations $R$ satisfying $2E < \sigma_A R$, where $E$ is the gluonic energy of the produced ``gluelump'' state, $\sigma_A = C_A/C_F \sigma_F$ is the string tension in the adjoint representation, valid at intermediate distances. For numerical studies of the adjoint string tensions, see e.g. Ref.~\cite{Kratochvila:2003zj}. While the precise form of the N-ality dependence is not known there are several models widely used in the literature. Among them, for instance, the ``Casimir scaling'' proposal assumes that the string tension for the lowest dimensional representation ($k$-string tension), behaves asymptotically for the $SU(N)$ gauge theory as
\begin{eqnarray}
 \sigma_r = \frac{k(N-k)}{N-1}\sigma_F \,.
 \label{Casimir-sigma}
\end{eqnarray}

If the true confinement phenomenon implies the formation of an electric flux tube in the form of a quantum Nambu-like string, typical predictions of the string model such as subleading deviations from linearity of the potential as well as the spectrum of string excitations should find their evidence in a first principle analysis of confining gauge theories. In particular, one such prediction is a subleading $1/R$ correction term to the static quark potential emerging due to transverse fluctuations of the string known as the {\it L\"uscher term} \cite{Luscher:1980ac,Alvarez:1981kc}
\begin{eqnarray}
 V(R) = \sigma_r R - \frac{\pi(D-2)}{24} \frac{1}{R} + {\rm const} \,.
\end{eqnarray}
such that the VEV of a large rectangular Wilson loop can be generically parameterised as
\begin{eqnarray}
 W_r(R,T_{\rm E}) = \exp[-\sigma_r RT_{\rm E} + \tau (R+T_{\rm E}) - 
 \xi (T_{\rm E}/R + R/T_{\rm E}) + \eta] \,,
 \label{rect-loop-VEV}
\end{eqnarray}
where the second term in the exponent is a self-energy contribution that diverges in the continuum limit as was mentioned above. On the lattice, one may extract the asymptotic string tension $\sigma$ as the following ratio computed at large loop areas
\begin{eqnarray}
 -\log\Big[ \frac{W(R,T_{\rm E}) W(R-1,T_{\rm E}-1)}{W(R-1,T_{\rm E})W(R,T_{\rm E}-1)} \Big] \to \sigma \qquad {\rm for} \qquad RT_{\rm E} \to \infty \,,
 \label{Creutz}
\end{eqnarray}
known as the {\it Creutz ratio}.

Another property of the Nambu string is that the cross section area of the string grows logarithmically with the quark separation, the effect knows as {\it roughening} \cite{Luscher:1980iy,Hasenfratz:1980ue}. An agreement with Nambu string model predictions was found earlier in the analysis of closed string excitations in the $D=2+1$ $SU(N)$ gauge theory in Ref.~\cite{Athenodorou:2007du}.

\section{Center vortex mechanism of confinement}
\label{Sect:vortex}

The center vortex mechanism of confinement is strongly supported by the fact that the static potential slope depends only on N-ality, while N-ality zero (or adjoint) string tensions vanish at asymptotically large distances. Also, when adopting a picture of pair creation of particles out of the vacuum at a certain distance causing the string to break, one implies a microscopic perturbative language of particle states in a particular configuration. While an extrapolation of perturbative particle states towards large distances may not necessarily work out well in confining theories, an effective particle picture of string breaking is still considered to adequately reflect the reality, at least qualitatively. In proper path integral computations, one sums over all possible field configurations that should provide the same result for the gauge-invariant observables (such as Wilson loop VEVs) as the phenomenologically successful effective particle picture of the string breaking. Ultimately, one would like to find out how the vacuum field fluctuations induce N-ality dependence of the asymptotic string tension and describe colour screening of the static sources \cite{Greensite:2011zz}.

While instantons \cite{Belavin:1975fg} are saddle points of the classical gauge-field action, vortices are interpreted to be saddle points of the effective one-loop action \cite{Ambjorn:1980ms,Diakonov:2002bx} that incorporates the vacuum polarisation effects, and hence have a pronounced fundamental meaning (see also Ref.~\cite{Greensite:2016pfc}). Fluctuations of center vortices that can be identified as solitonic objects in typical field configurations are known to give rise to an area law of Wilson loops. A remarkable property is that Wilson loops in different representations but with the same N-ality get the same contributions from center vortices, while loops of N-ality zero are not affected. This follows from the simple fact that the creation of a vortex linked to the loop $C$ affects the loop holonomy of a given N-ality $k$ as $U(C) \to zU(C)$ and its VEV as $W_r(C)\to z^k W_r(C)$ for $z$ from the center group $\mathbb{Z}_N$ of $SU(N)$.

In a more generic case, consider a set of vortices linked to a given loop $C$, with linking numbers $l_{1,2,3,\dots}$, having the center elements $z_{1,2,3,\dots}$. Then, creation of this set modifies the Wilson loop VEV as $W_r(C) \to Z^k(C) W_r(C)$, where $Z(C)=z_1^{l_1} z_2^{l_2} z_3^{l_3} \dots$. In the vortex picture of confinement \cite{tHooft:1977nqb,Ambjorn:1980ms,Nielsen:1979xu,Cornwall:1979hz} (see also Ref.~\cite{Greensite:2011zz} and references therein) the gauge-field vacuum configuration is considered to be a set of vortices superimposed on a non-confining configuration. Then random fluctuations in number of vortices in the system as well as in their linking numbers to a given Wilson loop $C$ induces the area law dependence of the corresponding Wilson loop VEV. The loop holonomy can be represented in a factorised form $U(C)=Z(C) u(C)$, where $u(C)$ is a contribution from a non-confining background, and $Z(C)\subset \mathbb{Z}_N$ is a center-valued holonomy. Then, the vortex mechanism implies factorisation of the Wilson loop VEV
\begin{eqnarray}
 \langle \chi_r[U(C)] \rangle \simeq \langle Z^k(C) \rangle\langle \chi_r[u(C)] \rangle \simeq
 \exp[-\sigma_r A(C)] \exp[-\mu_r P(C)] \,.
\end{eqnarray}
A detailed proof relies on a weak correlation between $Z(C)$ and $U(C)$, as well as between $Z(C_1)$ and $Z(C_2)$ for any large loops $C,C_{1,2}$, and can be found for instance in Ref.~\cite{Greensite:2011zz}. It manifestly demonstrates that the string tension computed for smaller loops is the same as that for the larger ones provided that the above assumptions hold and $Z(C_i)$ experience independent fluctuations.

Numerical estimates \cite{Kovacs:2000sy} suggest that the thickness of the vortex is close to one fermi so, in principle, the Wilson loops with an extension below this scale may get affected. As was demonstrated in Refs.~\cite{Faber:1997rp,Greensite:2006sm}, such a vortex thickness plays an important role for generating the Casimir scaling at intermediate distances. At large distances dominated by large Wilson loops the N-ality dependence of the linear static potential is reproduced as expected. From this point of view, vortices are non-local objects that represent specific field configurations that lead to an asymptotic string tension as a function of N-ality.

The link configurations ${\cal U}_\mu(x)=g(x)z_\mu(x)g^{-1}(x+\hat\mu)$ that produce $Z(C)$ holonomies can be transformed into the link configurations $z_\mu(x)$ of $\mathbb{Z}_N$ lattice gauge theory responsible for confinement by means of a specific $SU(N)$ gauge transformation $g(x)$. The thin vortices then have a meaning of excitations of the center-group $\mathbb{Z}_N$ lattice gauge theory. The original link variables $U_\mu(x)$ get separated into a product of center elements $z_\mu(x)$ and the link variables of the non-confining background $V_\mu(x)$ by the $g(x)$ transform
\begin{eqnarray}
    U_\mu(x)=g(x)z_\mu(x)V_\mu(x)g^{-1}(x+\mu) \,.
\end{eqnarray}
The main aim of the vortex mechanism of confinement is to find a specific $g(x)$ for a given non-confining background $V_\mu(x)$ typically assumed to be a small fluctuation about the unity. Locations of center vortices then can be extracted from $z_\mu(x)$ after the above factorisation $U\to zV$ has been performed \cite{DelDebbio:1998luz}. One such $g(x)$ transforms the DoFs into a specific gauge known as the direct maximal center gauge where the deviation of the links in the adjoint representation from the identity matrix is minimal, or where the quantity 
\begin{eqnarray}
K=\sum_{x,\mu} {\rm Tr}[U^A_\mu(x)]=\sum_{x,\mu} {\rm Tr}[U_\mu(x)]{\rm Tr}[U^\dagger_\mu(x)]-1 \,,
\end{eqnarray}
with the adjoint link $U^A_\mu(x)$, is maximal. Locations of center vortices then can be extracted in a dedicated Monte-Carlo procedure from identified center elements $z_\mu(x)$ once the center mapping (projection) $U_\mu(x)\to z_\mu(x)$ has been performed. If the product $Z(p)$ of $z_\mu(x)$ on the projected $\mathbb{Z}_N$ lattice around a plaquette $p$ satisfies $Z(p)\not=1$, a thin vortex (or $P$-vortex) is then located on that plaquette. The vortex picture of confinement then reduces to a consideration of $P$-vortices as random surfaces percolating through the spacetime volume. Uncorrelated piercings by the $P$-vortices on a given planar surface correspond to uncorrelated large center-projected loops. The numerical procedures, however, may fix the projected lattice to only one out of a large amount of local maxima of the gauge-fixing functional $K$ known as the {\it Gribov copies} \cite{Gribov:1977wm}, not straight to its global maximum, which is considered to be a problem in several widely used center-gauge fixing approaches.

The problem of Gribov copies is one of the main obstacles for a consistent treatment of the confinement problem. Considering a set of gauge-equivalent configurations of the gauge field known as a gauge orbit and imposing a gauge-fixing condition as a certain hypersurface in a space of gauge field configurations, the Gribov copies can be visualised as many possible intersections of the gauge orbit with the gauge-fixing hypersurface. Summing over all contributions from Gribov copies in a path integral, the latter may actually vanish since those contributions come with opposite signs and may mutually eliminate each other in a given observable. This is the statement of the Neuberger’s theorem \cite{Neuberger:1986xz} rendering BRST quantization not well defined in the non-perturbative regime of a gauge theory (see a detailed discussion in Ref.~\cite{Greensite:2011zz}). One possibility is to restrict the functional integral to a subspace of gauge configurations with positive Faddeev–Popov determinant, the so-called Gribov region, and its boundary containing also the lowest non-trivial eigenmode with zeroth eigenvalue is called the Gribov horizon. An instructive example of such a hypersurface restricted to the Gribov region is the Landau gauge fixing condition that minimises the functional
\begin{eqnarray}
R=-\sum_{x,\mu}{\rm ReTr} [U_\mu(x)] \,,
\end{eqnarray}
such that the corresponding Gribov region consists of all possible minima of $R=R[A]$ for a given gauge orbit. However, various gauge orbits might cross the Gribov region a different number of times leading to different weights assigned to different gauge orbits. A proposal to consider only unique global minima of $R[A]$ functional for each gauge orbit \cite{Zwanziger:1998ez} may be very difficult to realise in practical calculations. Also, there is no any reason to believe a particular Gribov copy with the global minimum for $R[A]$ is more physical than other local minima. Lattice procedures, in general, assume that a particular choice of a Gribov copy would not make a big difference on the numerical results.

In order to establish a direct connection between the existence of $P$-vortices and magnetically disordered phase, following the reasoning of Ref.~\cite{Greensite:2011zz}, let us first consider whether the center-projected $Z_{\mu}(x)$ link variables (extracted, for instance, in a maximal center gauge) are responsible for confinement. For this purpose, it is instructive to consider the VEV of the rectangular $R\times T_{\rm E}$ Wilson loop, $W(R,T_{\rm E})$, defined in Eq.~(\ref{rect-loop-VEV}). If such a loop is constructed from $Z_{\mu}(x)$ links on a center-projected lattice, the corresponding Creutz ratio (\ref{Creutz}) appears to converge much faster to $\sigma$ than for the unprojected Wilson loop VEV. Already at $R=2$ the static potential becomes linear -- the property of the so-called precocious linearity. The fact that the asymptotic string tensions extracted from center-projected and unprojected Wilson loop VEVs at large $R$ are the same is known as the {\it center dominance}. There is also an excellent agreement of the Creutz ratios on the center-projected lattice with the well known predictions of the asymptotic freedom for large $\beta$ (small gauge couplings). 

A slow convergence of the Creutz ratio (\ref{Creutz}) to the string tension at large $R$ in the unprojected case means that here we deal with thick vortices linked to large Wilson loops. The center projection effectively shrinks the thickness of the vortices down to a single lattice spacing, so the linking appears to be relevant already for small center-projected loops. Indeed, as was deduced earlier, $P$-vortex piercings are totally uncorrelated on a planar surface causing the linearity of the potential already at small distances. One naturally wishes to establish that each thin $P$-vortex in the projected configurations matches a thick center vortex in unprojected lattice in order to prove that the $P$-vortices do not carry artefacts of the gauge fixing procedure and indeed are responsible for the underlined physics of magnetic disorder (and hence confinement).

As thoroughly described in Ref.~\cite{Greensite:2011zz}, one way of proving the relevant correlation of $P$-vortices with gauge-invariant observables (unprojected Wilson loops) is to compute a so-called vortex-limited Wilson loop VEV defined as an expectation value of an ordinary unprojected loop holonomy $W_n(C)$ but taken in the ensemble of configurations where the minimal surface area of the loop is pierced by $n$ $P$-vortices. Then, considering for simplicity the $SU(2)$ theory, if the ratios asymptotically behave as $W_n(C)/W_0(C)\to (-1)^n$ provided that $\langle Z(C) \rangle = (-1)^n$ ($-1$ per each vortex piercing) then the procedure of finding thin $P$-vortices on the center-projected lattice effectively locates thick center vortices on the unprojected lattice. This indeed has been confirmed by lattice simulations, see e.g. Ref.~\cite{Faber:2001zs}.

Another test proposed in Refs.~\cite{deForcrand:1999our} suggests to insert a thin vortex found by the center projection operation into a thick vertex on the unprojected lattice and then to check if their disordering effects, due to center dominance, cancel out asymptotically at large distances. Indeed, an explicit calculation shows that this procedure eliminates the string tension and hence the disorder effect. It was also checked in Ref.~\cite{Engelhardt:1998wu} that $P$-vortex density is independent on the lattice spacing in the continuum limit as expected for physical objects. An additional observation of Ref.~\cite{Gubarev:2002ek} revealed that the continuum action density appears to be singular at the location of $P$-vortices which, together with their constant density, signals an intricate cancellation between action and entropy at a surface of infinite action associated with a vortex.

As was discussed above, at finite temperatures $T$ in a time-periodic lattice the Polyakov loop VEVs determine the quark free energy $F_q$. In $SU(2)$ gauge theory, at $T>T_c=220$ MeV a deconfinement transition occurs when $F_q$ becomes finite and the static quark potential goes flat. However, even at large $T>T_c$ space-like Wilson loops retain their area-law falloff such that vacuum fluctuations inherit some of the key properties of the confined phase. 

This observation fits well with the center-vortex mechanism of confinement \cite{Greensite:2011zz}. At low $T$, due to uncorrelated piercings of the minimal loop areas, one finds $\langle P(\vec x) \rangle=0$ and an exponential falloff of the Polyakov loop correlators for large interquark separation $\langle P(\vec x)P(\vec x+\vec R)\rangle \sim \exp[-\sigma(T)L_tR]$, with $\sigma(T)$ the $T$-dependent string tension of a flux tube stretched between $q$ and $\bar q$. Since the vortices running in spacelike directions have a finite diameter, as temperature rises, they get squeezed by the reduced finite lattice extension in time $L_t$ until they effectively stop percolating eliminating the exponential falloff of Polyakov loop correlator and hence $\langle P(\vec x) \rangle$ is no longer zero \cite{Kovacs:2000sy,deForcrand:2001nd}. The asymptotic behaviour of the space-like Wilson loop, however, is determined by the piercings of center vortices oriented in periodic time (i.e. running in timelike directions), and their cross-section is not limited by a small extension in the time direction at large $T$. Thus, the corresponding $P$-vortices keep percolating on a time slice in the spacial directions such that the exponential falloff of spacelike Wilson loops remains unaffected in the deconfined regime \cite{Engelhardt:1999fd,Langfeld:1998cz}.

As we already discussed above, the center symmetry turns out to be explicitly broken by the dynamical fields in the fundamental representation. The center dominance in the confinement region in $SU(2)$ gauge-Higgs theory has been tested in Ref.~\cite{Greensite:2006ng}. In a region where the screening effects by the matter fields become important the center vortices do not disappear but somehow rearrange themselves in order to allow for asymptotically vanishing string tension but still generating a linear slope in the potential at intermediate distances (no signature of linearity has been found in the Higgs region at any scale). In the presence of matter fields, Dirac volume shrinks and the vortex piercings of the Wilson loop minimal area are expected to become correlated at large distances, but to the best of our knowledge there is no full consensus on exactly how this occurs.

\section{Chiral symmetry breaking and topological charge}
\label{Sect:condensates}

The global chiral symmetry of QCD light $u,d$ quark sector $SU(N_f)_{\rm R}\times SU(N_f)_{\rm L}$ (with the number of flavors, say, $N_f=2$) is broken spontaneously by the order parameter known as the quark (or chiral) condensate $\langle \bar q q \rangle\not = 0$. In addition, it is also broken explicitly by light current quark mass turning the Goldstone bosons, the pions, into massive pseudo-Goldstone states. Another less known mechanism based upon the linear sigma model of effective quark-meson interactions introduces yet another source of the global chiral symmetry breaking, through a linear term in $\sigma$-field proportional to the quark condensate. Such a breaking is also explicit and as such it provides an additional finite contribution to the pion mass. A symmetry breaking due to quark condensation phenomenon is often referred to as {\it dynamical symmetry breaking} and is considered a baseline for Technicolor models of EW symmetry breaking \cite{Weinberg:1975gm,Susskind:1978ms} (for a detailed review of existing concepts, see e.g.~Ref.~\cite{Hill:2002ap}).

As was discussed earlier in the case of Ising model, in order to get a nontrivial value of the order parameter one should perform two limits in a certain order -- first, take volume to infinity, and then set the quark masses to zero. This procedure leads to the well-known Banks-Casher relation \cite{Banks:1979yr} between the chiral condensate as the trace of the quark propagator and the value of the density of the close-to-zero eigenvalues of the Dirac operator characterised by vacuum field configurations. The latter density receives no perturbative contributions, and hence the dynamical chiral symmetry breaking is an intrinsically non-perturbative phenomenon. 

Provided that in real QCD with light quarks the string tension vanishes asymptotically due to colour-screening and string breaking, the chiral condensate by itself is not tied to the area-law falloff of large Wilson loop VEVs and does not even require the presence of gauge fields, in analogy to the effective Nambu--Jona-Lasinio model \cite{Nambu:1961tp}. Naively, one might think that these observations indicate no immediate connection between the chiral symmetry breaking mechanism and the confinement phenomenon. As was emphasised in Ref.~\cite{Suganuma:2014wya}, the low-lying Dirac eigenmodes which are crucial for chiral symmetry breaking provide vanishingly small contributions to the string tension and to the Polyakov loop in both confined and deconfined phases. These observations provided no indication of an immediate correspondence between chiral symmetry breaking and confinement.

Interestingly enough though, the critical temperatures of chiral and deconfinement phase transitions appear to be the same or close to each other as suggested by lattice simulations, motivating a further search for possible hidden connections between the two transitions. In particular, a connection between the Polyakov loop, center symmetry, and the chiral condensate may be due to the fact that, after integrating out fermions, the chiral condensate is basically a complex expectation value of many Wilson loops, including those wrapping around compact dimensions. As was elaborated in detail in Ref.~\cite{Gattringer:2006ci}, the spectral properties of the Dirac operator are affected by confinement, in particular, causing the correlators of Dirac eigenvector densities to decay exponentially instead of a power law in the deconfined phase. Ultimately, one would need to establish a link between the spectral properties of the Dirac operator in the infrared regime presumably, responsible for chiral symmetry breaking with those in the ultraviolet regime tightly connected to confinement.

Remarkably, in Ref.~\cite{Alexandrou:1999vx,deForcrand:1999our} it was shown that the chiral condensate vanishes as soon as vortices are removed from the underlined field configurations, while the chiral condensate values are notably larger in center-projected configurations than those on the unmodified lattice. This observation shows that the center vortices are responsible not only for magnetic disorder but also determine the chiral symmetry breaking -- thus, both phenomena are tightly connected \cite{Trewartha:2015nna}.

It is well known that the axial symmetry $U(1)_{\rm A}$ of the classical QCD action is broken by chiral anomaly at quantum level. The topological charge given by the integral of the divergence of the axial current,
\begin{eqnarray}
 Q=\frac{1}{32\pi^2} \int d^4x \epsilon^{\mu\nu\alpha\beta}\, {\rm Tr}[F_{\mu\nu}F_{\alpha\beta}] \,,
\end{eqnarray}
receives contributions from finite action configurations known as instantons \cite{Belavin:1975fg}. Due to the Atiyah-Singer Index theorem, integer $Q$ value has a meaning of a difference of numbers of zero modes of the Dirac operator with positive and negative chiralities. The $\eta’$ meson, which would have been a (pseudo-)Goldstone boson of $U(1)_{\rm A}$ breaking, appears to be way too heavy phenomenologically (above 1 GeV). Its mass is found to be proportional to the topological susceptibility found in pure gauge theory in the chiral and large-$N_c$ limits, i.e.
\begin{eqnarray}
 m_{\eta'}^2\simeq \frac{2N_f}{f_\pi^2}\chi\,, \qquad 
 \chi = \frac{\langle Q \rangle}{V} \,,
\end{eqnarray}
-- the relation known as the Veneziano-Witten formula \cite{Witten:1979vv,Veneziano:1979ec}. Here, $V\to \infty$ is a large volume, and $f_\pi$ is the pion decay constant. For lattice calculations of the topological susceptibility and tests of the Veneziano-Witten formula, see e.g. Refs.~\cite{DelDebbio:2004ns,Cichy:2015jra}.

The topological susceptibility $\chi$ is characterised by the vacuum quantum-field fluctuations in a pure gauge theory, without any quark fields. Like the chiral condensate, the density of topological charge may not seem to immediately connect to the IR property of confinement, and naively one would guess that it may be determined by non-confining configurations such as instantons in the standard picture. However, as was shown in Ref.~\cite{Engelhardt:2000wc} a $P$-vortex acquires a fractional topological charge at ``writhing'' points, and it is possible to get a correct topological susceptability in certain vortex models \cite{Engelhardt:2010ft}. Moreover, the results of Ref.~\cite{deForcrand:1999our} actually demonstrate that the topological charge tends to vanish upon vortex removal, while in Ref.~\cite{Bertle:2001xd} it was shown that $\chi$ computed from $P$-vortices appears to be consistent with the measurements. So the initial naive guess do appear to be wrong, and confinement plays a crucial role here as well.
\begin{figure}[hbt]
\begin{center}
\includegraphics[height=25em]{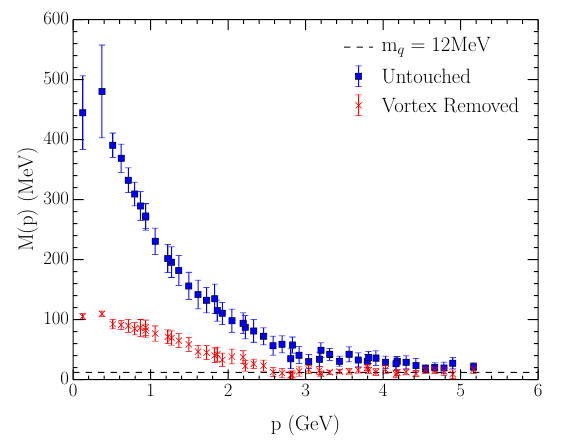}
\caption{Mass function of the effective quark propagator computed accounting for the full and vortex-removed configurations. The figure is taken from Ref.~\cite{Kamleh:2017lij}.
}
\label{fig:Mp}
\end{center}
\end{figure}

Yet another, more recent, test of the vortex mechanism considering the effective quark propagator in Landau gauge in the following IR form
\begin{eqnarray}
 S(k) = \frac{Z(k)}{i\slashed{k} + M(k)}
\end{eqnarray}
has been performed in Refs.~\cite{Trewartha:2015nna,Trewartha:2015ida} (see also Refs.~\cite{Greensite:2011zz,Greensite:2016pfc} for a pedagogical discussion). With an appropriate smoothing (``cooling'') procedure in the $SU(3)$ gauge theory that eliminates short-distance fluctuations, the effective mass $M(k)$ and renormalisation $Z(k)$ functions have been computed for the full, vortex-only and vortex-removed configurations and compared to each other. Removing the vortices causes the mass function to plummet dramatically -- see Fig.~\ref{fig:Mp}, while the full and vortex-only results have appeared to be essentially the same, hence demonstrating a critical role of the vortices in dynamical mass generation and chiral symmetry breaking. The maxima of the action for vortex-only configurations appear to resemble those of instantons, while the number density of those objects is notably similar for the full and vortex-only configurations and by far much larger than that for the vortex-removed case. It seems likely that vortices and instantons are indeed connected in some very non-trivial way.

Remarkably, the center vortices thus appear to describe a number of fundamentally important IR phenomena in non-abelian gauge theories in a gauge-invariant way. Nevertheless, there are also weak points in the vortex mechanism of confinement that require further clarification and, in a perspective, a more complete understanding of, for instance, the Gribov copies problem and a lack of natural explanation of the L\"uscher term. A further, more complete theory of vortices should address these issues, hopefully, on a first-principle basis. 

A lack of a perfect consistency of the vortex scenario with full numerical results for the $SU(3)$ gauge theory has also emerged in the literature. For instance, center projection in $SU(3)$ case yields $2/3$rd of the asymptotic string action computed on the full lattice \cite{Langfeld:2003ev}. However, a consistency has been substantially improved by means of a certain gauge-field smoothing procedure \cite{Trewartha:2015ida}, so this may not be regarded as a critical problem.

In order to make the next step in our understanding of the vortex dynamics, it may enlightening to suggest an EFT of vortices as non-local dynamical objects -- fluctuating surfaces -- in $D$ dimensions where all the IR phenomena described above would emerge naturally among its key predictions. Such a theory known as the {\it random surface model} that resembles a string theory on lattice has been proposed and elaborated e.g. in Refs.~\cite{Engelhardt:1999wr,Engelhardt:2000wc,Engelhardt:2002qs,Quandt:2004gy,Engelhardt:2010ft}.

In order to build the simplest $D=4$ action density of vortices in this framework, one considers an extrinsic curvature of the vortex worldsheet multiplied essentially by a single coupling while the additional Nambu-like string term proportional to the area of the vortex worldsheet appears to be redundant and can be omitted. In the $SU(2)$ version of this model, one assigns $(-1)^n$ to the Wilson loop holonomy for the number of vortex piercings $n$ per minimal loop area and then one averages it over an ensemble of center vortex configurations. The latter can be generated by Monte-Carlo methods for a lattice action density given by the number of cases when a single link is shared by two adjacent orthogonal vortex plaquettes. 

In order to compute the topological charge density in this model, for instance, one employs a weighted stochastic procedure of introducing the monopole lines to the surface of each vortex plaquette (see Sect.~\ref{Sect:monopoles} below for a brief description of the magnetic monopoles' scenario of confinement). The topological susceptibility appears to be insensitive to the monopole lines' density —- a sign of strong predictive power of the model. Besides, the model correctly predicts the emergence of the chiral condensate at $T<T_c$ and restoration of the chiral symmetry at $T>T_c$, with a critical temperature of the transition $T_c$. A variation in the lattice time extension can provide a temperature dependence, and the second-order deconfinement phase transition has been found. The single dimensionless coupling and the lattice spacing $a$ determine a wide range of long-distance non-perturbative phenomena and were fixed through a matching to the physical $T_c/\sqrt{\sigma}$ and $\sigma/a^2=(440 MeV)^2$, in terms of the string tension $\sigma$. Upon such a matching, the temperature-dependent values of $\sigma$, the chiral condensate and $\chi$ are shown to be in agreement with the full theory. Remarkably, in the case of $SU(3)$ gauge theory the random surface model predicts the electric flux tubes in a form of $Y$-shaped string junctions for baryons (three-quark systems) \cite{Engelhardt:2004qq}, also in agreement with numerical results of Refs.~\cite{Alexandrou:2002sn,Takahashi:2004rw}. 

An alternative EFT approach to dynamics of vortices was suggested in Ref.~\cite{Cornwall:1979hz} that is based upon a gauge theory with an adjoint matter field and its gauge-invariant mass term which provides a mass for the gauge field via the Higgs mechanism. Besides the vortex solutions it also naturally reveals another type of solutions with magnetic monopoles running along the vortex sheets that are necessary to generate a topological charge.

For a more thorough discussion on the existing vortex-based scenarios, we refer the reader to Ref.~\cite{Greensite:2011zz}. Now, we turn to alternative scenarios of confinement, yet trying to connect them with the existence of vortices whenever possible.

\section{Gribov-Zwanziger scenario, non-perturbative propagators and gluon chains}
\label{Sect:GZ-scenario}

Starting from the Coulomb gauge, in Refs.~\cite{Gribov:1977wm,Zwanziger:1991gz,Zwanziger:1998ez} it has been suggested that very small eigenvalues of the Faddeev-Popov operator that are located close to the Gribov horizon contribute the most to the Coulomb potential $V_{C}(R)$ and could in principle enhance it to a linear form (see also Ref.~\cite{Greensite:2004ur}). This is the so-called Gribov-Zwanziger scenario of confinement. As it should be for a confined phase, numerical analysis on the lattice demonstrates the linear rise of Coulomb potential $V_{C}(R)$ which is basically a separation-dependent part of the interaction energy of the physical $q\bar q$ state defined as
\begin{eqnarray}
&& R\to \infty \,, \qquad V_{C}(R) \to V(R,T=0) \,, \qquad 
V(R,T)=-\frac{d}{dT}\log[G(R,T)] \,, \nonumber \\ 
&& G(R,T)=\langle \Psi_{q\bar q} | e^{(H-E_0)T} | \Psi_{q\bar q}\rangle\,, \qquad 
\Psi_{q\bar q}=\bar{q}(0)q(R)\Psi_0 \,, 
\label{Coul}
\end{eqnarray}
with the ground-state of the theory $\Psi_0$, the vacuum energy $E_0$ and with self-energy contribution neglected at large $R$. However, the slope of the extracted Coulomb potential $V_{C}(R)$ is significantly (for a factor of 2-3, depending on the gauge coupling) larger than that of the static quark potential $V(R)\simeq \lim_{T\to \infty} V(R,T)$ obtained by gauge-invariant methods \cite{Greensite:2004ke}. Although the latter is in agreement with Zwanziger inequality \cite{Zwanziger:2002sh}, $V(R)\leq V_{C}(R)$, the potential is overconfining prompting discussions in the literature on whether the Coulomb potential in this formulation actually is the full story of confinement or some crucial ingredients are still missing. It is worth mentioning however that the asymptotic string tension of the Coulomb potential appears to vanish as soon as vortices are removed from the underlined gauge field configurations rendering the importance of the vortices for understanding the confinement phenomenon in the Coulomb gauge \cite{Greensite:2003xf}. Such configurations without vortices in fact behave as perturbations of the free gauge theory, in consistency with expectations.

In the confined phase, the Coulomb self-energy of an isolated static charge ${\cal E}$ is expected to be infinite, and the main condition for that reads
\begin{eqnarray}
{\cal E}\propto \int d\lambda \Big\langle \frac{\rho(\lambda)F(\lambda)}{\lambda} 
\Big\rangle \to \infty \,, \qquad 
\lim_{\lambda \to 0} \frac{\rho(\lambda)F(\lambda)}{\lambda} > 0 \,, \qquad 
F(\lambda) = \langle \phi_{\lambda} | (-\nabla^2) | \phi_{\lambda} \rangle \,,
\label{Coulomb-crit}
\end{eqnarray}
where the first relation relies on the continuum limit of small eigenvalues $\lambda\to 0$ of the Faddeev-Popov operator, with the corresponding eigenstates $\phi_{\lambda}$ and density of the eigenvalue distribution $\rho(\lambda)$. By using the lattice methods it was found that  \cite{Greensite:2004ur}
\begin{eqnarray}
\rho(\lambda)\sim \lambda^{0.25} \,, \qquad F(\lambda) \sim \lambda^{0.38} \,,
\end{eqnarray}
yielding a divergent ${\cal E}\to \infty$ and hence satisfying the confinement criterion (\ref{Coulomb-crit}). An enhancement of $\rho(\lambda)$ and $F(\lambda)$ close to the Gribov horizon $\lambda\to 0$ seems to be associated with the role of a center vortex ensemble. However, as was advocated in Ref.~\cite{Greensite:2004ke} the Coulomb force appears to be confining also at temperatures above the deconfinement phase transition temperature which contradicts to the fact that a confining potential must be associated with a phase of magnetic disorder.

The linear confining Coulomb potential in the Gribov-Zwanziger scenario can be associated with the instantaneous part of the two-gluon correlator. So, confinement could be effectively considered as an emergent property due to a gluon exchange with a non-perturbative (dressed) gluon propagator. A naive calculation shows that a linear potential may arise if the propagator of the gluon exchange scales with momentum transfer as $\sim 1/k^4$ at $k\to 0$, at least, in one of the possible gauges \cite{West:1982bt}. One typically attempts to analyse the IR behavior of the effective gluon and ghost propagators and vertices using the formalism of the Dyson–Schwinger equations following from the disappearance of the functional integral of a total derivative,
\begin{eqnarray}
 \Big\langle -\frac{\delta S}{\delta \phi_i(x)} + j_i(x) \Big\rangle = 0 \,,
\end{eqnarray}
with subsequent differentiation over the sources $\{j_k\}$. For a review on phenomenological implications of the Dyson-Schwinger approach, see e.g. Ref.~\cite{Eichmann:2013afa} and references therein.

The full gluon and ghost propagators in Euclidean spacetime are conventionally represented in terms of form factors as
\begin{eqnarray}
 D_{\mu\nu}^{ab}(k) = \delta^{ab}\Big( \delta_{\mu\nu} - \frac{k_\mu k_\nu}{k^2} \Big) \frac{Z(k^2)}{k^2}\,, \qquad
 G^{ab}(k) = \delta^{ab} \frac{J(k^2)}{k^2} \,,
\end{eqnarray}
respectively, such that their IR behavior, as the virtuality of the exchange 
vanishes $k^2\to 0$, is controlled by
\begin{eqnarray}
 Z(k^2) \propto (k^2)^{-\kappa_{\rm gl}} \,, \qquad J(k^2) \propto (k^2)^{-\kappa_{\rm gh}} \,,
\end{eqnarray}
where $\kappa_{\rm gh}$ and $\kappa_{\rm gl}$ are the so-called IR critical exponents (or anomalous dimensions), to be determined in the calculations.

A necessary condition for the Kugo–Ojima confinement criterion is that the ghost propagator features an enhanced (stronger than $1/k^2$) IR singularity, i.e. $\lim_{k\to 0}[J(k^2)]^{-1}=0$, known as the horizon condition \cite{Zwanziger:2001kw}. The second condition is the vanishing gluon propagator, $\lim_{k\to 0}[Z(k^2)/k^2]=0$. This is the exactly case for the so-called scaling solution \cite{Fischer:2006vf,Alkofer:2008jy,Fischer:2009tn} that implies a specific relation between $\kappa_{\rm gl}$ and $\kappa_{\rm gh}$ in $D$-dimensions \cite{Zwanziger:2001kw,Lerche:2002ep,Fischer:2006vf,Fischer:2009tn}
\begin{eqnarray}
 \kappa_{\rm gl} + 2 \kappa_{\rm gh} = - \frac{4-D}{2} \,.
\end{eqnarray}
For $D=4$ case, the values are found to be $\kappa_{\rm gh}\simeq 0.595$ and $\kappa_{\rm gl}\simeq -1.19$, such that the gluon propagator indeed tends to vanish at $k\to 0$. In order to explain confinement, it was argued in Ref.~\cite{Alkofer:2008tt} that the quark-gluon vertex should be sufficiently singular in the long-distance limit, such that its combination with a non-singular gluon propagator gives rise to the confining potential. The scaling solution has been confirmed by a lattice analysis of Ref.~\cite{Maas:2007uv} in $SU(2)$ gauge theory in Landau gauge and only in $D=2$ dimensions, but it was not observed for $D>2$ \cite{Cucchieri:2007rg,Bogolubsky:2009dc}. 

Another well-known solution, the so-called decoupling solution, with
\begin{eqnarray}
 \kappa_{\rm gl} = -1 \,, \qquad \kappa_{\rm gh} = 0 \,,
\end{eqnarray}
has been proposed e.g. in Refs.~\cite{Boucaud:2008ji,Aguilar:2008xm,Dudal:2008sp}. This solution corresponds to a saturated form of the IR gluon propagator tending to a constant and, hence, effectively decouples from the dynamics in analogy to a massive particle. It is worth noticing here that the non-perturbative gluon propagator does not behave as a propagator for a massive state. Indeed, from numerical simulations one observes indications for violation of positivity, in consistency with the fact that no colored gluons exist in the asymptotic spectrum of a gauge theory that is traditionally connected to gluon confinement \cite{Fischer:2008uz,Cucchieri:2004mf}. Besides, the decoupling solution implies a simple $1/k^2$ pole for the ghost propagator. This solution appears to be favoured by known lattice simulations for $D>2$ which also indicate a disagreement with the Kugo-Ojima criterion. A more generic criterion for quark confinement applicable in arbitrary gauges relying on the IR behaviour of ghost and gluon propagators has been proposed in Ref.~\cite{Braun:2007bx}.

One would remark here that the primary probe for the magnetic disorder phase is, of course, the area-law falloff of gauge-invariant observables, Wilson loop VEVs, and not the gluon propagator itself which is not a gauge-invariant object. So one should be extra careful in interpreting the IR behavior of the propagator in order to avoid spurious results. For recent comprehensive effort to obtain a linear static potential in the framework of Dyson-Schwinger formalism in Coulomb gauge, see Ref.~\cite{Cooper:2018xyn}. For a thorough analysis of the Polyakov line VEVs and effective potential based upon the formalism of Functional Renormalisation Group \cite{Wetterich:1992yh} has been performed in Ref.~\cite{Fister:2013bh,Marhauser:2008fz}, and an agreement with lattice results has been found. However, the search for the area-law dependence of large Wilson loops' VEVs with these methods has not been successful so far.

The picture of strongly collimated color-electric flux tubes stretched between the color-charged static sources does not seem to apply to the distribution of color-electric field in the Coulomb gauge \cite{Greensite:2011zz}. Indeed, there is a significant long-range dipole contribution to the Coulomb electric field that would cause rather strong van der Waals type forces between hadrons at large distances. This would immediately contradict to the mass gap existence \cite{Cornwall:1981zr} that requires only short-range forces between composite color-neutral states. This problem generically emerges in any confinement scenario such as the Dyson-Schwinger type approaches where a confining force is associated with a single (dressed) gluon exchange at large distances. While providing a linear potential, such one-gluon exchange scenarios (including the Coulomb confinement one) imply a spread out of the electric field towards large distances, possibly with flux collimation to some extent \cite{Chung:2017ref}. 

A possible development that may eventually address the shortcomings of the Coulomb confinement scenario discussed earlier is to notice that the $q\bar q$ state defined in Eq.~(\ref{Coul}) is not necessarily a minimum-energy state of a system containing a single $q\bar q$ pair, and lower energy states could in principle be constructed using operators $Q_i^j$ -- functionals of the lattice links -- that effectively create ``constituent'' coupled gluons as
\begin{eqnarray}
\tilde{\Psi}_{q\bar q}=\bar{q}^i(0)Q_i^jq_j(R)\Psi_0 \,,
\label{Coul-new-state}
\end{eqnarray}
where schematically,
\begin{eqnarray}
 Q_i^j = a_0\delta_{i}^{j} + a_1A_{i}^{j} + a_2A_{i}^{k}A_{k}^{j} + \dots \,.
\end{eqnarray}
The resulting state effectively represents a chain of gluons bound by attractive forces, with a $q$ and $\bar q$ at the end of the chain, at large $R$, that could, in principle, provide a necessary suppression of the long-range dipole fields. Hence, such a gluon chain may be viewed as a color-electric flux tube itself \cite{Tiktopoulos:1976sj,Greensite:2001nx}. Indeed, as $q$ and $\bar q$ get separated, more and more constituent gluons get pulled out of the vacuum to minimise the energy of the system \cite{Greensite:2014bua,Greensite:2015nea}. This picture rather naturally emerges by expanding the Wilson line stretched between $q$ and $\bar q$ in powers of the gluon field and actually implies the absence of dipole fields at large $R$. In the limit of large number of colors $N$ in the $SU(N)$ theory, such a chain of gluons on a given time slice is dominated by a high-order planar Feynman amplitude that can be, in principle, tackled by analytic methods. 

Among remarkable features of the gluon chain model are the Casimir scaling in the leading order of $1/N$ expansion and a subleading $1/N^2$ string breaking effect at some critical length-scale leading to a correct N-ality dependence of the string tension asymptotically. In the case of heavy (static) charges in the adjoint representation of $SU(N)$, for instance, in the limit $N\to \infty$ two gluon chains instead of one are formed between the charges, leading to twice larger adjoint string tension compared to the one in the fundamental representation, i.e. $\sigma_A=2\sigma_F$. The latter is defined only at intermediate distances, but must disappear at asymptotic distances due to color screening by N-ality zero gluons in the vacuum. Although gluons do not break the center symmetry as such, they take part in the color screening on the same footing as light quarks in QCD such that both quarks and gluons are absent in the asymptotic spectrum in the virtue of $C$-confinement and the string hadronisation model. This suggests a non-trivial but less explored and speculative possibility that the non-perturbative gauge-field vacuum somehow rearranges itself at large distances in such a way that the center symmetry might get broken somehow even without the presence of matter fields in the fundamental representation\footnote{By construction, a pure gauge theory does not contain any fundamental-representation charges. So instead of heavy quarks, the use of ``constituent gluons'' as static color charges to probe the formation and properties of the flux tubes of finite lengths, color screening, string breaking mechanism and the phases of the theory would be the most natural approach to study confinement in pure non-abelian theories.}.

Indeed, pulling the two gluons (or adjoint matter states) apart from each other, eventually the virtual gluons from the QCD vacuum are prompted to bind to the octet-charged sources yielding color-singlet states -- gluelumps -- at asymptotically large distances. Such a gluon color-screening mechanism is very similar to that driven by dynamical virtual quarks being brought on mass-shell to screen the charge of heavy static quarks as the latter move apart, and the energy accumulated in the string is partially spent for that purpose. So, the color-screening and hence the string-breaking phenomenon is not particularly sensitive to N-ality but rather to the color charge itself being the necessary prerequisite for $C$-confinement. While formation of a flux tube between the two gluons at intermediate distances applies for the confining phase in the strongly-coupled regime, $C$-confinement as an asymptotic phenomenon occurs also in the Higgs phase but without formation an intermediate flux tube.

Note, an adjoint string breaks via $1/N^2$ suppressed but very important (at large $R$) interaction between the gluon chains enabling them to transform into a pair of gluelumps as described above (see also Ref.~\cite{Greensite:2001nx}). This correctly generalises for sources in arbitrary gauge group representation giving rise to N-ality dependence of the asymptotic string tension. Such an important string-like property of the gluon chain as the L\"uscher term appears due to fluctuations in the gluons' positions on the chain \cite{Greensite:2001nx}. As was demonstrated in Ref.~\cite{Greensite:2009mi}, introducing up two gluons in a chain preserves the linearity of the Coulomb potential but that is already enough to bring its slope much closer to the true static potential (i.e. obtained by gauge-invariant methods). In this calculation, it was shown that multi-gluon configurations in the chain appear to be increasingly important at large $R$, also strongly reducing the sensitivity of the results to the lattice volume. This means that the long-range dipole field becomes strongly suppressed indicating a possible formation of a localised color flux tube (for more discussion on this aspect, see Ref.~\cite{Greensite:2011zz}).

\section{Dual superconductivity and magnetic monopoles}
\label{Sect:monopoles}

As was proposed a long ago in Refs.~\cite{Nambu:1974zg,tHooft1975,Mandelstam:1974pi,Polyakov:1975rs,Polyakov:1976fu}, the QCD vacuum could be viewed as a ``dual'' superconductor an analog of type-II superconductor where the electric and magnetic fields are interchanged. These studies have pioneered the developments of a beautiful theory of what is sometimes called the {\it dual superconductor picture of confinement}. In a usual superconductor one deals with a condensate of electric charges (in fact, bosonic Cooper pairs) and, due to repulsion (or confinement) the magnetic fields get squeezed into magnetic flux tubes with a constant energy density (Abrikosov vortices). In a ``dual'' superconductor one instead works with a condensate of magnetic charges known as {\it magnetic monopoles} where electric field of static charges would be squeezed (confined) into electric flux tubes. The latter realisation is what we often regard as ordinary confinement in QCD. Both, the static potential of magnetic monopoles in type-II superconductor and the static potential of color-electric charges in ``dual'' superconductor would rise linearly with the charge separation.

This effect gives rise to a very simple picture of confinement essentially based upon a suitable generalization of the Landau-Ginzburg superconductivity theory. Indeed, starting from relativistic abelian Higgs model 
\begin{eqnarray}
S=\int d^Dx\Big( \frac14 F_{\mu\nu}F^{\mu\nu} + |D_\mu \phi|^2 + 
 \frac{\lambda}{4}(\phi^\dagger \phi - v^2)^2 \Big) \,, \qquad 
 D_\mu \phi = \partial_\mu + ieA_\mu \,,
\end{eqnarray}
on recovers the magnetic flux-tube Abrikosov-like solutions dubbed as the Nielsen–Olesen vortices \cite{Nielsen:1973cs}. Attributing a non-trivial winding number $n$ to the Higgs complex phase, a Nielsen–Olesen vortex carries the magnetic flux $2\pi n/e$. In the dual version, such vortex carries an electric flux that confine the electric charges. A particular model, where the dual abelian Higgs model with confinement is realised, is the $N=2$ supersymmetric YM theory known as the Seiberg–Witten model \cite{Seiberg:1994rs,Seiberg:1994aj} having several distinct types of electric flux tubes. In this model, a continuous set of distinct vacua is spanned by the ``moduli'' space of certain scalar field operators. Soft supersymmetry breaking then reduces the theory down to an effective $N=1$ theory where confinement of electric charge is realised due to condensation of the monopole field and electric flux tube formation. This happens in full analogy to confinement of magnetic charge due to magnetic flux tube formation in usual type II superconductors and in the ordinary abelian Higgs model. The duality transformation in the Seiberg–Witten model inverts a certain combination of the effective coupling constant and the $\theta$ angle enabling one to obtain the effective action of light fields at any value of the gauge coupling from the detailed knowledge about the weak-coupling regime of the theory and its infrared singularities (for a detailed review of the underlined concepts and formalism, see e.g. Refs.~\cite{Gomez:1995rk,Bilal:1995hc,DHoker:1999yni}). Such a duality is due to an exact symmetry of the abelian effective theory manifest at low energies, and not of the original $SU(2)$ theory. In fact, this duality is a proper generalisation of the famous electric-magnetic duality of the Dirac formulation of Maxwell electrodynamics (with magnetic monopoles) exchanging the electric charge $q_e$ and its magnetic counterpart $q_m=2\pi/q_e$. Hence, by means of such duality transformation one hopes to learn about strong-coupling (or long-distance) dynamics of a given from the weak-coupling regime of its dual formulation. In Ref.~\cite{Douglas:1995nw} it was shown that $k$-string tensions in $SU(N)$ version of the Seiberg–Witten model obey the Sine law
\begin{eqnarray}
\sigma_r=\frac{\sin(\pi k/N) }{\sin(\pi/N)}\sigma_F \,,
\label{Sine-law}
\end{eqnarray}
being numerically not very different to that of the Casimir scaling (c.f. Eq.~(\ref{Casimir-sigma})).

In itself, the superconductivity picture of confinement is an abelian mechanism which has been explored originally by Polyakov \cite{Polyakov:1987ez} in the context of confinement of electric charges in $D=2+1$ compact $U(1)$ gauge theory. This theory turns out to be an important starting point to approach QCD confinement. While in $D=2+1$ case, the compact QED features monopoles (topological excitations), in $D=3+1$ those monopoles are point-like defects in spacetime, i.e. they are also instantons. Effectively integrating out all the DoFs except monopoles in $D=2+1$ compact QED, it was shown in Refs.~\cite{Polyakov:1987ez,Polyakov:1975rs,Polyakov:1976fu} that the action of the monopole gas interacting by means of Coulomb force on the lattice reads 
\begin{eqnarray}
S_{\rm m} = \frac{2\pi^2}{g^2a} 
\Big[ \sum_{i\not=j}m_im_j G(r_i - r_j) + G(0) \sum_{i}m_i^2 \Big] \,,
\end{eqnarray}
with $i,j = 1\dots N$ for $N$ monopoles, and the lattice Coulomb propagator $G$ at large distances behaves as $G \sim 1/4\pi|r_i-r_j|$. A Wilson loop in this approach can be expressed in terms of the monopole density and appears as a current loop that generates its magnetic field being effectively screened away by the (anti)monopoles from the background. Such an effect causes the area-law falloff for the Wilson loop VEVs. Polyakov has explicitly demonstrated that even arbitrarily low density of these monopoles is sufficient to produce confinement and the mass gap of the theory. This happens in a regime when the entropy related to the size and shape of large Wilson loops wins over the cost in the monopole action for a large loop. The latter effect occurs at any coupling for $D=3$ QED, but only for large enough couplings in the $D=4$ case.

In the case of YM theories one needs to extract an abelian subgroup from the gauge group e.g. by means of an adjoint Higgs field. An important realisation in the case of $SU(2)$ gauge theory is the Georgi–Glashow model where in the minimum of the Higgs potential and in unitary gauge there is a residual $U(1)$ local gauge symmetry. Due to this symmetry, the model exhibits magnetic (‘t Hooft–Polyakov) monopoles \cite{tHooft:1974kcl,Polyakov:1974ek} as instanton solutions of the classical equations of motion in $D=3$, or as static solutions (solitons) in $D=3+1$. The Higgs field that is used to fix the unitary gauge necessarily vanishes at the center of each ‘t Hooft–Polyakov monopole making the unitary gauge fixing ambiguous at those sites. The Wilson loop VEVs are then computed in a similar way as was done in compact $D=3$ QED resulting in a finite string tension $\sigma \sim \exp(-S_{\rm m})$ \cite{Polyakov:1976fu}. In $D=4$, the Georgi–Glashow theory has both confining and non-confining phases, however, stable monopole solutions exist only in the non-confining phase where they do not form a Coulomb plasma.

An important caveat in $D=3$ theory is that one cannot simply neglect of effects of $W$ bosons at large distances (and hence in the analysis of confinement) in the long-range effective action. Indeed, the string tensions cannot acquire a correct N-ality dependence without $W$ bosons. The Coulomb monopole gas approximation can be justified in a certain intermediate range below a string-breaking length-scale where a $W$ bosons carrying two units of electric charge are pair-produced and screen the charges of the static sources, also possessing two units of electric charge. Analogically, the dual abelian Higgs model that ignores the effect of $W$ bosons predicts a wrong N-ality dependence of the Wilson loop VEVs. Thus, it is unable to consistently describe long-range physics of vacuum fluctuations at characteristic distances exceeding the color screening length-scale. Non-abelian supersymmetric versions of the dual Higgs model have been proposed in a number of existing works yielding specific non-abelian vortex solutions; see for a detailed review on these aspects e.g. Refs.~\cite{Shifman:2007ce,Shifman:2010jp} and references therein.

Dynamical ``abelization'' of $SU(N)$ gauge fields can be achieved even without an adjoint Higgs field. Instead of using an adjoint Higgs field, another way to extract a Cartan (abelian) subgroup $U(1)^{N-1}$ of $SU(N)$ suggested in Ref.~\cite{tHooft:1981bkw} is the so-called {\it abelian projection}, using a composite operator which transforms like a matter field in adjoint representation and fixing a gauge in which this operator is diagonal. The same effect emerges also with adjoint fermions fields \cite{Unsal:2007vu,Shifman:2008ja}, or by adding a trace deformation term to the action \cite{Unsal:2008ch}, both methods have been successfully explored by lattice simulations (see e.g.~Refs.~\cite{Cossu:2013ora,Bergner:2018unx,Bonati:2020lal}).

The gluons from the coset of abelian projection are charged under $U(1)^{N-1}$ while the monopole condensation would describe their confinement in a way similar to the dual abelian Higgs model. The basic idea then is to look for a specific gauge in which the quantum fluctuations of the $U(1)^{N-1}$-charged gluons are strongly suppressed compared to the fluctuations of ``photons'' from the Cartan subgroup $U(1)^{N-1}$. In such a gauge called {\it maximal abelian gauge} \cite{Kronfeld:1987ri} the link variables would be close to a diagonal form. For instance, in the $SU(2)$ gauge theory this is achieved by means of requiring $\sum {\rm Tr}[U_\mu(x)\sigma_3U^\dagger_\mu(x)\sigma_3]$ to be maximal while leaving the residual $U(1)$ symmetry w.r.t. gauge transformations
\begin{eqnarray}
U_\mu(x) \to e^{i\phi(x)\sigma_3}U_\mu(x)e^{-i\phi(x+\hat\mu)\sigma_3} \,.
\end{eqnarray}
This enables one to decompose $U_\mu(x)=C_\mu(x)u_\mu(x)$, where $C_\mu(x)$ matrix is expressed in terms of a ``matter'' field $c_\mu(x)$ with two units of $U(1)$ charge, while the diagonal $u_\mu(x)={\rm diag}(\exp(i\theta_\mu(x)),\exp(-i\theta_\mu(x)))$ is given in terms of abelian $U(1)$ gauge field $\theta_\mu(x)$, the ``photon'', coupled to the ``matter'' field $c_\mu(x)$. One therefore obtains the abelian-projected lattice by means of $U_\mu(x)\to \exp(i\theta_\mu(x))$ projection. Note in the case of $SU(3)$ theory the maximal abelian projection is not unambiguously defined as has been discussed for instance in Ref.~\cite{Stack:2002ysv}.

In the {\it monopole dominance approximation} \cite{Shiba:1994ab,Stack:1994wm}, one then replaces the link variables by the monopole links constructed from the Dirac string variables and the Coulomb propagator and then one computes the VEVs of the Wilson loops over an ensemble of such monopoles. This procedure leads to (almost) the same values for the asymptotic string tensions of the single-charged Wilson loops in the $SU(2)$ lattice gauge theory as in the gauge-invariant approach. Also, the single-charged Polyakov loops computed in the abelian-projected configurations and in the monopole dominance approximation agree with each other and both vanish below the critical temperature of the deconfinement transition, in consistency with expectations. However, these results do not agree for double-charged Polyakov loops. Vanishing VEVs of the latter, and hence the confining disorder, are found in the monopole dominance approximation which is inconsistent with the charge screening effect that must be in place for double-charged static sources, and for that matter -- with the N-ality requirement. This means that in the case of magnetic disorder dominated by abelian gauge field configurations the abelian flux can not be distributed according to the Coulomb monopole-gas approximation.
\begin{figure}[hbt]
\begin{center}
\includegraphics[height=25em]{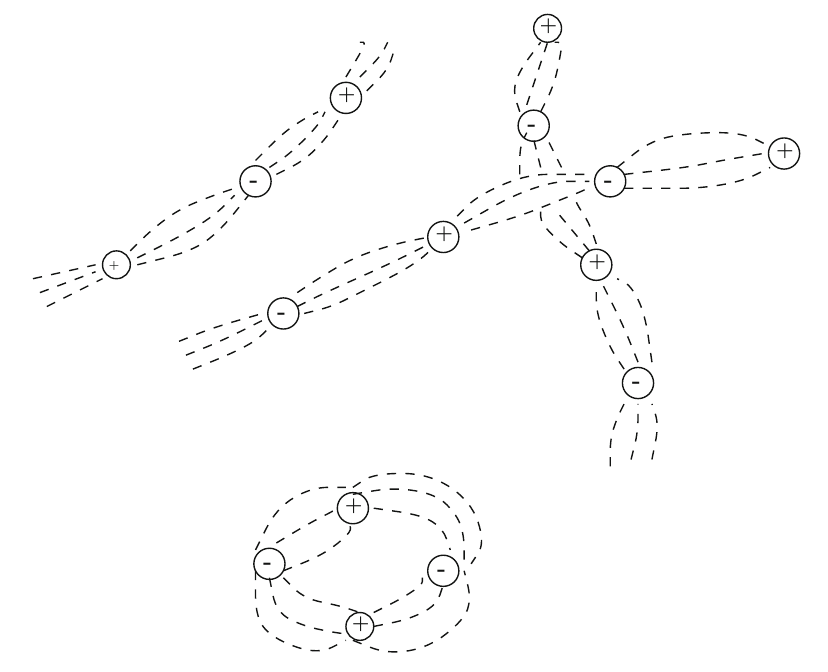}
\caption{Formation of a center vortex through a collimation of the monopole/antimonopole flux. The figure is taken from Ref.~\cite{Greensite:2016pfc}.
}
\label{fig:monopole-vortex}
\end{center}
\end{figure}

The latter problem is not present in the abelian-projected configurations yielding a correct asymptotic behavior of large Wilson loop VEVs in the fundamental representation. Fixing an abelian projection gauge in the $SU(2)$ gauge theory arranges the monopoles and antimonopoles coupled to each other into a chain with the total monopole flux of $\pm 2\pi$. At a certain fixed time, such a flux can be squeezed into center vortex structures on the abelian-projected lattice \cite{Ambjorn:1998qp} -- for an illustration of this effect, see Fig.~\ref{fig:monopole-vortex}. Indeed, the numerical analysis that locates both the (anti)monopoles through abelian projection and the center vortices through center projection showed that almost all (anti)monopoles are located on the vortex sheets arranging themselves into alternating order in a chain (for an inspiring discussion, see Refs.~\cite{Greensite:2011zz,Greensite:2016pfc}). The fact that the double-charged (Wilson and Polyakov) loops do not get contributions from a linking with such vortices on the abelian-projected lattice is reflected in a vanishing asymptotic string tension in this case, in agreement with the charge screening effect.

It is instructive to introduce a specific order parameter, the VEV of a monopole creation operator $\langle \mu(\vec x) \rangle$, that would signal an emergence of the dual superconducting phase in a non-abelian gauge theory \cite{DelDebbio:1994sx,DiGiacomo:1999yas}. The operator $\mu(\vec x)$ effectively inserts a monopole configuration at a certain position into the system such that it does not commute with the total magnetic charge operator, and hence its VEV would break the corresponding dual $U(1)$ gauge symmetry, a remnant of the gauge symmetry. According to the monopole condensation mechanism of confinement, the system is in a confining phase if and only if $\langle \mu(\vec x) \rangle\not = 0$ while a transition to a non-confining configuration occurs when $\langle \mu(\vec x) \rangle \to 0$ which indeed coincides with a more generic numerical analysis in the full theory (also at finite temperatures). There are, however, severe ambiguities such that $\langle \mu(\vec x) \rangle$ may vanish also in the absence of any thermodynamic transition to a deconfined phase \cite{Greensite:2008ss}. Indeed, as was already briefly discussed earlier the breaking of a gauge symmetry remnant cannot be utilised as a correct signature of the magnetic disorder phase.

In a pure non-abelian gauge theory in $D=4$, classical instanton solutions can not be responsible for magnetic disorder of the vacuum field configurations since their field strength falls off too fast at large distances. However, at finite temperature the instanton solutions as saddle points of the Euclidean gauge fields' action called calorons can be relevant for confinement. The latter solutions were found in Refs.~\cite{Kraan:1998pm,Kraan:1998kp,Lee:1998bb} and are known in the literature as {\it KvBLL solutions}. They may contain monopole constituents sourcing both electric and magnetic fields, also known as dyons or Bogolmolny-Prasad-Sommerfield (BPS) monopoles \cite{Bogomolny:1975de,Prasad:1975kr}, which can be widely separated. The thermal approach to pure 4D YM theories based upon nonperturbative results on a thermal ground state in the deconfined phase derived from (anti)caloron ensemble has been thoroughly discussed in Ref.~\cite{Hofmann-book} and in references therein. Among important corollaries to this approach is, for instance, the derivation of the 3D critical exponent of the Ising model for correlation length cirticality.

The early work of Ref.~\cite{Harrington:1978ve} made important contributions to understanding the trivial-holonomy calorons in $SU(2)$ Euclidean gauge theory based on Ref.~\cite{tHooft:1976snw}, while nontrivial holonomy solutions have been studied e.g. in Refs.~\cite{Kraan:1998pm,Lee:1998bb,Nahm:1983sv,Garland:1988bv,Nahm:1979yw}. Considering for instance the maximally non-trivial Polyakov loop holonomy $P(\vec x)$ introduced in $SU(N)$ in Eq.~(\ref{Polyakov-holonomy}), where $\mu_j$ are ordered and spaced with a maximal distance from each other
\begin{eqnarray}
\mu_n^{\rm max} = -\frac12 -\frac{1}{2N} +\frac{n}{N} \,,
\end{eqnarray}
the probability density of calorons in the vacuum would be peaked at ${\rm Tr}P(\vec x)=0$. As was discussed earlier, in the center vortex mechanism of confinement the vanishing Polyakov loop expectation value computed on an ensemble, where positive and negative fluctuations in vortex configurations cancel out, is a signature of unbroken center symmetry and, hence, that of the confinement property. Notably, in the caloron configurations the maximally non-trivial Polyakov loop holonomy vanishes by itself before any averaging as the basic property of such configurations.

Due to this property, a system of widely separated dyons, the dyon gas, whose free energy is minimal for ${\rm Tr}\,P(\vec x)=0$, has been considered as the basis for description of the magnetic disorder in YM theories \cite{Diakonov:2007nv}. Indeed, it was shown that the $k$-string tensions extracted from space-like Wilson loops are in agreement with those that determine the asymptotic behavior of Polyakov loop correlators, and follow the Sine law (\ref{Sine-law}). In the high temperature regime, a phase transition to the deconfinement phase occurs with $T_c/\sqrt{\sigma}$ values being in perfect agreement with numerical lattice results. Despite of such tremendous success, the path integral measure for the multi-dyon configurations appears to be not positively definite thus violating the basic property of the exact measure \cite{Bruckmann:2009nw}. However, numerical simulations of Refs.~\cite{Gerhold:2006sk} with a suitable parameterisation of the integration measure confirmed that the confining static potential indeed emerges in the dyon gas approximation. Thus, one can conclude that the monopole mechanism based upon the caloron classical solutions is one of the most promising scenarios of confinement in non-abelian gauge theories. 

There are some critical points to be made regarding the dyon gas picture of confinement neatly summarised in Ref.~\cite{Greensite:2011zz}. The same question as for the monopole Coulomb gas applies also for the dyon gas regarding the asymptotic string tension of double-charged Wilson loops that should disappear due to a screening by gluons. Another question concerns the probability distribution of Polyakov loop holonomies which is peaked for a vanishing maximally-nontrivial holonomy in the fundamental representation i.e. ${\rm Tr}\, P(\vec x)=0$. If this is indeed true, it implies a negative expectation value of the Polyakov loop in the adjoint representation. However, if the latter is positive, the probability distribution would be peaked at the center-element holonomy as suggested by the center vortex scenario of confinement. Remarkably, the expectation value of the adjoint Polyakov line in the phase of magnetic disorder has been found to be positive in the $SU(3)$ theory in Ref.~\cite{Gupta:2007ax}.

Besides, considering the asymptotic behavior of double-winding Wilson loops VEVs' (i.e. Wilson loops winding around closed co-planar loops $C_1$ and $C_2$), one reveals a dramatic difference in predictions of the monopole and vortex mechanisms of confinement \cite{Greensite:2014gra} (see also \cite{Greensite:2011zz,Greensite:2016pfc}). The vortex scenario provides the ``difference-of-areas'' law behavior for such loops, $\sim \exp[-\sigma|A(C_1)-A(C_2)|$, where the ``$-$'' sign is due to a vortex linking to the largest loop, correctly reproducing the full lattice results. In this case, the monopole scenario predicts the ``sum-of-areas'' falloff as $\sim \exp[-\sigma(A(C_1)+A(C_2))]$, which is disfavoured by numerical simulations. The latter observation indicates that the vacuum cannot be in a dual-superconducting state of monopole/dyon plasma. In fact, the ``difference-of-areas'' law is restored by heavy $W$ bosons that are present in the full YM theory, but not in an abelain part of it. As was suggested in Ref.~\cite{Greensite:2016pfc}, upon integrating out the $W$ bosons' states, one expects the monopole-antimonopole lines to get collimated into $\mathbb{Z}_2$ vortices, as is illustrated in Fig.~\ref{fig:monopole-vortex}, i.e. in a similar fashion to what has been seen on an abelian-projected lattice. This effectively turns the monopole ensemble into a configuration of $\mathbb{Z}_2$ vortices, offering an intricate connection between the two pictures of confinement.

Another observation of Ref.~\cite{Diakonov:2009jq} in the case of $G_2$ gauge theory has suggested that the Polyakov loop expectation value exactly vanishes in the dyon gas picture. The color screening in $G_2$, however, requires to bind a static source in the fundamental representation to a minimum of three gluons likely leading to a very small, but non-zero, Polyakov loop expectation value that would be difficult to identify numerically \cite{Greensite:2011zz}. If the color screening mechanism is a valid approach, the center symmetry breaking at large distances would be manifest in $G_2$, whereas the dyon gas approximation, like the Coulomb gas approximation discussed earlier, might be lacking something relevant in the asymptotic regime.

\section{Separation-of-charge confinement criterion}
\label{Sect:separation}

A clear symmetry-based distinction between the confining and Higgs phases in a gauge theory with fundamental representation matter fields has been recently proposed in Refs.~\cite{Greensite:2017ajx,Greensite:2018mhh,Greensite:2020nhg}. An important generalised criterion of confinement valid in {\it both} pure YM theories and YM theories with matter in fundamental representation states that
\begin{eqnarray}
E_V(R)\equiv \langle \Psi_V | H | \Psi_V \rangle - {\cal E}_{\rm vac} \geq E_0(R) \,, \label{Sc-conf}
\end{eqnarray}
with $\Psi_V$ being the $q\bar q$ state connected by a Wilson line,
\begin{eqnarray}
\Psi_V \equiv \bar q^a(\vec x)V^{ab}(\vec x,\vec y;A)q^b(\vec y)\Psi_0 \,,
\end{eqnarray}
for {\it any} choice of the gauge bi-covariant non-local operator $V^{ab}(\vec x,\vec y;A)$. The latter depends only on the gauge field, thus, eliminating any possibility for a string breaking by means of dynamical matter fields. This criterion is a necessary and sufficient condition for {\it separation-of-charge confinement} (or $S_c$-confinement, for short), which is meaningful only in gauge theories with a non-trivial center symmetry. In Eq.~(\ref{Sc-conf}), $H$ is the Hamiltonian, ${\cal E}_{\rm vac}$ is the vacuum energy, $E_0(R) \sim \sigma R$ at $R\to \infty$ is an asymptotically linear function which has the meaning of the ground-state energy of the $q\bar q$ in a pure $SU(N)$ gauge theory (but not in the one with matter fields) where the above criterion is equivalent to the area-law falloff of Wilson loop VEVs. 

In the $SU(2)$ gauge-Higgs theory, the confining phase is found for $\gamma \ll \beta \ll 1$ and $\gamma \ll 1/10$ where the $S_c$-confinement condition (\ref{Sc-conf}) is satisfied. But deeply in the Higgs phase, for other couplings' ranges, this criterion is not fulfilled, hence we deal with only a weaker $C$-confinement situation there. In Refs.~\cite{Greensite:2017ajx,Greensite:2018mhh,Greensite:2020nhg} it has been shown that a transition between the $C$- and $S_c$-confinement phases must take place in the gauge-Higgs theory, and the unbroken custodial symmetry has been found to separate the $S_c$-confining (if not massless) phase from the Higgs phase corresponding to a $C$-confined spin glass state where the custodial symmetry is actually broken. It would be very interesting to see how such a new concept of $S_c$-confinement can be applied for more realistic theories such as QCD.

\section{Separating the Higgs and confinement phases: vortex holonomy phase}
\label{Sect:Hconf-transitions}

As was discussed earlier, the results of Refs.~\cite{Osterwalder:1977pc,Fradkin:1978dv,Banks:1979fi} state that it is possible to identify a continuous path between the Higgs and confinement regimes where no first-order phase transitions occur like what is believed to happen in physics of high-to-low temperature QCD (smooth crossover) transition and in several other specific models. Indeed, one would naively expect that such a continuity always takes place unless the phases are separated by different realisations of global symmetries.

An important counter-example to this statement has been recently explored in a whole class of models in Refs.~\cite{Cherman:2018jir,Cherman:2020hbe}. Namely, it has been demonstrated that even in the case of a spontaneous global $U(1)$ symmetry breaking in both Higgs and confinement regimes (an analog to the baryon symmetry breaking in dense QCD at low $T$) it is still possible to identify a novel {\it non-local} order parameter (a vortex holonomy phase) that separates the two phases leading to a thermodynamical phase transition. This proof provides an important argument against the quark-hadron continuity (Sch\"afer-Wilczek) conjecture in dense QCD mentioned above in Sect.~\ref{Sect:QCDphases}.

For this purpose, the authors of Ref.~\cite{Cherman:2020hbe} started with the Polyakov's $D=4$ compact $U(1)$ gauge theory discussed in the previous section, then imposed a single global $U(1)_{\rm G}$ symmetry (an analog to $U(1)_{\rm B}$ of QCD) and added three complex scalar fields that effectively mimic dynamical quark fields in QCD, with the following assignments under $U(1)\times U(1)_{\rm G}$ group
\begin{equation}
    \phi^+=\{+1,-1\} \,, \qquad \phi^-=\{-1,-1\} \,, \qquad \phi^0=\{0,+2\} \,.
\end{equation}
Then, the product $\phi^+\phi^-$ appears to be an analog to the gauge invariant baryon operator, $\phi^0$ can couple to $\phi^+\phi^-$ product and can be considered as a baryon interpolating field or a source for baryons. A VEV in $\phi^0$ would bring the theory into a superfluid phase of spontaneously broken $U(1)_{\rm G}$, such that the global symmetries' realisation is the same in the confinement and Higgs phases. In the monopole-driven confinement picture, the monopoles in this theory would have a finite action that can be UV-completed through an $SU(2)$ symmetry just as in the original Polyakov's description. Finally, additional discrete $\mathbb{Z}_2$ (charge conjugation and flavor-flip) symmetries have also been introduced, and can be considered as analogous to the flavor symmetry in QCD. The effect of monopoles in this model induces an additional term in the Lagrangian \cite{Polyakov:1976fu}
\begin{eqnarray}
V_m(\sigma) \propto e^{-S_I}\cos\sigma \,,
\end{eqnarray}
which is a potential for the ``dual photon'' -- a periodic scalar field $\sigma \to \sigma + 2\pi$ related to the field strength by the abelian duality relation
\begin{eqnarray}
F_{\mu\nu}=\frac{ie^2}{2\pi} \epsilon_{\mu\nu\lambda}\partial^\lambda \sigma \,.
\end{eqnarray}

As one varies the adjustable mass parameters three different regimes of the theory emerge. One of them corresponds to compact $3D$ $U(1)$ theory with heavy scalar quarks and confinement where no symmetries are spontaneously broken (``gapped confined'' regime). The second regime features the Higgs mechanism with a non-zero VEV $\langle\phi^+\phi^-\rangle\not=0$, such that a cubic $\phi^+\phi^-\phi^0$ term in the potential drives condensation of $\phi_0$, $\langle\phi^0\rangle\not=0$ and hence spontaneous breaking of $U(1)_{\rm G}$ (Higgs phase with a single massless Goldstone boson). The third regime has monopole-driven confinement, while $\langle\phi^0\rangle\not=0$ spontaneously breaks $U(1)_{\rm G}$ symmetry and the heavy charged scalars are very heavy and may be disregarded. A key question here is that whether the two phases with spontaneously broken $U(1)_{\rm G}$ (``Higgs'' and ``confining'') with no distinguishing local order parameter are really two distinct phases or might be continuously connected. 

The main claim of Ref.~\cite{Cherman:2020hbe} is that these phases are distinct and can be distinguished {\it only} by a new non-local order parameter that is connected with topological excitations. Physically, both phases with spontaneously broken $U(1)_{\rm G}$ should be considered as superfluids that have vortices. As a consequence of $U(1)_{\rm G}$ breaking, the theory possesses a gapless Nambu-Goldstone mode which is a phase of $\langle\phi^0\rangle$ condensate, as well as topologically stable vortex excitations when the phase of the $\langle\phi^0\rangle$ condensate winds around the unit circle when one goes around a given loop. In ordinary superfluids in three spacial dimensions this provides vortex loops, but in $D=2+1$ vortices act like point particles, the point around which the condensate phase winds. Winding of that phase, in the language of superfluids, is exactly what is called quantized minimal ``circulation''. The winding number can then be found in terms of a contour integral of the gradient of the phase, or simply as
\begin{eqnarray}
w \propto \oint_C \frac{d\langle\phi^0\rangle}{|\langle\phi^0\rangle|} \,.
\end{eqnarray}
The charge particles in the superfluid phase would then interact with (minimal-energy) vortices through acquiring an Aharonov-Bohm phase as one sends a charged particle into a loop that links with the worldline of the vortex. This phase is measured by the Wilson loop holonomy
\begin{equation}
    \Omega(C)= e^{i\oint_C A} \,,
\end{equation}
whose expectation value 
\begin{equation}
    \langle\Omega(C)\rangle \sim e^{-mP(C)}\,.
\end{equation}
This means that short-range quantum fluctuations in the phase (with dynamical fundamental representation charges) automatically lead to a perimeter-law decay of the VEV of a large Wilson loop. In fact, this represents the same physics as that of string breaking that turns the area-law behavior of a pure YM theory into a perimeter-law behavior present in real QCD and in the Higgs phase as was discussed in detail in previous sections.

Let us consider a Wilson loop expectation value in the presence of a minimal-energy vortex which can be thought in terms of a constrained functional integral where one integrates over all the field configurations in the theory but with a constraint forcing the presence of a vortex along some large worldline $C$ in $D=3$ spacetime. The same short-range physics guarantees it is going to have the same perimeter-law behavior but it can have an additional phase factor
\begin{equation}
    \langle\Omega(C)\rangle_{w=1} \sim e^{i\Phi} e^{-mP(C)} \,.
\end{equation}
In order to extract the phase, one defines the ratio \cite{Cherman:2020hbe} -- the {\it vortex order parameter},
\begin{equation}
    O_{\Omega} \equiv \lim_{r_v\to \infty} \frac{\langle\Omega(C)\rangle_{w=1}}{\langle\Omega(C)\rangle} \,,
\end{equation}
taking the size of the Wilson loop and its separation from the vortex $r_v$ arbitrarily large simultaneously. The symmetries of the theory guarantee that Wilson loop VEV $\langle\Omega(C)\rangle$ is real, and it can be made positive. While the charge conjugation and reflection symmetries are broken, the flavor-flip symmetry is preserved in the presence of a vortex. The latter also flips the sign of the gauge field and hence conjugates the holonomy guaranteeing that the vortex-constrained expectation value of the Wilson loop is also real, but can be either positive or negative, i.e. $O_{\Omega}=\pm 1$. This means that the vortex order parameter cannot vary smoothly under variations of model parameters when moving between the two phases. 

In Ref.~\cite{Cherman:2020hbe} it was demonstrated by means of a semi-classical analysis and through minimization of the vortex energy that in the Higgs phase the vortex order parameter must be equal to $-1$. Integrating out the heavy charged scalars in the confining phase, one can conclude that the vortex-constrained Wilson loop expectation value hardly knows about the presence of the vortex, thus $O_{\Omega}=1$. It was argued in Ref.~\cite{Cherman:2020hbe} that if one varies parameters of the theory and at some point the magnetic flux carried by vortices suddenly jumps, which is what the vortex order parameter is really probing, that surely is going to change the core energy density of the vortex. In this case, it does change the probability of having vortex excitations in the ground state wave function affecting the ground state energy density. In other words, a sudden change in the vortex properties really should be reflected in genuine thermodynamic phase transition.

Given the close analogies of the considered model with QCD, by construction of the model above, this conclusion may be straightforwardly generalised to a $D=4$ non-abelian theory with fundamental-representation matter charged under a given global symmetry. This is the case of dense QCD with broken $U(1)_{\rm B}$, where the $SU(3)$ vortex order parameter can be shown to take two distinct values in two phases \cite{Cherman:2020hbe}
\begin{equation}
    O_{\Omega} \equiv \lim_{r_v\to \infty} 
    \frac{\langle {\rm Tr}\Omega(C)\rangle_{w=1}}
    {\langle{\rm Tr}\Omega(C)\rangle} = 
    \begin{cases} 
      e^{2\pi i/3} & {\rm CFL/Higgs} \; {\rm phase} \\
      +1 & {\rm nuclear/confining} \; {\rm phase}
   \end{cases} \,,
\end{equation}
such that the latter $O_{\Omega}=+1$ is understood as the characteristic signature of confined QCD phase with spontaneously broken baryon symmetry.

This indeed illustrates the main point suggesting that the quark-hadron continuity between the nuclear and quark-matter phases may not hold. Ref.~\cite{Alford:2018mqj} has argued that despite the noted discontinuity in the vortex order parameter the continuity of phases may still be intact due to a continuous connection between the vortex in the CSC/CFL phase and the corresponding one in the nuclear phase. In response to this claim, Ref.~\cite{Cherman:2020hbe} has explicitly proven the existence of thermodynamical phase transition at the interface between the two phases connected to the manifest discontinuity in the non-local vortex order parameter. While the debate about this important issue will likely continue in the literature, it once again reveals the surprising underlined complexity of the non-perturbative QCD vacuum and the associated approaches to the confinement problem may still not be in their final form. It would be very instructive to find possible connections between the vortex holonomy phase and its discontinuity with the $S_c$-confinement criterion briefly discussed in the previous section, both pursuing the same goal of sharply separating the Higgs and confining phases.

\section{Summary}
\label{Sect:Summary}

To summarise, the mass gap and color confinement that are realised already in a gauge-Higgs theory may not be connected to an asymptotically rising static potential and Regge trajectories and, hence, they do not necessarily represent an emergence of the magnetic disorder state. On the other hand, the magnetic disorder and the associated area-law behaviour of Wilson line VEVs imply color confinement and the mass gap automatically. In this sense, color confinement and the mass gap represent only a small part of a bigger picture of confinement and should be considered as a consequence of the confined magnetic disorder state and flux tubes formation corresponding to a phase with unbroken non-trivial center symmetry. If there is no a non-analytic boundary between the massive and magnetic disorder phases at finite values of coupling constants and at some critical length-scale, i.e. a first-order phase transition, one should talk about a single massive phase at all scales, like for instance in the gauge-Higgs theories. The flux tubes formation is only an approximate picture in this case, roughly consistent with reality at some intermediate distances, but it does not necessarily represent an emergence of a new phase. 

One of the big questions for real QCD though, i.e. with physical quarks and gluons, which would distinguish it from EW theory is then whether a magnetic disorder phase really exists within some finite interval of characteristic length-scales that would abruptly transit to a massive phase at asymptotically large length-scales (due to string-breaking), or not. If not, then real QCD would always be considered on the same footing with a gauge-Higgs theory as existing in the massive phase only which is one of the basic options actively discussed in the literature. One thing, however, that distinguishes real QCD from the EW theory is the existence of experimentally observed Regge trajectories in QCD with light quarks that, in fact, may indicate the presence of a non-analytic phase boundary at moderately large distances in QCD, in variance to EW theory, while the both would be in the massive phase asymptotically. Numerical values of the coupling constant here should play decisive role here, and for weak couplings the magnetic disorder may not emerge at all.

In fact, light ``sea'' quarks, i.e. with masses way below the confinement energy scale of QCD, emerge due to gluon splitting $G_a\to q\bar q$ such that correlated $q\bar q$ pairs could be viewed as effective gluons as long as the resolution length-scale is above the wave-length of such a pair. If at such length-scales the strong-coupling constant is large enough, one can view physics in such a regime as that of an effective pure YM theory in a magnetically disordered phase with unbroken center symmetry. By pulling $q$ and $\bar q$ apart from each other at length-scales larger than the resolution scale, the center symmetry gets effectively broken and the theory enters the massive phase. In the infinite quark mass limit, however, the string-breaking length-scale grows indefinitely making the magnetic disorder phase valid for asymptotically large distances.

Depending on the values of the gauge coupling constant, the same can occur in a gauge-Higgs theory in a strongly coupled regime which is supported by lattice simulations. Thus, we arrive to radically different phase structure of a gauge theory depending on whether it is in a strongly-coupled or in a weakly-coupled regime. But even without matter fields involved, the major problem of confinement remains, namely, to understand why pure YM theories with a non-trivial center symmetry in $D\leq 4$ dimensions can only exist in a state of magnetic disorder. Once this key problem is solved, it will become clearer under what conditions realistic theories like a gauge-Higgs theory or real QCD may exist the magnetically disordered phase, and a first-order phase transition towards the massive phase may occur, if at all.

As was elaborated in this review, the phase structure and properties of the quantum QCD vacuum is still under intense explorations, both experimentally and theoretically, numerically and analytically, and is far from its complete and satisfactory description. However, a tremendous progress has been made and some basic contours of the fundamental picture of confinement have started to emerge. We do understand a confining phase as an asymptotic magnetic disorder phase with unbroken non-trivial center symmetry that manifests itself through the area-law behavior of large Wilson loop VEVs and, hence, a linear rise of the corresponding static (string) potential. However, real QCD features such a phase only pre-asymptotically where color-electric flux tubes exist at not-too-large distances, while they break apart at length-scales beyond an inverse to the lightest meson mass (pion) scale yielding a massive phase asymptotically. Quanta of magnetic flux, vortices, have proven to play a crucial role at all stages, from formation of a flux tube to its breaking. Given the overwhelming qualitative and quantitative evidence collected in vast amounts of studies in the literature, the center vortex mechanism remains among the most favored scenarios of confinement so far. Other ideas such as the monopole scenario highlight the underlined complexity of confining phase and phase transitions and offer different perspectives, but in one way or another connect to the vortex picture. Various order parameters briefly described in this review probe the confining phase and are capable of separating it from a non-confining (Higgs) phase, with the latter remaining under a continuous debate in the literature.

\section*{Acknowledgements}

R.P.~is supported in part by the Swedish Research Council grant, contract number 2016-05996, as well as by the European Research Council (ERC) under the European Union's Horizon 2020 research and innovation programme (grant agreement No 668679). M. \v{S}. is partially supported by the grants LTT17018 and LTT18002 of the Ministry of Education of the Czech Republic.

\bibliographystyle{mdpi}
\bibliography{bib}

\begin{thebibliography}{-------}
\providecommand{\natexlab}[1]{#1}

\bibitem[Brambilla \em{et~al.}(2014)Brambilla et~al.]{Brambilla:2014jmp}
Brambilla, N.; others.
\newblock {QCD and Strongly Coupled Gauge Theories: Challenges and
  Perspectives}.
\newblock {\em Eur. Phys. J. C} {\bf 2014}, {\em 74},~2981,
  \href{http://xxx.lanl.gov/abs/1404.3723}{{\normalfont
  [arXiv:hep-ph/1404.3723]}}.

\bibitem['t~Hooft(1974)]{tHooft:1974pnl}
't~Hooft, G.
\newblock {A Two-Dimensional Model for Mesons}.
\newblock {\em Nucl. Phys. B} {\bf 1974}, {\em 75},~461--470.

\bibitem[Greensite(2020)]{Greensite:2011zz}
Greensite, J.
\newblock {\em {An introduction to the confinement problem}}; Vol. 972,
  Springer Nature,  2020.

\bibitem[Greensite(2017)]{Greensite:2016pfc}
Greensite, J.
\newblock {Confinement from Center Vortices: A review of old and new results}.
\newblock {\em EPJ Web Conf.} {\bf 2017}, {\em 137},~01009,
  \href{http://xxx.lanl.gov/abs/1610.06221}{{\normalfont
  [arXiv:hep-lat/1610.06221]}}.

\bibitem[Gross and Wilczek(1973)]{Gross:1973id}
Gross, D.J.; Wilczek, F.
\newblock {Ultraviolet Behavior of Nonabelian Gauge Theories}.
\newblock {\em Phys. Rev. Lett.} {\bf 1973}, {\em 30},~1343--1346.

\bibitem[Politzer(1973)]{Politzer:1973fx}
Politzer, H.D.
\newblock {Reliable Perturbative Results for Strong Interactions?}
\newblock {\em Phys. Rev. Lett.} {\bf 1973}, {\em 30},~1346--1349.

\bibitem[Collins and Perry(1975)]{Collins:1974ky}
Collins, J.C.; Perry, M.J.
\newblock {Superdense Matter: Neutrons Or Asymptotically Free Quarks?}
\newblock {\em Phys. Rev. Lett.} {\bf 1975}, {\em 34},~1353.

\bibitem[Cabibbo and Parisi(1975)]{Cabibbo:1975ig}
Cabibbo, N.; Parisi, G.
\newblock {Exponential Hadronic Spectrum and Quark Liberation}.
\newblock {\em Phys. Lett. B} {\bf 1975}, {\em 59},~67--69.

\bibitem[Shuryak(1978{\natexlab{a}})]{Shuryak:1977ut}
Shuryak, E.V.
\newblock {Theory of Hadronic Plasma}.
\newblock {\em Sov. Phys. JETP} {\bf 1978}, {\em 47},~212--219.

\bibitem[Shuryak(1978{\natexlab{b}})]{Shuryak:1978ij}
Shuryak, E.V.
\newblock {Quark-Gluon Plasma and Hadronic Production of Leptons, Photons and
  Psions}.
\newblock {\em Phys. Lett. B} {\bf 1978}, {\em 78},~150.

\bibitem[Freedman and McLerran(1977)]{Freedman:1976ub}
Freedman, B.A.; McLerran, L.D.
\newblock {Fermions and Gauge Vector Mesons at Finite Temperature and Density.
  3. The Ground State Energy of a Relativistic Quark Gas}.
\newblock {\em Phys. Rev. D} {\bf 1977}, {\em 16},~1169.

\bibitem[Polyakov(1978)]{Polyakov:1978vu}
Polyakov, A.M.
\newblock {Thermal Properties of Gauge Fields and Quark Liberation}.
\newblock {\em Phys. Lett. B} {\bf 1978}, {\em 72},~477--480.

\bibitem[Kapusta(1979)]{Kapusta:1979fh}
Kapusta, J.I.
\newblock {Quantum Chromodynamics at High Temperature}.
\newblock {\em Nucl. Phys. B} {\bf 1979}, {\em 148},~461--498.

\bibitem[Witten(1984)]{Witten:1984rs}
Witten, E.
\newblock {Cosmic Separation of Phases}.
\newblock {\em Phys. Rev. D} {\bf 1984}, {\em 30},~272--285.

\bibitem[Arsene \em{et~al.}(2005)Arsene et~al.]{BRAHMS:2004adc}
Arsene, I.; others.
\newblock {Quark gluon plasma and color glass condensate at RHIC? The
  Perspective from the BRAHMS experiment}.
\newblock {\em Nucl. Phys. A} {\bf 2005}, {\em 757},~1--27,
  \href{http://xxx.lanl.gov/abs/nucl-ex/0410020}{{\normalfont
  [nucl-ex/0410020]}}.

\bibitem[Back \em{et~al.}(2005)Back et~al.]{PHOBOS:2004zne}
Back, B.B.; others.
\newblock {The PHOBOS perspective on discoveries at RHIC}.
\newblock {\em Nucl. Phys. A} {\bf 2005}, {\em 757},~28--101,
  \href{http://xxx.lanl.gov/abs/nucl-ex/0410022}{{\normalfont
  [nucl-ex/0410022]}}.

\bibitem[Adams \em{et~al.}(2005)Adams et~al.]{STAR:2005gfr}
Adams, J.; others.
\newblock {Experimental and theoretical challenges in the search for the quark
  gluon plasma: The STAR Collaboration's critical assessment of the evidence
  from RHIC collisions}.
\newblock {\em Nucl. Phys. A} {\bf 2005}, {\em 757},~102--183,
  \href{http://xxx.lanl.gov/abs/nucl-ex/0501009}{{\normalfont
  [nucl-ex/0501009]}}.

\bibitem[Adcox \em{et~al.}(2005)Adcox et~al.]{PHENIX:2004vcz}
Adcox, K.; others.
\newblock {Formation of dense partonic matter in relativistic nucleus-nucleus
  collisions at RHIC: Experimental evaluation by the PHENIX collaboration}.
\newblock {\em Nucl. Phys. A} {\bf 2005}, {\em 757},~184--283,
  \href{http://xxx.lanl.gov/abs/nucl-ex/0410003}{{\normalfont
  [nucl-ex/0410003]}}.

\bibitem[Braun-Munzinger \em{et~al.}(2016)Braun-Munzinger, Koch, Sch\"afer, and
  Stachel]{Braun-Munzinger:2015hba}
Braun-Munzinger, P.; Koch, V.; Sch\"afer, T.; Stachel, J.
\newblock {Properties of hot and dense matter from relativistic heavy ion
  collisions}.
\newblock {\em Phys. Rept.} {\bf 2016}, {\em 621},~76--126,
  \href{http://xxx.lanl.gov/abs/1510.00442}{{\normalfont
  [arXiv:nucl-th/1510.00442]}}.

\bibitem[Pasechnik and \v{S}umbera(2017)]{Pasechnik:2016wkt}
Pasechnik, R.; \v{S}umbera, M.
\newblock {Phenomenological Review on Quark\textendash{}Gluon Plasma: Concepts
  vs. Observations}.
\newblock {\em Universe} {\bf 2017}, {\em 3},~7,
  \href{http://xxx.lanl.gov/abs/1611.01533}{{\normalfont
  [arXiv:hep-ph/1611.01533]}}.

\bibitem[Kapusta \em{et~al.}(2003)Kapusta, Muller, and
  Rafelski]{Rafelski:2003zz}
Kapusta, J.; Muller, B.; Rafelski, J.
\newblock {\em {Quark-Gluon Plasma: Theoretical Foundations}}; Elsevier,  2003.

\bibitem[Shuryak(2017)]{Shuryak:2014zxa}
Shuryak, E.
\newblock {Strongly coupled quark-gluon plasma in heavy ion collisions}.
\newblock {\em Rev. Mod. Phys.} {\bf 2017}, {\em 89},~035001,
  \href{http://xxx.lanl.gov/abs/1412.8393}{{\normalfont
  [arXiv:hep-ph/1412.8393]}}.

\bibitem[Zyla \em{et~al.}(2020)Zyla et~al.]{ParticleDataGroup:2020ssz}
Zyla, P.A.; others.
\newblock {Review of Particle Physics}.
\newblock {\em PTEP} {\bf 2020}, {\em 2020},~083C01.

\bibitem[Gelis \em{et~al.}(2010)Gelis, Iancu, Jalilian-Marian, and
  Venugopalan]{Gelis:2010nm}
Gelis, F.; Iancu, E.; Jalilian-Marian, J.; Venugopalan, R.
\newblock {The Color Glass Condensate}.
\newblock {\em Ann. Rev. Nucl. Part. Sci.} {\bf 2010}, {\em 60},~463--489,
  \href{http://xxx.lanl.gov/abs/1002.0333}{{\normalfont
  [arXiv:hep-ph/1002.0333]}}.

\bibitem[Fujii and Kharzeev(1999)]{Fujii:1999xn}
Fujii, H.; Kharzeev, D.
\newblock {Long range forces of QCD}.
\newblock {\em Phys. Rev. D} {\bf 1999}, {\em 60},~114039,
  \href{http://xxx.lanl.gov/abs/hep-ph/9903495}{{\normalfont
  [hep-ph/9903495]}}.

\bibitem[Lacey \em{et~al.}(2007)Lacey, Ajitanand, Alexander, Chung, Holzmann,
  Issah, Taranenko, Danielewicz, and Stoecker]{Lacey:2006bc}
Lacey, R.A.; Ajitanand, N.N.; Alexander, J.M.; Chung, P.; Holzmann, W.G.;
  Issah, M.; Taranenko, A.; Danielewicz, P.; Stoecker, H.
\newblock {Has the QCD Critical Point been Signaled by Observations at RHIC?}
\newblock {\em Phys. Rev. Lett.} {\bf 2007}, {\em 98},~092301,
  \href{http://xxx.lanl.gov/abs/nucl-ex/0609025}{{\normalfont
  [nucl-ex/0609025]}}.

\bibitem[Heinz and Snellings(2013)]{Heinz:2013th}
Heinz, U.; Snellings, R.
\newblock {Collective flow and viscosity in relativistic heavy-ion collisions}.
\newblock {\em Ann. Rev. Nucl. Part. Sci.} {\bf 2013}, {\em 63},~123--151,
  \href{http://xxx.lanl.gov/abs/1301.2826}{{\normalfont
  [arXiv:nucl-th/1301.2826]}}.

\bibitem[Adcox \em{et~al.}(2002)Adcox et~al.]{PHENIX:2001hpc}
Adcox, K.; others.
\newblock {Suppression of hadrons with large transverse momentum in central
  Au+Au collisions at $\sqrt{s_{NN}}$ = 130-GeV}.
\newblock {\em Phys. Rev. Lett.} {\bf 2002}, {\em 88},~022301,
  \href{http://xxx.lanl.gov/abs/nucl-ex/0109003}{{\normalfont
  [nucl-ex/0109003]}}.

\bibitem[Adler \em{et~al.}(2003)Adler et~al.]{STAR:2002svs}
Adler, C.; others.
\newblock {Disappearance of back-to-back high $p_{T}$ hadron correlations in
  central Au+Au collisions at $\sqrt{s_{NN}}$ = 200-GeV}.
\newblock {\em Phys. Rev. Lett.} {\bf 2003}, {\em 90},~082302,
  \href{http://xxx.lanl.gov/abs/nucl-ex/0210033}{{\normalfont
  [nucl-ex/0210033]}}.

\bibitem[Thoma(2006)]{Thoma:2005aw}
Thoma, M.H.
\newblock {Complex plasmas as a model for the quark-gluon-plasma liquid}.
\newblock {\em Nucl. Phys. A} {\bf 2006}, {\em 774},~307--314,
  \href{http://xxx.lanl.gov/abs/hep-ph/0509154}{{\normalfont
  [hep-ph/0509154]}}.

\bibitem[Ioffe \em{et~al.}(2010)Ioffe, Fadin, and Lipatov]{Ioffe:2010zz}
Ioffe, B.L.; Fadin, V.S.; Lipatov, L.N.
\newblock {\em {Quantum chromodynamics: Perturbative and nonperturbative
  aspects}}; Cambridge Univ. Press,  2010.

\bibitem[Campbell \em{et~al.}(2017)Campbell, Huston, and
  Krauss]{Campbell:2017hsr}
Campbell, J.; Huston, J.; Krauss, F.
\newblock {\em {The Black Book of Quantum Chromodynamics}: {A Primer for the
  LHC Era}}; Oxford University Press,  2017.

\bibitem[Gribov \em{et~al.}(1983)Gribov, Levin, and Ryskin]{Gribov:1983ivg}
Gribov, L.V.; Levin, E.M.; Ryskin, M.G.
\newblock {Semihard Processes in QCD}.
\newblock {\em Phys. Rept.} {\bf 1983}, {\em 100},~1--150.

\bibitem[Kharzeev(2002)]{Kharzeev:2002np}
Kharzeev, D.
\newblock {Classical chromodynamics of relativistic heavy ion collisions}.
\newblock  {Cargese Summer School on QCD Perspectives on Hot and Dense Matter},
   2002,  \href{http://xxx.lanl.gov/abs/hep-ph/0204014}{{\normalfont
  [hep-ph/0204014]}}.

\bibitem[Berges \em{et~al.}(2020)Berges, Heller, Mazeliauskas, and
  Venugopalan]{Berges:2020fwq}
Berges, J.; Heller, M.P.; Mazeliauskas, A.; Venugopalan, R.
\newblock {Thermalization in QCD: theoretical approaches, phenomenological
  applications, and interdisciplinary connections} {\bf 2020}.
\newblock  \href{http://xxx.lanl.gov/abs/2005.12299}{{\normalfont
  [arXiv:hep-th/2005.12299]}}.

\bibitem[McLerran(2008)]{McLerran:2008es}
McLerran, L.
\newblock {A Brief Introduction to the Color Glass Condensate and the Glasma}.
\newblock  {38th International Symposium on Multiparticle Dynamics},  2008,
  \href{http://xxx.lanl.gov/abs/0812.4989}{{\normalfont
  [arXiv:hep-ph/0812.4989]}}.

\bibitem[McLerran and Venugopalan(1994)]{McLerran:1993ni}
McLerran, L.D.; Venugopalan, R.
\newblock {Computing quark and gluon distribution functions for very large
  nuclei}.
\newblock {\em Phys. Rev. D} {\bf 1994}, {\em 49},~2233--2241,
  \href{http://xxx.lanl.gov/abs/hep-ph/9309289}{{\normalfont
  [hep-ph/9309289]}}.

\bibitem[Kovner \em{et~al.}(1995)Kovner, McLerran, and Weigert]{Kovner:1995ja}
Kovner, A.; McLerran, L.D.; Weigert, H.
\newblock {Gluon production from nonAbelian Weizsacker-Williams fields in
  nucleus-nucleus collisions}.
\newblock {\em Phys. Rev. D} {\bf 1995}, {\em 52},~6231--6237,
  \href{http://xxx.lanl.gov/abs/hep-ph/9502289}{{\normalfont
  [hep-ph/9502289]}}.

\bibitem[Braun-Munzinger and Wambach(2009)]{Braun-Munzinger:2008szb}
Braun-Munzinger, P.; Wambach, J.
\newblock {The Phase Diagram of Strongly-Interacting Matter}.
\newblock {\em Rev. Mod. Phys.} {\bf 2009}, {\em 81},~1031--1050,
  \href{http://xxx.lanl.gov/abs/0801.4256}{{\normalfont
  [arXiv:hep-ph/0801.4256]}}.

\bibitem[Fukushima and Hatsuda(2011)]{Fukushima:2010bq}
Fukushima, K.; Hatsuda, T.
\newblock {The phase diagram of dense QCD}.
\newblock {\em Rept. Prog. Phys.} {\bf 2011}, {\em 74},~014001,
  \href{http://xxx.lanl.gov/abs/1005.4814}{{\normalfont
  [arXiv:hep-ph/1005.4814]}}.

\bibitem[Addazi \em{et~al.}(2019)Addazi, Marcian\`o, and
  Pasechnik]{Addazi:2018ctp}
Addazi, A.; Marcian\`o, A.; Pasechnik, R.
\newblock {Time-crystal ground state and production of gravitational waves from
  QCD phase transition}.
\newblock {\em Chin. Phys. C} {\bf 2019}, {\em 43},~065101,
  \href{http://xxx.lanl.gov/abs/1812.07376}{{\normalfont
  [arXiv:hep-th/1812.07376]}}.

\bibitem[Huang \em{et~al.}(2020)Huang, Reichert, Sannino, and
  Wang]{Huang:2020mso}
Huang, W.C.; Reichert, M.; Sannino, F.; Wang, Z.W.
\newblock {Testing the Dark Confined Landscape: From Lattice to Gravitational
  Waves} {\bf 2020}.
\newblock  \href{http://xxx.lanl.gov/abs/2012.11614}{{\normalfont
  [arXiv:hep-ph/2012.11614]}}.

\bibitem[Pasechnik \em{et~al.}(2017)Pasechnik, Prokhorov, and
  Teryaev]{Pasechnik:2016twe}
Pasechnik, R.; Prokhorov, G.; Teryaev, O.
\newblock {Mirror QCD and Cosmological Constant}.
\newblock {\em Universe} {\bf 2017}, {\em 3},~43,
  \href{http://xxx.lanl.gov/abs/1609.09249}{{\normalfont
  [arXiv:hep-ph/1609.09249]}}.

\bibitem[Pasechnik \em{et~al.}(2013{\natexlab{a}})Pasechnik, Beylin, and
  Vereshkov]{Pasechnik:2013poa}
Pasechnik, R.; Beylin, V.; Vereshkov, G.
\newblock {Dark Energy from graviton-mediated interactions in the QCD vacuum}.
\newblock {\em JCAP} {\bf 2013}, {\em 1306},~011,
  \href{http://xxx.lanl.gov/abs/1302.6456}{{\normalfont
  [arXiv:gr-qc/1302.6456]}}.

\bibitem[Pasechnik \em{et~al.}(2013{\natexlab{b}})Pasechnik, Beylin, and
  Vereshkov]{Pasechnik:2013sga}
Pasechnik, R.; Beylin, V.; Vereshkov, G.
\newblock {Possible compensation of the QCD vacuum contribution to the dark
  energy}.
\newblock {\em Phys. Rev.} {\bf 2013}, {\em D88},~023509,
  \href{http://xxx.lanl.gov/abs/1302.5934}{{\normalfont
  [arXiv:gr-qc/1302.5934]}}.

\bibitem[Addazi \em{et~al.}(2019)Addazi, Marcianò, Pasechnik, and
  Prokhorov]{Addazi:2019mlo}
Addazi, A.; Marcianò, A.; Pasechnik, R.; Prokhorov, G.
\newblock {Mirror Symmetry of quantum Yang-Mills vacua and cosmological
  implications}.
\newblock {\em Eur. Phys. J.} {\bf 2019}, {\em C79},~251,
  \href{http://xxx.lanl.gov/abs/1804.09826}{{\normalfont
  [arXiv:hep-th/1804.09826]}}.

\bibitem[Pasechnik(2016)]{Pasechnik:2016sbh}
Pasechnik, R.
\newblock {Quantum Yang\textendash{}Mills Dark Energy}.
\newblock {\em Universe} {\bf 2016}, {\em 2},~4,
  \href{http://xxx.lanl.gov/abs/1605.07610}{{\normalfont
  [arXiv:gr-qc/1605.07610]}}.

\bibitem[Stephanov \em{et~al.}(1998)Stephanov, Rajagopal, and
  Shuryak]{Stephanov:1998dy}
Stephanov, M.A.; Rajagopal, K.; Shuryak, E.V.
\newblock {Signatures of the tricritical point in QCD}.
\newblock {\em Phys. Rev. Lett.} {\bf 1998}, {\em 81},~4816--4819,
  \href{http://xxx.lanl.gov/abs/hep-ph/9806219}{{\normalfont
  [hep-ph/9806219]}}.

\bibitem[Gupta \em{et~al.}(2011)Gupta, Luo, Mohanty, Ritter, and
  Xu]{Gupta:2011wh}
Gupta, S.; Luo, X.; Mohanty, B.; Ritter, H.G.; Xu, N.
\newblock {Scale for the Phase Diagram of Quantum Chromodynamics}.
\newblock {\em Science} {\bf 2011}, {\em 332},~1525--1528,
  \href{http://xxx.lanl.gov/abs/1105.3934}{{\normalfont
  [arXiv:hep-ph/1105.3934]}}.

\bibitem[Adamczyk \em{et~al.}(2017)Adamczyk et~al.]{STAR:2017sal}
Adamczyk, L.; others.
\newblock {Bulk Properties of the Medium Produced in Relativistic Heavy-Ion
  Collisions from the Beam Energy Scan Program}.
\newblock {\em Phys. Rev. C} {\bf 2017}, {\em 96},~044904,
  \href{http://xxx.lanl.gov/abs/1701.07065}{{\normalfont
  [arXiv:nucl-ex/1701.07065]}}.

\bibitem[Bzdak \em{et~al.}(2020)Bzdak, Esumi, Koch, Liao, Stephanov, and
  Xu]{Bzdak:2019pkr}
Bzdak, A.; Esumi, S.; Koch, V.; Liao, J.; Stephanov, M.; Xu, N.
\newblock {Mapping the Phases of Quantum Chromodynamics with Beam Energy Scan}.
\newblock {\em Phys. Rept.} {\bf 2020}, {\em 853},~1--87,
  \href{http://xxx.lanl.gov/abs/1906.00936}{{\normalfont
  [arXiv:nucl-th/1906.00936]}}.

\bibitem[Bellwied \em{et~al.}(2015)Bellwied, Borsanyi, Fodor, G\"unther, Katz,
  Ratti, and Szabo]{Bellwied:2015rza}
Bellwied, R.; Borsanyi, S.; Fodor, Z.; G\"unther, J.; Katz, S.D.; Ratti, C.;
  Szabo, K.K.
\newblock {The QCD phase diagram from analytic continuation}.
\newblock {\em Phys. Lett. B} {\bf 2015}, {\em 751},~559--564,
  \href{http://xxx.lanl.gov/abs/1507.07510}{{\normalfont
  [arXiv:hep-lat/1507.07510]}}.

\bibitem[Ding \em{et~al.}(2016)Ding, Karsch, and Mukherjee]{Ding:2016qdj}
Ding, H.T.; Karsch, F.; Mukherjee, S., {Thermodynamics of Strong-Interaction
  Matter from Lattice QCD}.
\newblock In {\em {Quark-Gluon Plasma 5}}; Wang, X.N., Ed.;  2016.

\bibitem[Bazavov \em{et~al.}(2017)Bazavov et~al.]{Bazavov:2017dus}
Bazavov, A.; others.
\newblock {The QCD Equation of State to $\mathcal{O}(\mu_B^6)$ from Lattice
  QCD}.
\newblock {\em Phys. Rev. D} {\bf 2017}, {\em 95},~054504,
  \href{http://xxx.lanl.gov/abs/1701.04325}{{\normalfont
  [arXiv:hep-lat/1701.04325]}}.

\bibitem[Philipsen(2019)]{Philipsen:2019rjq}
Philipsen, O.
\newblock {Constraining the phase diagram of QCD at finite temperature and
  density}.
\newblock {\em PoS} {\bf 2019}, {\em LATTICE2019},~273,
  \href{http://xxx.lanl.gov/abs/1912.04827}{{\normalfont
  [arXiv:hep-lat/1912.04827]}}.

\bibitem[Dean and Hjorth-Jensen(2003)]{Dean:2002zx}
Dean, D.J.; Hjorth-Jensen, M.
\newblock {Pairing in nuclear systems: From neutron stars to finite nuclei}.
\newblock {\em Rev. Mod. Phys.} {\bf 2003}, {\em 75},~607--656,
  \href{http://xxx.lanl.gov/abs/nucl-th/0210033}{{\normalfont
  [nucl-th/0210033]}}.

\bibitem[Gandolfi \em{et~al.}(2015)Gandolfi, Gezerlis, and
  Carlson]{Gandolfi:2015jma}
Gandolfi, S.; Gezerlis, A.; Carlson, J.
\newblock {Neutron Matter from Low to High Density}.
\newblock {\em Ann. Rev. Nucl. Part. Sci.} {\bf 2015}, {\em 65},~303--328,
  \href{http://xxx.lanl.gov/abs/1501.05675}{{\normalfont
  [arXiv:nucl-th/1501.05675]}}.

\bibitem[Cherman \em{et~al.}(2020)Cherman, Jacobson, Sen, and
  Yaffe]{Cherman:2020hbe}
Cherman, A.; Jacobson, T.; Sen, S.; Yaffe, L.G.
\newblock {Higgs-confinement phase transitions with fundamental representation
  matter}.
\newblock {\em Phys. Rev. D} {\bf 2020}, {\em 102},~105021,
  \href{http://xxx.lanl.gov/abs/2007.08539}{{\normalfont
  [arXiv:hep-th/2007.08539]}}.

\bibitem[Barrois(1977)]{Barrois:1977xd}
Barrois, B.C.
\newblock {Superconducting Quark Matter}.
\newblock {\em Nucl. Phys. B} {\bf 1977}, {\em 129},~390--396.

\bibitem[Bailin and Love(1984)]{Bailin:1983bm}
Bailin, D.; Love, A.
\newblock {Superfluidity and Superconductivity in Relativistic Fermion
  Systems}.
\newblock {\em Phys. Rept.} {\bf 1984}, {\em 107},~325.

\bibitem[Alford \em{et~al.}(2008)Alford, Schmitt, Rajagopal, and
  Sch\"afer]{Alford:2007xm}
Alford, M.G.; Schmitt, A.; Rajagopal, K.; Sch\"afer, T.
\newblock {Color superconductivity in dense quark matter}.
\newblock {\em Rev. Mod. Phys.} {\bf 2008}, {\em 80},~1455--1515,
  \href{http://xxx.lanl.gov/abs/0709.4635}{{\normalfont
  [arXiv:hep-ph/0709.4635]}}.

\bibitem[Baym \em{et~al.}(2018)Baym, Hatsuda, Kojo, Powell, Song, and
  Takatsuka]{Baym:2017whm}
Baym, G.; Hatsuda, T.; Kojo, T.; Powell, P.D.; Song, Y.; Takatsuka, T.
\newblock {From hadrons to quarks in neutron stars: a review}.
\newblock {\em Rept. Prog. Phys.} {\bf 2018}, {\em 81},~056902,
  \href{http://xxx.lanl.gov/abs/1707.04966}{{\normalfont
  [arXiv:astro-ph.HE/1707.04966]}}.

\bibitem[Alford \em{et~al.}(1999)Alford, Rajagopal, and Wilczek]{Alford:1998mk}
Alford, M.G.; Rajagopal, K.; Wilczek, F.
\newblock {Color flavor locking and chiral symmetry breaking in high density
  QCD}.
\newblock {\em Nucl. Phys. B} {\bf 1999}, {\em 537},~443--458,
  \href{http://xxx.lanl.gov/abs/hep-ph/9804403}{{\normalfont
  [hep-ph/9804403]}}.

\bibitem[Cherman \em{et~al.}(2019)Cherman, Sen, and Yaffe]{Cherman:2018jir}
Cherman, A.; Sen, S.; Yaffe, L.G.
\newblock {Anyonic particle-vortex statistics and the nature of dense quark
  matter}.
\newblock {\em Phys. Rev. D} {\bf 2019}, {\em 100},~034015,
  \href{http://xxx.lanl.gov/abs/1808.04827}{{\normalfont
  [arXiv:hep-th/1808.04827]}}.

\bibitem[Alford and Reddy(2003)]{Alford:2002rj}
Alford, M.; Reddy, S.
\newblock {Compact stars with color superconducting quark matter}.
\newblock {\em Phys. Rev. D} {\bf 2003}, {\em 67},~074024,
  \href{http://xxx.lanl.gov/abs/nucl-th/0211046}{{\normalfont
  [nucl-th/0211046]}}.

\bibitem[Steiner \em{et~al.}(2002)Steiner, Reddy, and Prakash]{Steiner:2002gx}
Steiner, A.W.; Reddy, S.; Prakash, M.
\newblock {Color neutral superconducting quark matter}.
\newblock {\em Phys. Rev. D} {\bf 2002}, {\em 66},~094007,
  \href{http://xxx.lanl.gov/abs/hep-ph/0205201}{{\normalfont
  [hep-ph/0205201]}}.

\bibitem[Sch\"afer and Wilczek(1999{\natexlab{a}})]{Schafer:1998ef}
Sch\"afer, T.; Wilczek, F.
\newblock {Continuity of quark and hadron matter}.
\newblock {\em Phys. Rev. Lett.} {\bf 1999}, {\em 82},~3956--3959,
  \href{http://xxx.lanl.gov/abs/hep-ph/9811473}{{\normalfont
  [hep-ph/9811473]}}.

\bibitem[Sch\"afer and Wilczek(1999{\natexlab{b}})]{Schafer:1999pb}
Sch\"afer, T.; Wilczek, F.
\newblock {Quark description of hadronic phases}.
\newblock {\em Phys. Rev. D} {\bf 1999}, {\em 60},~074014,
  \href{http://xxx.lanl.gov/abs/hep-ph/9903503}{{\normalfont
  [hep-ph/9903503]}}.

\bibitem[Sch\"afer and Wilczek(1999{\natexlab{c}})]{Schafer:1999jg}
Sch\"afer, T.; Wilczek, F.
\newblock {Superconductivity from perturbative one gluon exchange in high
  density quark matter}.
\newblock {\em Phys. Rev. D} {\bf 1999}, {\em 60},~114033,
  \href{http://xxx.lanl.gov/abs/hep-ph/9906512}{{\normalfont
  [hep-ph/9906512]}}.

\bibitem[Alford \em{et~al.}(2019)Alford, Baym, Fukushima, Hatsuda, and
  Tachibana]{Alford:2018mqj}
Alford, M.G.; Baym, G.; Fukushima, K.; Hatsuda, T.; Tachibana, M.
\newblock {Continuity of vortices from the hadronic to the color-flavor locked
  phase in dense matter}.
\newblock {\em Phys. Rev. D} {\bf 2019}, {\em 99},~036004,
  \href{http://xxx.lanl.gov/abs/1803.05115}{{\normalfont
  [arXiv:hep-ph/1803.05115]}}.

\bibitem[Wan and Wang(2020)]{Wan:2019oax}
Wan, Z.; Wang, J.
\newblock {Higher anomalies, higher symmetries, and cobordisms III: QCD matter
  phases anew}.
\newblock {\em Nucl. Phys. B} {\bf 2020}, {\em 957},~115016,
  \href{http://xxx.lanl.gov/abs/1912.13514}{{\normalfont
  [arXiv:hep-th/1912.13514]}}.

\bibitem[Alford \em{et~al.}(2019)Alford, Han, and Schwenzer]{Alford:2019oge}
Alford, M.G.; Han, S.; Schwenzer, K.
\newblock {Signatures for quark matter from multi-messenger observations}.
\newblock {\em J. Phys. G} {\bf 2019}, {\em 46},~114001,
  \href{http://xxx.lanl.gov/abs/1904.05471}{{\normalfont
  [arXiv:nucl-th/1904.05471]}}.

\bibitem[McLerran and Pisarski(2007)]{McLerran:2007qj}
McLerran, L.; Pisarski, R.D.
\newblock {Phases of cold, dense quarks at large N(c)}.
\newblock {\em Nucl. Phys. A} {\bf 2007}, {\em 796},~83--100,
  \href{http://xxx.lanl.gov/abs/0706.2191}{{\normalfont
  [arXiv:hep-ph/0706.2191]}}.

\bibitem[McLerran and Reddy(2019)]{McLerran:2018hbz}
McLerran, L.; Reddy, S.
\newblock {Quarkyonic Matter and Neutron Stars}.
\newblock {\em Phys. Rev. Lett.} {\bf 2019}, {\em 122},~122701,
  \href{http://xxx.lanl.gov/abs/1811.12503}{{\normalfont
  [arXiv:nucl-th/1811.12503]}}.

\bibitem[Shifman(2010)]{Shifman:2010jp}
Shifman, M.
\newblock {Understanding Confinement in QCD: Elements of a Big Picture}.
\newblock {\em Int. J. Mod. Phys. A} {\bf 2010}, {\em 25},~4015--4031,
  \href{http://xxx.lanl.gov/abs/1007.0531}{{\normalfont
  [arXiv:hep-th/1007.0531]}}.

\bibitem[Ogilvie(2011)]{Ogilvie:2010vx}
Ogilvie, M.C.
\newblock {Quark Confinement and the Renormalization Group}.
\newblock {\em Phil. Trans. Roy. Soc. Lond. A} {\bf 2011}, {\em 369},~2718,
  \href{http://xxx.lanl.gov/abs/1010.1942}{{\normalfont
  [arXiv:hep-lat/1010.1942]}}.

\bibitem[Reinhardt(2018)]{Reinhardt:2018roz}
Reinhardt, H.
\newblock {Effective Approaches to QCD}.
\newblock  {53rd Winter School of Theoretical Physics}: {Understanding the
  Origin of Matter from QCD},  2018,
  \href{http://xxx.lanl.gov/abs/1804.03875}{{\normalfont
  [arXiv:hep-th/1804.03875]}}.

\bibitem[Wegner(1971)]{Wegner:1971app}
Wegner, F.J.
\newblock {Duality in Generalized Ising Models and Phase Transitions Without
  Local Order Parameters}.
\newblock {\em J. Math. Phys.} {\bf 1971}, {\em 12},~2259--2272.

\bibitem[DeGrand and Detar(2006)]{DeGrand:2006zz}
DeGrand, T.; Detar, C.E.
\newblock {\em {Lattice methods for quantum chromodynamics}}; World Scientific,
   2006.

\bibitem[Ghiglieri \em{et~al.}(2020)Ghiglieri, Kurkela, Strickland, and
  Vuorinen]{Ghiglieri:2020dpq}
Ghiglieri, J.; Kurkela, A.; Strickland, M.; Vuorinen, A.
\newblock {Perturbative Thermal QCD: Formalism and Applications}.
\newblock {\em Phys. Rept.} {\bf 2020}, {\em 880},~1--73,
  \href{http://xxx.lanl.gov/abs/2002.10188}{{\normalfont
  [arXiv:hep-ph/2002.10188]}}.

\bibitem[Lundberg and Pasechnik(2021)]{Lundberg:2020mwu}
Lundberg, T.; Pasechnik, R.
\newblock {Thermal Field Theory in real-time formalism: concepts and
  applications for particle decays}.
\newblock {\em Eur. Phys. J. A} {\bf 2021}, {\em 57},~71,
  \href{http://xxx.lanl.gov/abs/2007.01224}{{\normalfont
  [arXiv:hep-th/2007.01224]}}.

\bibitem[Celik \em{et~al.}(1983)Celik, Engels, and Satz]{Celik:1983wz}
Celik, T.; Engels, J.; Satz, H.
\newblock {The Order of the Deconfinement Transition in SU(3) Yang-Mills
  Theory}.
\newblock {\em Phys. Lett. B} {\bf 1983}, {\em 125},~411--414.

\bibitem[Maiani and Testa(1990)]{Maiani:1990ca}
Maiani, L.; Testa, M.
\newblock {Final state interactions from Euclidean correlation functions}.
\newblock {\em Phys. Lett. B} {\bf 1990}, {\em 245},~585--590.

\bibitem[Luscher(1991)]{Luscher:1990ux}
Luscher, M.
\newblock {Two particle states on a torus and their relation to the scattering
  matrix}.
\newblock {\em Nucl. Phys. B} {\bf 1991}, {\em 354},~531--578.

\bibitem[Hansen and Sharpe(2019)]{Hansen:2019nir}
Hansen, M.T.; Sharpe, S.R.
\newblock {Lattice QCD and Three-particle Decays of Resonances}.
\newblock {\em Ann. Rev. Nucl. Part. Sci.} {\bf 2019}, {\em 69},~65--107,
  \href{http://xxx.lanl.gov/abs/1901.00483}{{\normalfont
  [arXiv:hep-lat/1901.00483]}}.

\bibitem[Aarts(2016)]{Aarts:2015tyj}
Aarts, G.
\newblock {Introductory lectures on lattice QCD at nonzero baryon number}.
\newblock {\em J. Phys. Conf. Ser.} {\bf 2016}, {\em 706},~022004,
  \href{http://xxx.lanl.gov/abs/1512.05145}{{\normalfont
  [arXiv:hep-lat/1512.05145]}}.

\bibitem[Bollweg \em{et~al.}(2021)Bollweg, Karsch, Mukherjee, and
  Schmidt]{Bollweg:2020yum}
Bollweg, D.; Karsch, F.; Mukherjee, S.; Schmidt, C.
\newblock {Higher order cumulants of net baryon-number distributions at
  non-zero $\mu_B$}.
\newblock {\em Nucl. Phys. A} {\bf 2021}, {\em 1005},~121835,
  \href{http://xxx.lanl.gov/abs/2002.01837}{{\normalfont
  [arXiv:hep-lat/2002.01837]}}.

\bibitem[Bazavov \em{et~al.}(2019)Bazavov, Karsch, Mukherjee, and
  Petreczky]{Bazavov:2019lgz}
Bazavov, A.; Karsch, F.; Mukherjee, S.; Petreczky, P.
\newblock {Hot-dense Lattice QCD: USQCD whitepaper 2018}.
\newblock {\em Eur. Phys. J. A} {\bf 2019}, {\em 55},~194,
  \href{http://xxx.lanl.gov/abs/1904.09951}{{\normalfont
  [arXiv:hep-lat/1904.09951]}}.

\bibitem[Bors\'anyi \em{et~al.}(2021)Bors\'anyi, Fodor, Guenther, Kara, Katz,
  Parotto, P\'asztor, Ratti, and Szab\'o]{Borsanyi:2021sxv}
Bors\'anyi, S.; Fodor, Z.; Guenther, J.N.; Kara, R.; Katz, S.D.; Parotto, P.;
  P\'asztor, A.; Ratti, C.; Szab\'o, K.K.
\newblock {Lattice QCD equation of state at finite chemical potential from an
  alternative expansion scheme}.
\newblock {\em Phys. Rev. Lett.} {\bf 2021}, {\em 126},~232001,
  \href{http://xxx.lanl.gov/abs/2102.06660}{{\normalfont
  [arXiv:hep-lat/2102.06660]}}.

\bibitem[Narayanan and Neuberger(2006)]{Narayanan:2006rf}
Narayanan, R.; Neuberger, H.
\newblock {Infinite N phase transitions in continuum Wilson loop operators}.
\newblock {\em JHEP} {\bf 2006}, {\em 03},~064,
  \href{http://xxx.lanl.gov/abs/hep-th/0601210}{{\normalfont
  [hep-th/0601210]}}.

\bibitem[Halpern(1979)]{Halpern:1978ik}
Halpern, M.B.
\newblock {Field Strength and Dual Variable Formulations of Gauge Theory}.
\newblock {\em Phys. Rev. D} {\bf 1979}, {\em 19},~517.

\bibitem[Batrouni and Halpern(1984)]{Batrouni:1984rb}
Batrouni, G.G.; Halpern, M.B.
\newblock {String, Corner and Plaquette Formulation of Finite Lattice Gauge
  Theory}.
\newblock {\em Phys. Rev. D} {\bf 1984}, {\em 30},~1782.

\bibitem[Intriligator and Seiberg(1996)]{Intriligator:1995er}
Intriligator, K.A.; Seiberg, N.
\newblock {Phases of N=1 supersymmetric gauge theories and electric - magnetic
  triality}.
\newblock {\em Nucl. Phys. B Proc. Suppl.} {\bf 1996}, {\em 39},~1,
  \href{http://xxx.lanl.gov/abs/hep-th/9506084}{{\normalfont
  [hep-th/9506084]}}.

\bibitem[Arefeva(1980)]{Arefeva:1979dp}
Arefeva, I.
\newblock {NonAbelian Stokes formula}.
\newblock {\em Theor. Math. Phys.} {\bf 1980}, {\em 43},~353.

\bibitem[Fishbane \em{et~al.}(1981)Fishbane, Gasiorowicz, and
  Kaus]{Fishbane:1980eq}
Fishbane, P.M.; Gasiorowicz, S.; Kaus, P.
\newblock {Stokes' Theorems for Nonabelian Fields}.
\newblock {\em Phys. Rev. D} {\bf 1981}, {\em 24},~2324.

\bibitem[Diakonov and Petrov(1989)]{Diakonov:1989fc}
Diakonov, D.; Petrov, V.Y.
\newblock {A Formula for the Wilson Loop}.
\newblock {\em Phys. Lett. B} {\bf 1989}, {\em 224},~131--135.

\bibitem[Karp \em{et~al.}(1999)Karp, Mansouri, and Rno]{Karp:1999vq}
Karp, R.L.; Mansouri, F.; Rno, J.S.
\newblock {Product integral formalism and nonAbelian Stokes theorem}.
\newblock {\em J. Math. Phys.} {\bf 1999}, {\em 40},~6033--6043,
  \href{http://xxx.lanl.gov/abs/hep-th/9910173}{{\normalfont
  [hep-th/9910173]}}.

\bibitem[Hirayama and Ueno(2000)]{Hirayama:1999ar}
Hirayama, M.; Ueno, M.
\newblock {NonAbelian Stokes theorem for Wilson loops associated with general
  gauge groups}.
\newblock {\em Prog. Theor. Phys.} {\bf 2000}, {\em 103},~151--159,
  \href{http://xxx.lanl.gov/abs/hep-th/9907063}{{\normalfont
  [hep-th/9907063]}}.

\bibitem[Diakonov and Petrov(2001)]{Diakonov:2000kw}
Diakonov, D.; Petrov, V.
\newblock {NonAbelian Stokes theorems in Yang-Mills and gravity theories}.
\newblock {\em J. Exp. Theor. Phys.} {\bf 2001}, {\em 92},~905--920,
  \href{http://xxx.lanl.gov/abs/hep-th/0008035}{{\normalfont
  [hep-th/0008035]}}.

\bibitem[Kondo and Taira(2000{\natexlab{a}})]{Kondo:2000pp}
Kondo, K.I.; Taira, Y.
\newblock {NonAbelian Stokes Theorem and Quark confinement in SU(3) Yang-Mills
  gauge theory}.
\newblock {\em Mod. Phys. Lett. A} {\bf 2000}, {\em 15},~367--377,
  \href{http://xxx.lanl.gov/abs/hep-th/9906129}{{\normalfont
  [hep-th/9906129]}}.

\bibitem[Kondo and Taira(2000{\natexlab{b}})]{Kondo:1999tj}
Kondo, K.I.; Taira, Y.
\newblock {NonAbelian Stokes theorem and quark confinement in SU(N) Yang-Mills
  gauge theory}.
\newblock {\em Prog. Theor. Phys.} {\bf 2000}, {\em 104},~1189--1265,
  \href{http://xxx.lanl.gov/abs/hep-th/9911242}{{\normalfont
  [hep-th/9911242]}}.

\bibitem[Di~Giacomo \em{et~al.}(2002)Di~Giacomo, Dosch, Shevchenko, and
  Simonov]{DiGiacomo:2000irz}
Di~Giacomo, A.; Dosch, H.G.; Shevchenko, V.I.; Simonov, Y.A.
\newblock {Field correlators in QCD: Theory and applications}.
\newblock {\em Phys. Rept.} {\bf 2002}, {\em 372},~319--368,
  \href{http://xxx.lanl.gov/abs/hep-ph/0007223}{{\normalfont
  [hep-ph/0007223]}}.

\bibitem[Kuzmenko \em{et~al.}(2004)Kuzmenko, Shevchenko, and
  Simonov]{Kuzmenko:2004hk}
Kuzmenko, D.S.; Shevchenko, V.I.; Simonov, Y.A.
\newblock {The QCD vacuum, confinement and strings in the vacuum correlator
  method}.
\newblock {\em Phys. Usp.} {\bf 2004}, {\em 47},~1--15,
  \href{http://xxx.lanl.gov/abs/hep-ph/0310190}{{\normalfont
  [hep-ph/0310190]}}.

\bibitem[Collins(2009)]{Collins:1977jy}
Collins, P.D.B.
\newblock {\em {An Introduction to Regge Theory and High-Energy Physics}};
  Cambridge Monographs on Mathematical Physics, Cambridge Univ. Press:
  Cambridge, UK,  2009.

\bibitem[Philipsen and Wittig(1998)]{Philipsen:1998de}
Philipsen, O.; Wittig, H.
\newblock {String breaking in nonAbelian gauge theories with fundamental matter
  fields}.
\newblock {\em Phys. Rev. Lett.} {\bf 1998}, {\em 81},~4056--4059,
  \href{http://xxx.lanl.gov/abs/hep-lat/9807020}{{\normalfont
  [hep-lat/9807020]}}.
\newblock [Erratum: Phys.Rev.Lett. 83, 2684 (1999)].

\bibitem[Duncan \em{et~al.}(2001)Duncan, Eichten, and Thacker]{Duncan:2000kr}
Duncan, A.; Eichten, E.; Thacker, H.
\newblock {String breaking in four-dimensional lattice QCD}.
\newblock {\em Phys. Rev. D} {\bf 2001}, {\em 63},~111501,
  \href{http://xxx.lanl.gov/abs/hep-lat/0011076}{{\normalfont
  [hep-lat/0011076]}}.

\bibitem[Bernard \em{et~al.}(2001)Bernard, DeGrand, Detar, Lacock, Gottlieb,
  Heller, Hetrick, Orginos, Toussaint, and Sugar]{Bernard:2001tz}
Bernard, C.W.; DeGrand, T.A.; Detar, C.E.; Lacock, P.; Gottlieb, S.A.; Heller,
  U.M.; Hetrick, J.; Orginos, K.; Toussaint, D.; Sugar, R.L.
\newblock {Zero temperature string breaking in lattice quantum chromodynamics}.
\newblock {\em Phys. Rev. D} {\bf 2001}, {\em 64},~074509,
  \href{http://xxx.lanl.gov/abs/hep-lat/0103012}{{\normalfont
  [hep-lat/0103012]}}.

\bibitem[Frohlich \em{et~al.}(1981)Frohlich, Morchio, and
  Strocchi]{Frohlich:1981yi}
Frohlich, J.; Morchio, G.; Strocchi, F.
\newblock {HIGGS PHENOMENON WITHOUT SYMMETRY BREAKING ORDER PARAMETER}.
\newblock {\em Nucl. Phys. B} {\bf 1981}, {\em 190},~553--582.

\bibitem[Fradkin and Shenker(1979)]{Fradkin:1978dv}
Fradkin, E.H.; Shenker, S.H.
\newblock {Phase Diagrams of Lattice Gauge Theories with Higgs Fields}.
\newblock {\em Phys. Rev. D} {\bf 1979}, {\em 19},~3682--3697.

\bibitem[Greensite and Matsuyama(2017)]{Greensite:2017ajx}
Greensite, J.; Matsuyama, K.
\newblock {Confinement criterion for gauge theories with matter fields}.
\newblock {\em Phys. Rev. D} {\bf 2017}, {\em 96},~094510,
  \href{http://xxx.lanl.gov/abs/1708.08979}{{\normalfont
  [arXiv:hep-lat/1708.08979]}}.

\bibitem[Lang \em{et~al.}(1981)Lang, Rebbi, and Virasoro]{Lang:1981qg}
Lang, C.B.; Rebbi, C.; Virasoro, M.
\newblock {The Phase Structure of a Nonabelian Gauge Higgs Field System}.
\newblock {\em Phys. Lett. B} {\bf 1981}, {\em 104},~294.

\bibitem[Elitzur(1975)]{Elitzur:1975im}
Elitzur, S.
\newblock {Impossibility of Spontaneously Breaking Local Symmetries}.
\newblock {\em Phys. Rev. D} {\bf 1975}, {\em 12},~3978--3982.

\bibitem[Osterwalder and Seiler(1978)]{Osterwalder:1977pc}
Osterwalder, K.; Seiler, E.
\newblock {Gauge Field Theories on the Lattice}.
\newblock {\em Annals Phys.} {\bf 1978}, {\em 110},~440.

\bibitem[Banks and Rabinovici(1979)]{Banks:1979fi}
Banks, T.; Rabinovici, E.
\newblock {Finite Temperature Behavior of the Lattice Abelian Higgs Model}.
\newblock {\em Nucl. Phys. B} {\bf 1979}, {\em 160},~349--379.

\bibitem[Bonati \em{et~al.}(2010)Bonati, Cossu, D'Elia, and
  Di~Giacomo]{Bonati:2009pf}
Bonati, C.; Cossu, G.; D'Elia, M.; Di~Giacomo, A.
\newblock {Phase diagram of the lattice SU(2) Higgs model}.
\newblock {\em Nucl. Phys. B} {\bf 2010}, {\em 828},~390--403,
  \href{http://xxx.lanl.gov/abs/0911.1721}{{\normalfont
  [arXiv:hep-lat/0911.1721]}}.

\bibitem[Andersson \em{et~al.}(1983)Andersson, Gustafson, Ingelman, and
  Sjostrand]{Andersson:1983ia}
Andersson, B.; Gustafson, G.; Ingelman, G.; Sjostrand, T.
\newblock {Parton Fragmentation and String Dynamics}.
\newblock {\em Phys. Rept.} {\bf 1983}, {\em 97},~31--145.

\bibitem[Sjostrand \em{et~al.}(2006)Sjostrand, Mrenna, and
  Skands]{Sjostrand:2006za}
Sjostrand, T.; Mrenna, S.; Skands, P.Z.
\newblock {PYTHIA 6.4 Physics and Manual}.
\newblock {\em JHEP} {\bf 2006}, {\em 05},~026,
  \href{http://xxx.lanl.gov/abs/hep-ph/0603175}{{\normalfont
  [hep-ph/0603175]}}.

\bibitem[Sj\"ostrand \em{et~al.}(2015)Sj\"ostrand, Ask, Christiansen, Corke,
  Desai, Ilten, Mrenna, Prestel, Rasmussen, and Skands]{Sjostrand:2014zea}
Sj\"ostrand, T.; Ask, S.; Christiansen, J.R.; Corke, R.; Desai, N.; Ilten, P.;
  Mrenna, S.; Prestel, S.; Rasmussen, C.O.; Skands, P.Z.
\newblock {An introduction to PYTHIA 8.2}.
\newblock {\em Comput. Phys. Commun.} {\bf 2015}, {\em 191},~159--177,
  \href{http://xxx.lanl.gov/abs/1410.3012}{{\normalfont
  [arXiv:hep-ph/1410.3012]}}.

\bibitem['t~Hooft(1974)]{tHooft:1973alw}
't~Hooft, G.
\newblock {A Planar Diagram Theory for Strong Interactions}.
\newblock {\em Nucl. Phys. B} {\bf 1974}, {\em 72},~461.

\bibitem[Sjostrand(1984)]{Sjostrand:1984ic}
Sjostrand, T.
\newblock {Jet Fragmentation of Nearby Partons}.
\newblock {\em Nucl. Phys. B} {\bf 1984}, {\em 248},~469--502.

\bibitem[Andersson \em{et~al.}(1980)Andersson, Gustafson, and
  Sjostrand]{Andersson:1980vk}
Andersson, B.; Gustafson, G.; Sjostrand, T.
\newblock {How to Find the Gluon Jets in e+ e- Annihilation}.
\newblock {\em Phys. Lett. B} {\bf 1980}, {\em 94},~211--215.

\bibitem[Andersson \em{et~al.}(1982)Andersson, Gustafson, and
  Sjostrand]{Andersson:1981ce}
Andersson, B.; Gustafson, G.; Sjostrand, T.
\newblock {A Model for Baryon Production in Quark and Gluon Jets}.
\newblock {\em Nucl. Phys. B} {\bf 1982}, {\em 197},~45--54.

\bibitem[Andersson \em{et~al.}(1985)Andersson, Gustafson, and
  Sjostrand]{Andersson:1984af}
Andersson, B.; Gustafson, G.; Sjostrand, T.
\newblock {Baryon Production in Jet Fragmentation and $\Upsilon$ Decay}.
\newblock {\em Phys. Scripta} {\bf 1985}, {\em 32},~574.

\bibitem[Kugo and Ojima(1979)]{Kugo:1979gm}
Kugo, T.; Ojima, I.
\newblock {Local Covariant Operator Formalism of Nonabelian Gauge Theories and
  Quark Confinement Problem}.
\newblock {\em Prog. Theor. Phys. Suppl.} {\bf 1979}, {\em 66},~1--130.

\bibitem[Kugo(1995)]{Kugo:1995km}
Kugo, T.
\newblock {The Universal renormalization factors Z(1) / Z(3) and color
  confinement condition in nonAbelian gauge theory}.
\newblock  {International Symposium on BRS Symmetry on the Occasion of Its 20th
  Anniversary},  1995,
  \href{http://xxx.lanl.gov/abs/hep-th/9511033}{{\normalfont
  [hep-th/9511033]}}.

\bibitem[Hata(1982)]{Hata:1981nd}
Hata, H.
\newblock {Restoration of the Local Gauge Symmetry and Color Confinement in
  Nonabelian Gauge Theories}.
\newblock {\em Prog. Theor. Phys.} {\bf 1982}, {\em 67},~1607.

\bibitem[Hata(1983)]{Hata:1983cs}
Hata, H.
\newblock {RESTORATION OF THE LOCAL GAUGE SYMMETRY AND COLOR CONFINEMENT IN
  NONABELIAN GAUGE THEORIES. II}.
\newblock {\em Prog. Theor. Phys.} {\bf 1983}, {\em 69},~1524--1536.

\bibitem[Marinari \em{et~al.}(1993)Marinari, Paciello, Parisi, and
  Taglienti]{Marinari:1992kh}
Marinari, E.; Paciello, M.L.; Parisi, G.; Taglienti, B.
\newblock {The String tension in gauge theories: A Suggestion for a new
  measurement method}.
\newblock {\em Phys. Lett. B} {\bf 1993}, {\em 298},~400--404,
  \href{http://xxx.lanl.gov/abs/hep-lat/9210021}{{\normalfont
  [hep-lat/9210021]}}.

\bibitem[Greensite \em{et~al.}(2004)Greensite, Olejnik, and
  Zwanziger]{Greensite:2004ke}
Greensite, J.; Olejnik, S.; Zwanziger, D.
\newblock {Coulomb energy, remnant symmetry, and the phases of nonAbelian gauge
  theories}.
\newblock {\em Phys. Rev. D} {\bf 2004}, {\em 69},~074506,
  \href{http://xxx.lanl.gov/abs/hep-lat/0401003}{{\normalfont
  [hep-lat/0401003]}}.

\bibitem[Greensite and Matsuyama(2018)]{Greensite:2018ebg}
Greensite, J.; Matsuyama, K.
\newblock {On the distinction between color confinement, and confinement}.
\newblock {\em PoS} {\bf 2018}, {\em Confinement2018},~046,
  \href{http://xxx.lanl.gov/abs/1811.01512}{{\normalfont
  [arXiv:hep-lat/1811.01512]}}.

\bibitem[Caudy and Greensite(2008)]{Caudy:2007sf}
Caudy, W.; Greensite, J.
\newblock {On the ambiguity of spontaneously broken gauge symmetry}.
\newblock {\em Phys. Rev. D} {\bf 2008}, {\em 78},~025018,
  \href{http://xxx.lanl.gov/abs/0712.0999}{{\normalfont
  [arXiv:hep-lat/0712.0999]}}.

\bibitem[Polyakov(1975)]{Polyakov:1975rs}
Polyakov, A.M.
\newblock {Compact Gauge Fields and the Infrared Catastrophe}.
\newblock {\em Phys. Lett. B} {\bf 1975}, {\em 59},~82--84.

\bibitem[Harrington and Shepard(1978{\natexlab{a}})]{Harrington:1978ua}
Harrington, B.J.; Shepard, H.K.
\newblock {Thermodynamics of the {Yang-Mills} Gas}.
\newblock {\em Phys. Rev. D} {\bf 1978}, {\em 18},~2990.

\bibitem[Harrington and Shepard(1978{\natexlab{b}})]{Harrington:1978ve}
Harrington, B.J.; Shepard, H.K.
\newblock {Periodic Euclidean Solutions and the Finite Temperature Yang-Mills
  Gas}.
\newblock {\em Phys. Rev. D} {\bf 1978}, {\em 17},~2122.

\bibitem[McLerran and Svetitsky(1981)]{McLerran:1981pb}
McLerran, L.D.; Svetitsky, B.
\newblock {Quark Liberation at High Temperature: A Monte Carlo Study of SU(2)
  Gauge Theory}.
\newblock {\em Phys. Rev. D} {\bf 1981}, {\em 24},~450.

\bibitem['t~Hooft(1978)]{tHooft:1977nqb}
't~Hooft, G.
\newblock {On the Phase Transition Towards Permanent Quark Confinement}.
\newblock {\em Nucl. Phys. B} {\bf 1978}, {\em 138},~1--25.

\bibitem['t~Hooft(1979)]{tHooft:1979rtg}
't~Hooft, G.
\newblock {A Property of Electric and Magnetic Flux in Nonabelian Gauge
  Theories}.
\newblock {\em Nucl. Phys. B} {\bf 1979}, {\em 153},~141--160.

\bibitem[Tomboulis and Yaffe(1985)]{Tomboulis:1985ah}
Tomboulis, E.T.; Yaffe, L.G.
\newblock {FINITE TEMPERATURE SU(2) LATTICE GAUGE THEORY}.
\newblock {\em Commun. Math. Phys.} {\bf 1985}, {\em 100},~313.

\bibitem[Cornwall(1982)]{Cornwall:1981zr}
Cornwall, J.M.
\newblock {Dynamical Mass Generation in Continuum QCD}.
\newblock {\em Phys. Rev. D} {\bf 1982}, {\em 26},~1453.

\bibitem[Bachas(1986)]{Bachas:1985xs}
Bachas, C.
\newblock {Convexity of the Quarkonium Potential}.
\newblock {\em Phys. Rev. D} {\bf 1986}, {\em 33},~2723.

\bibitem[Ambjorn \em{et~al.}(1984)Ambjorn, Olesen, and
  Peterson]{Ambjorn:1984mb}
Ambjorn, J.; Olesen, P.; Peterson, C.
\newblock {Stochastic Confinement and Dimensional Reduction. 1.
  Four-Dimensional SU(2) Lattice Gauge Theory}.
\newblock {\em Nucl. Phys. B} {\bf 1984}, {\em 240},~189--212.

\bibitem[Bali(2000)]{Bali:2000un}
Bali, G.S.
\newblock {Casimir scaling of SU(3) static potentials}.
\newblock {\em Phys. Rev. D} {\bf 2000}, {\em 62},~114503,
  \href{http://xxx.lanl.gov/abs/hep-lat/0006022}{{\normalfont
  [hep-lat/0006022]}}.

\bibitem[Junior \em{et~al.}(2020)Junior, Oxman, and Sim\~oes]{Junior:2019fty}
Junior, D.R.; Oxman, L.E.; Sim\~oes, G.M.
\newblock {3D Yang-Mills confining properties from a non-Abelian ensemble
  perspective}.
\newblock {\em JHEP} {\bf 2020}, {\em 01},~180,
  \href{http://xxx.lanl.gov/abs/1911.10144}{{\normalfont
  [arXiv:hep-th/1911.10144]}}.

\bibitem[Greensite(1979)]{Greensite:1979yn}
Greensite, J.P.
\newblock {Calculation of the {Yang-Mills} Vacuum Wave Functional}.
\newblock {\em Nucl. Phys. B} {\bf 1979}, {\em 158},~469--496.

\bibitem[Greensite(1980)]{Greensite:1979ha}
Greensite, J.P.
\newblock {Large Scale Vacuum Structure and New Calculational Techniques in
  Lattice SU($N$) Gauge Theory}.
\newblock {\em Nucl. Phys. B} {\bf 1980}, {\em 166},~113--124.

\bibitem[Leigh \em{et~al.}(2007)Leigh, Minic, and Yelnikov]{Leigh:2006vg}
Leigh, R.G.; Minic, D.; Yelnikov, A.
\newblock {On the Glueball Spectrum of Pure Yang-Mills Theory in 2+1
  Dimensions}.
\newblock {\em Phys. Rev. D} {\bf 2007}, {\em 76},~065018,
  \href{http://xxx.lanl.gov/abs/hep-th/0604060}{{\normalfont
  [hep-th/0604060]}}.

\bibitem[Karabali \em{et~al.}(1998)Karabali, Kim, and Nair]{Karabali:1998yq}
Karabali, D.; Kim, C.j.; Nair, V.P.
\newblock {On the vacuum wave function and string tension of Yang-Mills
  theories in (2+1)-dimensions}.
\newblock {\em Phys. Lett. B} {\bf 1998}, {\em 434},~103--109,
  \href{http://xxx.lanl.gov/abs/hep-th/9804132}{{\normalfont
  [hep-th/9804132]}}.

\bibitem[Karabali \em{et~al.}(2010)Karabali, Nair, and
  Yelnikov]{Karabali:2009rg}
Karabali, D.; Nair, V.P.; Yelnikov, A.
\newblock {The Hamiltonian Approach to Yang-Mills (2+1): An Expansion Scheme
  and Corrections to String Tension}.
\newblock {\em Nucl. Phys. B} {\bf 2010}, {\em 824},~387--414,
  \href{http://xxx.lanl.gov/abs/0906.0783}{{\normalfont
  [arXiv:hep-th/0906.0783]}}.

\bibitem[Reinhardt and Feuchter(2005)]{Reinhardt:2004mm}
Reinhardt, H.; Feuchter, C.
\newblock {On the Yang-Mills wave functional in Coulomb gauge}.
\newblock {\em Phys. Rev. D} {\bf 2005}, {\em 71},~105002,
  \href{http://xxx.lanl.gov/abs/hep-th/0408237}{{\normalfont
  [hep-th/0408237]}}.

\bibitem[Feuchter and Reinhardt(2004)]{Feuchter:2004mk}
Feuchter, C.; Reinhardt, H.
\newblock {Variational solution of the Yang-Mills Schrodinger equation in
  Coulomb gauge}.
\newblock {\em Phys. Rev. D} {\bf 2004}, {\em 70},~105021,
  \href{http://xxx.lanl.gov/abs/hep-th/0408236}{{\normalfont
  [hep-th/0408236]}}.

\bibitem[Greensite and Olejnik(2008)]{Greensite:2007ij}
Greensite, J.; Olejnik, S.
\newblock {Dimensional Reduction and the Yang-Mills Vacuum State in 2+1
  Dimensions}.
\newblock {\em Phys. Rev. D} {\bf 2008}, {\em 77},~065003,
  \href{http://xxx.lanl.gov/abs/0707.2860}{{\normalfont
  [arXiv:hep-lat/0707.2860]}}.

\bibitem[Kratochvila and de~Forcrand(2003)]{Kratochvila:2003zj}
Kratochvila, S.; de~Forcrand, P.
\newblock {Observing string breaking with Wilson loops}.
\newblock {\em Nucl. Phys. B} {\bf 2003}, {\em 671},~103--132,
  \href{http://xxx.lanl.gov/abs/hep-lat/0306011}{{\normalfont
  [hep-lat/0306011]}}.

\bibitem[Luscher(1981)]{Luscher:1980ac}
Luscher, M.
\newblock {Symmetry Breaking Aspects of the Roughening Transition in Gauge
  Theories}.
\newblock {\em Nucl. Phys. B} {\bf 1981}, {\em 180},~317--329.

\bibitem[Alvarez(1981)]{Alvarez:1981kc}
Alvarez, O.
\newblock {The Static Potential in String Models}.
\newblock {\em Phys. Rev. D} {\bf 1981}, {\em 24},~440.

\bibitem[Luscher \em{et~al.}(1981)Luscher, Munster, and Weisz]{Luscher:1980iy}
Luscher, M.; Munster, G.; Weisz, P.
\newblock {How Thick Are Chromoelectric Flux Tubes?}
\newblock {\em Nucl. Phys. B} {\bf 1981}, {\em 180},~1--12.

\bibitem[Hasenfratz \em{et~al.}(1981)Hasenfratz, Hasenfratz, and
  Hasenfratz]{Hasenfratz:1980ue}
Hasenfratz, A.; Hasenfratz, E.; Hasenfratz, P.
\newblock {Generalized Roughening Transition and Its Effect on the String
  Tension}.
\newblock {\em Nucl. Phys. B} {\bf 1981}, {\em 180},~353--367.

\bibitem[Athenodorou \em{et~al.}(2007)Athenodorou, Bringoltz, and
  Teper]{Athenodorou:2007du}
Athenodorou, A.; Bringoltz, B.; Teper, M.
\newblock {The Closed string spectrum of SU(N) gauge theories in 2+1
  dimensions}.
\newblock {\em Phys. Lett. B} {\bf 2007}, {\em 656},~132--140,
  \href{http://xxx.lanl.gov/abs/0709.0693}{{\normalfont
  [arXiv:hep-lat/0709.0693]}}.

\bibitem[Belavin \em{et~al.}(1975)Belavin, Polyakov, Schwartz, and
  Tyupkin]{Belavin:1975fg}
Belavin, A.A.; Polyakov, A.M.; Schwartz, A.S.; Tyupkin, Y.S.
\newblock {Pseudoparticle Solutions of the Yang-Mills Equations}.
\newblock {\em Phys. Lett. B} {\bf 1975}, {\em 59},~85--87.

\bibitem[Ambjorn and Olesen(1980)]{Ambjorn:1980ms}
Ambjorn, J.; Olesen, P.
\newblock {A Color Magnetic Vortex Condensate in QCD}.
\newblock {\em Nucl. Phys. B} {\bf 1980}, {\em 170},~265--282.

\bibitem[Diakonov and Maul(2002)]{Diakonov:2002bx}
Diakonov, D.; Maul, M.
\newblock {Center vortex solutions of the Yang-Mills effective action in three
  and four dimensions}.
\newblock {\em Phys. Rev. D} {\bf 2002}, {\em 66},~096004,
  \href{http://xxx.lanl.gov/abs/hep-lat/0204012}{{\normalfont
  [hep-lat/0204012]}}.

\bibitem[Nielsen and Olesen(1979)]{Nielsen:1979xu}
Nielsen, H.B.; Olesen, P.
\newblock {A Quantum Liquid Model for the QCD Vacuum: Gauge and Rotational
  Invariance of Domained and Quantized Homogeneous Color Fields}.
\newblock {\em Nucl. Phys. B} {\bf 1979}, {\em 160},~380--396.

\bibitem[Cornwall(1979)]{Cornwall:1979hz}
Cornwall, J.M.
\newblock {Quark Confinement and Vortices in Massive Gauge Invariant QCD}.
\newblock {\em Nucl. Phys. B} {\bf 1979}, {\em 157},~392--412.

\bibitem[Kovacs and Tomboulis(2000)]{Kovacs:2000sy}
Kovacs, T.G.; Tomboulis, E.T.
\newblock {Computation of the vortex free energy in SU(2) gauge theory}.
\newblock {\em Phys. Rev. Lett.} {\bf 2000}, {\em 85},~704--707,
  \href{http://xxx.lanl.gov/abs/hep-lat/0002004}{{\normalfont
  [hep-lat/0002004]}}.

\bibitem[Faber \em{et~al.}(1998)Faber, Greensite, and Olejnik]{Faber:1997rp}
Faber, M.; Greensite, J.; Olejnik, S.
\newblock {Casimir scaling from center vortices: Towards an understanding of
  the adjoint string tension}.
\newblock {\em Phys. Rev. D} {\bf 1998}, {\em 57},~2603--2609,
  \href{http://xxx.lanl.gov/abs/hep-lat/9710039}{{\normalfont
  [hep-lat/9710039]}}.

\bibitem[Greensite \em{et~al.}(2007)Greensite, Langfeld, Olejnik, Reinhardt,
  and Tok]{Greensite:2006sm}
Greensite, J.; Langfeld, K.; Olejnik, S.; Reinhardt, H.; Tok, T.
\newblock {Color Screening, Casimir Scaling, and Domain Structure in G(2) and
  SU(N) Gauge Theories}.
\newblock {\em Phys. Rev. D} {\bf 2007}, {\em 75},~034501,
  \href{http://xxx.lanl.gov/abs/hep-lat/0609050}{{\normalfont
  [hep-lat/0609050]}}.

\bibitem[Del~Debbio \em{et~al.}(1998)Del~Debbio, Faber, Giedt, Greensite, and
  Olejnik]{DelDebbio:1998luz}
Del~Debbio, L.; Faber, M.; Giedt, J.; Greensite, J.; Olejnik, S.
\newblock {Detection of center vortices in the lattice Yang-Mills vacuum}.
\newblock {\em Phys. Rev. D} {\bf 1998}, {\em 58},~094501,
  \href{http://xxx.lanl.gov/abs/hep-lat/9801027}{{\normalfont
  [hep-lat/9801027]}}.

\bibitem[Gribov(1978)]{Gribov:1977wm}
Gribov, V.N.
\newblock {Quantization of Nonabelian Gauge Theories}.
\newblock {\em Nucl. Phys. B} {\bf 1978}, {\em 139},~1.

\bibitem[Neuberger(1987)]{Neuberger:1986xz}
Neuberger, H.
\newblock {Nonperturbative {BRS} Invariance and the Gribov Problem}.
\newblock {\em Phys. Lett. B} {\bf 1987}, {\em 183},~337--340.

\bibitem[Zwanziger(1998)]{Zwanziger:1998ez}
Zwanziger, D.
\newblock {Renormalization in the Coulomb gauge and order parameter for
  confinement in QCD}.
\newblock {\em Nucl. Phys. B} {\bf 1998}, {\em 518},~237--272.

\bibitem[Faber \em{et~al.}(2001)Faber, Greensite, and Olejnik]{Faber:2001zs}
Faber, M.; Greensite, J.; Olejnik, S.
\newblock {Direct Laplacian center gauge}.
\newblock {\em JHEP} {\bf 2001}, {\em 11},~053,
  \href{http://xxx.lanl.gov/abs/hep-lat/0106017}{{\normalfont
  [hep-lat/0106017]}}.

\bibitem[de~Forcrand and D'Elia(1999)]{deForcrand:1999our}
de~Forcrand, P.; D'Elia, M.
\newblock {On the relevance of center vortices to QCD}.
\newblock {\em Phys. Rev. Lett.} {\bf 1999}, {\em 82},~4582--4585,
  \href{http://xxx.lanl.gov/abs/hep-lat/9901020}{{\normalfont
  [hep-lat/9901020]}}.

\bibitem[Engelhardt \em{et~al.}(1998)Engelhardt, Langfeld, Reinhardt, and
  Tennert]{Engelhardt:1998wu}
Engelhardt, M.; Langfeld, K.; Reinhardt, H.; Tennert, O.
\newblock {Interaction of confining vortices in SU(2) lattice gauge theory}.
\newblock {\em Phys. Lett. B} {\bf 1998}, {\em 431},~141--146,
  \href{http://xxx.lanl.gov/abs/hep-lat/9801030}{{\normalfont
  [hep-lat/9801030]}}.

\bibitem[Gubarev \em{et~al.}(2003)Gubarev, Kovalenko, Polikarpov, Syritsyn, and
  Zakharov]{Gubarev:2002ek}
Gubarev, F.V.; Kovalenko, A.V.; Polikarpov, M.I.; Syritsyn, S.N.; Zakharov,
  V.I.
\newblock {Fine tuned vortices in lattice SU(2) gluodynamics}.
\newblock {\em Phys. Lett. B} {\bf 2003}, {\em 574},~136--140,
  \href{http://xxx.lanl.gov/abs/hep-lat/0212003}{{\normalfont
  [hep-lat/0212003]}}.

\bibitem[de~Forcrand and von Smekal(2002)]{deForcrand:2001nd}
de~Forcrand, P.; von Smekal, L.
\newblock {'t Hooft loops, electric flux sectors and confinement in SU(2)
  Yang-Mills theory}.
\newblock {\em Phys. Rev. D} {\bf 2002}, {\em 66},~011504,
  \href{http://xxx.lanl.gov/abs/hep-lat/0107018}{{\normalfont
  [hep-lat/0107018]}}.

\bibitem[Engelhardt \em{et~al.}(2000)Engelhardt, Langfeld, Reinhardt, and
  Tennert]{Engelhardt:1999fd}
Engelhardt, M.; Langfeld, K.; Reinhardt, H.; Tennert, O.
\newblock {Deconfinement in SU(2) Yang-Mills theory as a center vortex
  percolation transition}.
\newblock {\em Phys. Rev. D} {\bf 2000}, {\em 61},~054504,
  \href{http://xxx.lanl.gov/abs/hep-lat/9904004}{{\normalfont
  [hep-lat/9904004]}}.

\bibitem[Langfeld \em{et~al.}(1999)Langfeld, Tennert, Engelhardt, and
  Reinhardt]{Langfeld:1998cz}
Langfeld, K.; Tennert, O.; Engelhardt, M.; Reinhardt, H.
\newblock {Center vortices of Yang-Mills theory at finite temperatures}.
\newblock {\em Phys. Lett. B} {\bf 1999}, {\em 452},~301,
  \href{http://xxx.lanl.gov/abs/hep-lat/9805002}{{\normalfont
  [hep-lat/9805002]}}.

\bibitem[Greensite and Olejnik(2006)]{Greensite:2006ng}
Greensite, J.; Olejnik, S.
\newblock {Vortices, symmetry breaking and temporary confinement in SU(2)
  gauge-Higgs theory}.
\newblock {\em Phys. Rev. D} {\bf 2006}, {\em 74},~014502,
  \href{http://xxx.lanl.gov/abs/hep-lat/0603024}{{\normalfont
  [hep-lat/0603024]}}.

\bibitem[Weinberg(1976)]{Weinberg:1975gm}
Weinberg, S.
\newblock {Implications of Dynamical Symmetry Breaking}.
\newblock {\em Phys. Rev. D} {\bf 1976}, {\em 13},~974--996.
\newblock [Addendum: Phys.Rev.D 19, 1277--1280 (1979)].

\bibitem[Susskind(1979)]{Susskind:1978ms}
Susskind, L.
\newblock {Dynamics of Spontaneous Symmetry Breaking in the Weinberg-Salam
  Theory}.
\newblock {\em Phys. Rev. D} {\bf 1979}, {\em 20},~2619--2625.

\bibitem[Hill and Simmons(2003)]{Hill:2002ap}
Hill, C.T.; Simmons, E.H.
\newblock {Strong Dynamics and Electroweak Symmetry Breaking}.
\newblock {\em Phys. Rept.} {\bf 2003}, {\em 381},~235--402,
  \href{http://xxx.lanl.gov/abs/hep-ph/0203079}{{\normalfont
  [hep-ph/0203079]}}.
\newblock [Erratum: Phys.Rept. 390, 553--554 (2004)].

\bibitem[Banks and Casher(1980)]{Banks:1979yr}
Banks, T.; Casher, A.
\newblock {Chiral Symmetry Breaking in Confining Theories}.
\newblock {\em Nucl. Phys. B} {\bf 1980}, {\em 169},~103--125.

\bibitem[Nambu and Jona-Lasinio(1961)]{Nambu:1961tp}
Nambu, Y.; Jona-Lasinio, G.
\newblock {Dynamical Model of Elementary Particles Based on an Analogy with
  Superconductivity. 1.}
\newblock {\em Phys. Rev.} {\bf 1961}, {\em 122},~345--358.

\bibitem[Suganuma \em{et~al.}(2016)Suganuma, Doi, and
  Iritani]{Suganuma:2014wya}
Suganuma, H.; Doi, T.M.; Iritani, T.
\newblock {Analytical formulae of the Polyakov and Wilson loops with Dirac
  eigenmodes in lattice QCD}.
\newblock {\em PTEP} {\bf 2016}, {\em 2016},~013B06,
  \href{http://xxx.lanl.gov/abs/1404.6494}{{\normalfont
  [arXiv:hep-lat/1404.6494]}}.

\bibitem[Gattringer(2006)]{Gattringer:2006ci}
Gattringer, C.
\newblock {Linking confinement to spectral properties of the Dirac operator}.
\newblock {\em Phys. Rev. Lett.} {\bf 2006}, {\em 97},~032003,
  \href{http://xxx.lanl.gov/abs/hep-lat/0605018}{{\normalfont
  [hep-lat/0605018]}}.

\bibitem[Alexandrou \em{et~al.}(2000)Alexandrou, de~Forcrand, and
  D'Elia]{Alexandrou:1999vx}
Alexandrou, C.; de~Forcrand, P.; D'Elia, M.
\newblock {The Role of center vortices in QCD}.
\newblock {\em Nucl. Phys. A} {\bf 2000}, {\em 663},~1031--1034,
  \href{http://xxx.lanl.gov/abs/hep-lat/9909005}{{\normalfont
  [hep-lat/9909005]}}.

\bibitem[Trewartha \em{et~al.}(2015)Trewartha, Kamleh, and
  Leinweber]{Trewartha:2015nna}
Trewartha, D.; Kamleh, W.; Leinweber, D.
\newblock {Evidence that centre vortices underpin dynamical chiral symmetry
  breaking in SU(3) gauge theory}.
\newblock {\em Phys. Lett. B} {\bf 2015}, {\em 747},~373--377,
  \href{http://xxx.lanl.gov/abs/1502.06753}{{\normalfont
  [arXiv:hep-lat/1502.06753]}}.

\bibitem[Witten(1979)]{Witten:1979vv}
Witten, E.
\newblock {Current Algebra Theorems for the U(1) Goldstone Boson}.
\newblock {\em Nucl. Phys. B} {\bf 1979}, {\em 156},~269--283.

\bibitem[Veneziano(1979)]{Veneziano:1979ec}
Veneziano, G.
\newblock {U(1) Without Instantons}.
\newblock {\em Nucl. Phys. B} {\bf 1979}, {\em 159},~213--224.

\bibitem[Del~Debbio \em{et~al.}(2005)Del~Debbio, Giusti, and
  Pica]{DelDebbio:2004ns}
Del~Debbio, L.; Giusti, L.; Pica, C.
\newblock {Topological susceptibility in the SU(3) gauge theory}.
\newblock {\em Phys. Rev. Lett.} {\bf 2005}, {\em 94},~032003,
  \href{http://xxx.lanl.gov/abs/hep-th/0407052}{{\normalfont
  [hep-th/0407052]}}.

\bibitem[Cichy \em{et~al.}(2015)Cichy, Garcia-Ramos, Jansen, Ottnad, and
  Urbach]{Cichy:2015jra}
Cichy, K.; Garcia-Ramos, E.; Jansen, K.; Ottnad, K.; Urbach, C.
\newblock {Non-perturbative Test of the Witten-Veneziano Formula from Lattice
  QCD}.
\newblock {\em JHEP} {\bf 2015}, {\em 09},~020,
  \href{http://xxx.lanl.gov/abs/1504.07954}{{\normalfont
  [arXiv:hep-lat/1504.07954]}}.

\bibitem[Engelhardt(2000)]{Engelhardt:2000wc}
Engelhardt, M.
\newblock {Center vortex model for the infrared sector of Yang-Mills theory:
  Topological susceptibility}.
\newblock {\em Nucl. Phys. B} {\bf 2000}, {\em 585},~614,
  \href{http://xxx.lanl.gov/abs/hep-lat/0004013}{{\normalfont
  [hep-lat/0004013]}}.

\bibitem[Engelhardt(2011)]{Engelhardt:2010ft}
Engelhardt, M.
\newblock {Center vortex model for the infrared sector of SU(3) Yang-Mills
  theory: Topological susceptibility}.
\newblock {\em Phys. Rev. D} {\bf 2011}, {\em 83},~025015,
  \href{http://xxx.lanl.gov/abs/1008.4953}{{\normalfont
  [arXiv:hep-lat/1008.4953]}}.

\bibitem[Bertle \em{et~al.}(2001)Bertle, Engelhardt, and Faber]{Bertle:2001xd}
Bertle, R.; Engelhardt, M.; Faber, M.
\newblock {Topological susceptibility of Yang-Mills center projection
  vortices}.
\newblock {\em Phys. Rev. D} {\bf 2001}, {\em 64},~074504,
  \href{http://xxx.lanl.gov/abs/hep-lat/0104004}{{\normalfont
  [hep-lat/0104004]}}.

\bibitem[Kamleh \em{et~al.}(2017)Kamleh, Leinweber, and
  Trewartha]{Kamleh:2017lij}
Kamleh, W.; Leinweber, D.B.; Trewartha, D.
\newblock {Centre vortices are the seeds of dynamical chiral symmetry
  breaking}.
\newblock {\em PoS} {\bf 2017}, {\em LATTICE2016},~353,
  \href{http://xxx.lanl.gov/abs/1701.03241}{{\normalfont
  [arXiv:hep-lat/1701.03241]}}.

\bibitem[Trewartha \em{et~al.}(2015)Trewartha, Kamleh, and
  Leinweber]{Trewartha:2015ida}
Trewartha, D.; Kamleh, W.; Leinweber, D.
\newblock {Connection between center vortices and instantons through
  gauge-field smoothing}.
\newblock {\em Phys. Rev. D} {\bf 2015}, {\em 92},~074507,
  \href{http://xxx.lanl.gov/abs/1509.05518}{{\normalfont
  [arXiv:hep-lat/1509.05518]}}.

\bibitem[Langfeld(2004)]{Langfeld:2003ev}
Langfeld, K.
\newblock {Vortex structures in pure SU(3) lattice gauge theory}.
\newblock {\em Phys. Rev. D} {\bf 2004}, {\em 69},~014503,
  \href{http://xxx.lanl.gov/abs/hep-lat/0307030}{{\normalfont
  [hep-lat/0307030]}}.

\bibitem[Engelhardt and Reinhardt(2000)]{Engelhardt:1999wr}
Engelhardt, M.; Reinhardt, H.
\newblock {Center vortex model for the infrared sector of Yang-Mills theory:
  Confinement and deconfinement}.
\newblock {\em Nucl. Phys. B} {\bf 2000}, {\em 585},~591--613,
  \href{http://xxx.lanl.gov/abs/hep-lat/9912003}{{\normalfont
  [hep-lat/9912003]}}.

\bibitem[Engelhardt(2002)]{Engelhardt:2002qs}
Engelhardt, M.
\newblock {Center vortex model for the infrared sector of Yang-Mills theory:
  Quenched Dirac spectrum and chiral condensate}.
\newblock {\em Nucl. Phys. B} {\bf 2002}, {\em 638},~81--110,
  \href{http://xxx.lanl.gov/abs/hep-lat/0204002}{{\normalfont
  [hep-lat/0204002]}}.

\bibitem[Quandt \em{et~al.}(2005)Quandt, Reinhardt, and
  Engelhardt]{Quandt:2004gy}
Quandt, M.; Reinhardt, H.; Engelhardt, M.
\newblock {Center vortex model for the infrared sector of SU(3) Yang-Mills
  theory - vortex free energy}.
\newblock {\em Phys. Rev. D} {\bf 2005}, {\em 71},~054026,
  \href{http://xxx.lanl.gov/abs/hep-lat/0412033}{{\normalfont
  [hep-lat/0412033]}}.

\bibitem[Engelhardt(2004)]{Engelhardt:2004qq}
Engelhardt, M.
\newblock {Center vortex model for the infrared sector of SU(3) Yang-Mills
  theory - baryonic potential}.
\newblock {\em Phys. Rev. D} {\bf 2004}, {\em 70},~074004,
  \href{http://xxx.lanl.gov/abs/hep-lat/0406022}{{\normalfont
  [hep-lat/0406022]}}.

\bibitem[Alexandrou \em{et~al.}(2003)Alexandrou, de~Forcrand, and
  Jahn]{Alexandrou:2002sn}
Alexandrou, C.; de~Forcrand, P.; Jahn, O.
\newblock {The Ground state of three quarks}.
\newblock {\em Nucl. Phys. B Proc. Suppl.} {\bf 2003}, {\em 119},~667--669,
  \href{http://xxx.lanl.gov/abs/hep-lat/0209062}{{\normalfont
  [hep-lat/0209062]}}.

\bibitem[Takahashi and Suganuma(2004)]{Takahashi:2004rw}
Takahashi, T.T.; Suganuma, H.
\newblock {Detailed analysis of the gluonic excitation in the three-quark
  system in lattice QCD}.
\newblock {\em Phys. Rev. D} {\bf 2004}, {\em 70},~074506,
  \href{http://xxx.lanl.gov/abs/hep-lat/0409105}{{\normalfont
  [hep-lat/0409105]}}.

\bibitem[Zwanziger(1991)]{Zwanziger:1991gz}
Zwanziger, D.
\newblock {Vanishing of zero momentum lattice gluon propagator and color
  confinement}.
\newblock {\em Nucl. Phys. B} {\bf 1991}, {\em 364},~127--161.

\bibitem[Greensite \em{et~al.}(2005)Greensite, Olejnik, and
  Zwanziger]{Greensite:2004ur}
Greensite, J.; Olejnik, S.; Zwanziger, D.
\newblock {Center vortices and the Gribov horizon}.
\newblock {\em JHEP} {\bf 2005}, {\em 05},~070,
  \href{http://xxx.lanl.gov/abs/hep-lat/0407032}{{\normalfont
  [hep-lat/0407032]}}.

\bibitem[Zwanziger(2003)]{Zwanziger:2002sh}
Zwanziger, D.
\newblock {No confinement without Coulomb confinement}.
\newblock {\em Phys. Rev. Lett.} {\bf 2003}, {\em 90},~102001,
  \href{http://xxx.lanl.gov/abs/hep-lat/0209105}{{\normalfont
  [hep-lat/0209105]}}.

\bibitem[Greensite and Olejnik(2003)]{Greensite:2003xf}
Greensite, J.; Olejnik, S.
\newblock {Coulomb energy, vortices, and confinement}.
\newblock {\em Phys. Rev. D} {\bf 2003}, {\em 67},~094503,
  \href{http://xxx.lanl.gov/abs/hep-lat/0302018}{{\normalfont
  [hep-lat/0302018]}}.

\bibitem[West(1982)]{West:1982bt}
West, G.B.
\newblock {Confinement, the Wilson Loop and the Gluon Propagator}.
\newblock {\em Phys. Lett. B} {\bf 1982}, {\em 115},~468--472.

\bibitem[Eichmann(2013)]{Eichmann:2013afa}
Eichmann, G.
\newblock {Hadron phenomenology in the Dyson-Schwinger approach}.
\newblock {\em J. Phys. Conf. Ser.} {\bf 2013}, {\em 426},~012014.

\bibitem[Zwanziger(2002)]{Zwanziger:2001kw}
Zwanziger, D.
\newblock {Nonperturbative Landau gauge and infrared critical exponents in
  QCD}.
\newblock {\em Phys. Rev. D} {\bf 2002}, {\em 65},~094039,
  \href{http://xxx.lanl.gov/abs/hep-th/0109224}{{\normalfont
  [hep-th/0109224]}}.

\bibitem[Fischer and Pawlowski(2007)]{Fischer:2006vf}
Fischer, C.S.; Pawlowski, J.M.
\newblock {Uniqueness of infrared asymptotics in Landau gauge Yang-Mills
  theory}.
\newblock {\em Phys. Rev. D} {\bf 2007}, {\em 75},~025012,
  \href{http://xxx.lanl.gov/abs/hep-th/0609009}{{\normalfont
  [hep-th/0609009]}}.

\bibitem[Alkofer \em{et~al.}(2010)Alkofer, Huber, and
  Schwenzer]{Alkofer:2008jy}
Alkofer, R.; Huber, M.Q.; Schwenzer, K.
\newblock {Infrared singularities in Landau gauge Yang-Mills theory}.
\newblock {\em Phys. Rev. D} {\bf 2010}, {\em 81},~105010,
  \href{http://xxx.lanl.gov/abs/0801.2762}{{\normalfont
  [arXiv:hep-th/0801.2762]}}.

\bibitem[Fischer and Pawlowski(2009)]{Fischer:2009tn}
Fischer, C.S.; Pawlowski, J.M.
\newblock {Uniqueness of infrared asymptotics in Landau gauge Yang-Mills theory
  II}.
\newblock {\em Phys. Rev. D} {\bf 2009}, {\em 80},~025023,
  \href{http://xxx.lanl.gov/abs/0903.2193}{{\normalfont
  [arXiv:hep-th/0903.2193]}}.

\bibitem[Lerche and von Smekal(2002)]{Lerche:2002ep}
Lerche, C.; von Smekal, L.
\newblock {On the infrared exponent for gluon and ghost propagation in Landau
  gauge QCD}.
\newblock {\em Phys. Rev. D} {\bf 2002}, {\em 65},~125006,
  \href{http://xxx.lanl.gov/abs/hep-ph/0202194}{{\normalfont
  [hep-ph/0202194]}}.

\bibitem[Alkofer \em{et~al.}(2009)Alkofer, Fischer, Llanes-Estrada, and
  Schwenzer]{Alkofer:2008tt}
Alkofer, R.; Fischer, C.S.; Llanes-Estrada, F.J.; Schwenzer, K.
\newblock {The Quark-gluon vertex in Landau gauge QCD: Its role in dynamical
  chiral symmetry breaking and quark confinement}.
\newblock {\em Annals Phys.} {\bf 2009}, {\em 324},~106--172,
  \href{http://xxx.lanl.gov/abs/0804.3042}{{\normalfont
  [arXiv:hep-ph/0804.3042]}}.

\bibitem[Maas(2007)]{Maas:2007uv}
Maas, A.
\newblock {Two and three-point Green's functions in two-dimensional
  Landau-gauge Yang-Mills theory}.
\newblock {\em Phys. Rev. D} {\bf 2007}, {\em 75},~116004,
  \href{http://xxx.lanl.gov/abs/0704.0722}{{\normalfont
  [arXiv:hep-lat/0704.0722]}}.

\bibitem[Cucchieri and Mendes(2008)]{Cucchieri:2007rg}
Cucchieri, A.; Mendes, T.
\newblock {Constraints on the IR behavior of the gluon propagator in Yang-Mills
  theories}.
\newblock {\em Phys. Rev. Lett.} {\bf 2008}, {\em 100},~241601,
  \href{http://xxx.lanl.gov/abs/0712.3517}{{\normalfont
  [arXiv:hep-lat/0712.3517]}}.

\bibitem[Bogolubsky \em{et~al.}(2009)Bogolubsky, Ilgenfritz, Muller-Preussker,
  and Sternbeck]{Bogolubsky:2009dc}
Bogolubsky, I.L.; Ilgenfritz, E.M.; Muller-Preussker, M.; Sternbeck, A.
\newblock {Lattice gluodynamics computation of Landau gauge Green's functions
  in the deep infrared}.
\newblock {\em Phys. Lett. B} {\bf 2009}, {\em 676},~69--73,
  \href{http://xxx.lanl.gov/abs/0901.0736}{{\normalfont
  [arXiv:hep-lat/0901.0736]}}.

\bibitem[Boucaud \em{et~al.}(2008)Boucaud, Leroy, Yaouanc, Micheli, Pene, and
  Rodriguez-Quintero]{Boucaud:2008ji}
Boucaud, P.; Leroy, J.P.; Yaouanc, A.L.; Micheli, J.; Pene, O.;
  Rodriguez-Quintero, J.
\newblock {IR finiteness of the ghost dressing function from numerical
  resolution of the ghost SD equation}.
\newblock {\em JHEP} {\bf 2008}, {\em 06},~012,
  \href{http://xxx.lanl.gov/abs/0801.2721}{{\normalfont
  [arXiv:hep-ph/0801.2721]}}.

\bibitem[Aguilar \em{et~al.}(2008)Aguilar, Binosi, and
  Papavassiliou]{Aguilar:2008xm}
Aguilar, A.C.; Binosi, D.; Papavassiliou, J.
\newblock {Gluon and ghost propagators in the Landau gauge: Deriving lattice
  results from Schwinger-Dyson equations}.
\newblock {\em Phys. Rev. D} {\bf 2008}, {\em 78},~025010,
  \href{http://xxx.lanl.gov/abs/0802.1870}{{\normalfont
  [arXiv:hep-ph/0802.1870]}}.

\bibitem[Dudal \em{et~al.}(2008)Dudal, Gracey, Sorella, Vandersickel, and
  Verschelde]{Dudal:2008sp}
Dudal, D.; Gracey, J.A.; Sorella, S.P.; Vandersickel, N.; Verschelde, H.
\newblock {A Refinement of the Gribov-Zwanziger approach in the Landau gauge:
  Infrared propagators in harmony with the lattice results}.
\newblock {\em Phys. Rev. D} {\bf 2008}, {\em 78},~065047,
  \href{http://xxx.lanl.gov/abs/0806.4348}{{\normalfont
  [arXiv:hep-th/0806.4348]}}.

\bibitem[Fischer \em{et~al.}(2009)Fischer, Maas, and Pawlowski]{Fischer:2008uz}
Fischer, C.S.; Maas, A.; Pawlowski, J.M.
\newblock {On the infrared behavior of Landau gauge Yang-Mills theory}.
\newblock {\em Annals Phys.} {\bf 2009}, {\em 324},~2408--2437,
  \href{http://xxx.lanl.gov/abs/0810.1987}{{\normalfont
  [arXiv:hep-ph/0810.1987]}}.

\bibitem[Cucchieri \em{et~al.}(2005)Cucchieri, Mendes, and
  Taurines]{Cucchieri:2004mf}
Cucchieri, A.; Mendes, T.; Taurines, A.R.
\newblock {Positivity violation for the lattice Landau gluon propagator}.
\newblock {\em Phys. Rev. D} {\bf 2005}, {\em 71},~051902,
  \href{http://xxx.lanl.gov/abs/hep-lat/0406020}{{\normalfont
  [hep-lat/0406020]}}.

\bibitem[Braun \em{et~al.}(2010)Braun, Gies, and Pawlowski]{Braun:2007bx}
Braun, J.; Gies, H.; Pawlowski, J.M.
\newblock {Quark Confinement from Color Confinement}.
\newblock {\em Phys. Lett. B} {\bf 2010}, {\em 684},~262--267,
  \href{http://xxx.lanl.gov/abs/0708.2413}{{\normalfont
  [arXiv:hep-th/0708.2413]}}.

\bibitem[Cooper and Zwanziger(2018)]{Cooper:2018xyn}
Cooper, P.; Zwanziger, D.
\newblock {Schwinger-Dyson Equations in Coulomb Gauge Consistent with Numerical
  Simulation}.
\newblock {\em Phys. Rev. D} {\bf 2018}, {\em 98},~114006,
  \href{http://xxx.lanl.gov/abs/1803.06597}{{\normalfont
  [arXiv:hep-th/1803.06597]}}.

\bibitem[Wetterich(1993)]{Wetterich:1992yh}
Wetterich, C.
\newblock {Exact evolution equation for the effective potential}.
\newblock {\em Phys. Lett. B} {\bf 1993}, {\em 301},~90--94,
  \href{http://xxx.lanl.gov/abs/1710.05815}{{\normalfont
  [arXiv:hep-th/1710.05815]}}.

\bibitem[Fister and Pawlowski(2013)]{Fister:2013bh}
Fister, L.; Pawlowski, J.M.
\newblock {Confinement from Correlation Functions}.
\newblock {\em Phys. Rev. D} {\bf 2013}, {\em 88},~045010,
  \href{http://xxx.lanl.gov/abs/1301.4163}{{\normalfont
  [arXiv:hep-ph/1301.4163]}}.

\bibitem[Marhauser and Pawlowski(2008)]{Marhauser:2008fz}
Marhauser, F.; Pawlowski, J.M.
\newblock {Confinement in Polyakov Gauge} {\bf 2008}.
\newblock  \href{http://xxx.lanl.gov/abs/0812.1144}{{\normalfont
  [arXiv:hep-ph/0812.1144]}}.

\bibitem[Chung and Greensite(2017)]{Chung:2017ref}
Chung, K.; Greensite, J.
\newblock {Coulomb flux tube on the lattice}.
\newblock {\em Phys. Rev. D} {\bf 2017}, {\em 96},~034512,
  \href{http://xxx.lanl.gov/abs/1704.08995}{{\normalfont
  [arXiv:hep-lat/1704.08995]}}.

\bibitem[Tiktopoulos(1977)]{Tiktopoulos:1976sj}
Tiktopoulos, G.
\newblock {Gluon Chains}.
\newblock {\em Phys. Lett. B} {\bf 1977}, {\em 66},~271--275.

\bibitem[Greensite and Thorn(2002)]{Greensite:2001nx}
Greensite, J.; Thorn, C.B.
\newblock {Gluon chain model of the confining force}.
\newblock {\em JHEP} {\bf 2002}, {\em 02},~014,
  \href{http://xxx.lanl.gov/abs/hep-ph/0112326}{{\normalfont
  [hep-ph/0112326]}}.

\bibitem[Greensite and Szczepaniak(2015)]{Greensite:2014bua}
Greensite, J.; Szczepaniak, A.P.
\newblock {Coulomb string tension, asymptotic string tension, and the gluon
  chain}.
\newblock {\em Phys. Rev. D} {\bf 2015}, {\em 91},~034503,
  \href{http://xxx.lanl.gov/abs/1410.3525}{{\normalfont
  [arXiv:hep-lat/1410.3525]}}.

\bibitem[Greensite and Szczepaniak(2016)]{Greensite:2015nea}
Greensite, J.; Szczepaniak, A.P.
\newblock {Constituent gluons and the static quark potential}.
\newblock {\em Phys. Rev. D} {\bf 2016}, {\em 93},~074506,
  \href{http://xxx.lanl.gov/abs/1505.05104}{{\normalfont
  [arXiv:hep-lat/1505.05104]}}.

\bibitem[Greensite and Olejnik(2009)]{Greensite:2009mi}
Greensite, J.; Olejnik, S.
\newblock {Constituent Gluon Content of the Static Quark-Antiquark State in
  Coulomb Gauge}.
\newblock {\em Phys. Rev. D} {\bf 2009}, {\em 79},~114501,
  \href{http://xxx.lanl.gov/abs/0901.0199}{{\normalfont
  [arXiv:hep-lat/0901.0199]}}.

\bibitem[Nambu(1974)]{Nambu:1974zg}
Nambu, Y.
\newblock {Strings, Monopoles and Gauge Fields}.
\newblock {\em Phys. Rev. D} {\bf 1974}, {\em 10},~4262.

\bibitem['t~Hooft(1975)]{tHooft1975}
't~Hooft, G.
\newblock {\em {High Energy Physics}}; Editorice Compositori, Bologna,  1975.

\bibitem[Mandelstam(1976)]{Mandelstam:1974pi}
Mandelstam, S.
\newblock {Vortices and Quark Confinement in Nonabelian Gauge Theories}.
\newblock {\em Phys. Rept.} {\bf 1976}, {\em 23},~245--249.

\bibitem[Polyakov(1977)]{Polyakov:1976fu}
Polyakov, A.M.
\newblock {Quark Confinement and Topology of Gauge Groups}.
\newblock {\em Nucl. Phys. B} {\bf 1977}, {\em 120},~429--458.

\bibitem[Nielsen and Olesen(1973)]{Nielsen:1973cs}
Nielsen, H.B.; Olesen, P.
\newblock {Vortex Line Models for Dual Strings}.
\newblock {\em Nucl. Phys. B} {\bf 1973}, {\em 61},~45--61.

\bibitem[Seiberg and Witten(1994{\natexlab{a}})]{Seiberg:1994rs}
Seiberg, N.; Witten, E.
\newblock {Electric - magnetic duality, monopole condensation, and confinement
  in N=2 supersymmetric Yang-Mills theory}.
\newblock {\em Nucl. Phys. B} {\bf 1994}, {\em 426},~19--52,
  \href{http://xxx.lanl.gov/abs/hep-th/9407087}{{\normalfont
  [hep-th/9407087]}}.
\newblock [Erratum: Nucl.Phys.B 430, 485--486 (1994)].

\bibitem[Seiberg and Witten(1994{\natexlab{b}})]{Seiberg:1994aj}
Seiberg, N.; Witten, E.
\newblock {Monopoles, duality and chiral symmetry breaking in N=2
  supersymmetric QCD}.
\newblock {\em Nucl. Phys. B} {\bf 1994}, {\em 431},~484--550,
  \href{http://xxx.lanl.gov/abs/hep-th/9408099}{{\normalfont
  [hep-th/9408099]}}.

\bibitem[Gomez and Hernandez(1995)]{Gomez:1995rk}
Gomez, C.; Hernandez, R.
\newblock {Electric - magnetic duality and effective field theories}.
\newblock  {Advanced School on Effective Theories},  1995,
  \href{http://xxx.lanl.gov/abs/hep-th/9510023}{{\normalfont
  [hep-th/9510023]}}.

\bibitem[Bilal(1997)]{Bilal:1995hc}
Bilal, A.
\newblock {Duality in N=2 SUSY SU(2) Yang-Mills theory: A Pedagogical
  introduction to the work of Seiberg and Witten}.
\newblock  {NATO Advanced Study Institute on Quantum Fields and Quantum Space
  Time},  1997, pp. 21--43,
  \href{http://xxx.lanl.gov/abs/hep-th/9601007}{{\normalfont
  [hep-th/9601007]}}.

\bibitem[D'Hoker and Phong(1999)]{DHoker:1999yni}
D'Hoker, E.; Phong, D.H.
\newblock {Lectures on supersymmetric Yang-Mills theory and integrable
  systems}.
\newblock  {9th CRM Summer School: Theoretical Physics at the End of the 20th
  Century},  1999, pp. 1--125,
  \href{http://xxx.lanl.gov/abs/hep-th/9912271}{{\normalfont
  [hep-th/9912271]}}.

\bibitem[Douglas and Shenker(1995)]{Douglas:1995nw}
Douglas, M.R.; Shenker, S.H.
\newblock {Dynamics of SU(N) supersymmetric gauge theory}.
\newblock {\em Nucl. Phys. B} {\bf 1995}, {\em 447},~271--296,
  \href{http://xxx.lanl.gov/abs/hep-th/9503163}{{\normalfont
  [hep-th/9503163]}}.

\bibitem[Polyakov(1987)]{Polyakov:1987ez}
Polyakov, A.M.
\newblock {\em Gauge fields and strings}; Vol.~3, {\em Contemporary Concepts in
  Physics}, Harwood Academic Publishers, Chur,  1987.

\bibitem['t~Hooft(1974)]{tHooft:1974kcl}
't~Hooft, G.
\newblock {Magnetic Monopoles in Unified Gauge Theories}.
\newblock {\em Nucl. Phys. B} {\bf 1974}, {\em 79},~276--284.

\bibitem[Polyakov(1974)]{Polyakov:1974ek}
Polyakov, A.M.
\newblock {Particle Spectrum in Quantum Field Theory}.
\newblock {\em JETP Lett.} {\bf 1974}, {\em 20},~194--195.

\bibitem[Shifman and Yung(2007)]{Shifman:2007ce}
Shifman, M.; Yung, A.
\newblock {Supersymmetric Solitons and How They Help Us Understand Non-Abelian
  Gauge Theories}.
\newblock {\em Rev. Mod. Phys.} {\bf 2007}, {\em 79},~1139,
  \href{http://xxx.lanl.gov/abs/hep-th/0703267}{{\normalfont
  [hep-th/0703267]}}.

\bibitem['t~Hooft(1981)]{tHooft:1981bkw}
't~Hooft, G.
\newblock {Topology of the Gauge Condition and New Confinement Phases in
  Nonabelian Gauge Theories}.
\newblock {\em Nucl. Phys. B} {\bf 1981}, {\em 190},~455--478.

\bibitem[Unsal(2008)]{Unsal:2007vu}
Unsal, M.
\newblock {Abelian duality, confinement, and chiral symmetry breaking in
  QCD(adj)}.
\newblock {\em Phys. Rev. Lett.} {\bf 2008}, {\em 100},~032005,
  \href{http://xxx.lanl.gov/abs/0708.1772}{{\normalfont
  [arXiv:hep-th/0708.1772]}}.

\bibitem[Shifman and Unsal(2008)]{Shifman:2008ja}
Shifman, M.; Unsal, M.
\newblock {QCD-like Theories on R(3) x S(1): A Smooth Journey from Small to
  Large r(S(1)) with Double-Trace Deformations}.
\newblock {\em Phys. Rev. D} {\bf 2008}, {\em 78},~065004,
  \href{http://xxx.lanl.gov/abs/0802.1232}{{\normalfont
  [arXiv:hep-th/0802.1232]}}.

\bibitem[Unsal and Yaffe(2008)]{Unsal:2008ch}
Unsal, M.; Yaffe, L.G.
\newblock {Center-stabilized Yang-Mills theory: Confinement and large N volume
  independence}.
\newblock {\em Phys. Rev. D} {\bf 2008}, {\em 78},~065035,
  \href{http://xxx.lanl.gov/abs/0803.0344}{{\normalfont
  [arXiv:hep-th/0803.0344]}}.

\bibitem[Cossu \em{et~al.}(2014)Cossu, Hatanaka, Hosotani, and
  Noaki]{Cossu:2013ora}
Cossu, G.; Hatanaka, H.; Hosotani, Y.; Noaki, J.I.
\newblock {Polyakov loops and the Hosotani mechanism on the lattice}.
\newblock {\em Phys. Rev. D} {\bf 2014}, {\em 89},~094509,
  \href{http://xxx.lanl.gov/abs/1309.4198}{{\normalfont
  [arXiv:hep-lat/1309.4198]}}.

\bibitem[Bergner \em{et~al.}(2018)Bergner, Piemonte, and
  \"Unsal]{Bergner:2018unx}
Bergner, G.; Piemonte, S.; \"Unsal, M.
\newblock {Adiabatic continuity and confinement in supersymmetric Yang-Mills
  theory on the lattice}.
\newblock {\em JHEP} {\bf 2018}, {\em 11},~092,
  \href{http://xxx.lanl.gov/abs/1806.10894}{{\normalfont
  [arXiv:hep-lat/1806.10894]}}.

\bibitem[Bonati \em{et~al.}(2021)Bonati, Cardinali, D'Elia, Giordano, and
  Mazziotti]{Bonati:2020lal}
Bonati, C.; Cardinali, M.; D'Elia, M.; Giordano, M.; Mazziotti, F.
\newblock {Reconfinement, localization and thermal monopoles in $SU(3)$
  trace-deformed Yang-Mills theory}.
\newblock {\em Phys. Rev. D} {\bf 2021}, {\em 103},~034506,
  \href{http://xxx.lanl.gov/abs/2012.13246}{{\normalfont
  [arXiv:hep-lat/2012.13246]}}.

\bibitem[Kronfeld \em{et~al.}(1987)Kronfeld, Laursen, Schierholz, and
  Wiese]{Kronfeld:1987ri}
Kronfeld, A.S.; Laursen, M.L.; Schierholz, G.; Wiese, U.J.
\newblock {Monopole Condensation and Color Confinement}.
\newblock {\em Phys. Lett. B} {\bf 1987}, {\em 198},~516--520.

\bibitem[Stack \em{et~al.}(2002)Stack, Tucker, and Wensley]{Stack:2002ysv}
Stack, J.D.; Tucker, W.W.; Wensley, R.J.
\newblock {The Maximal Abelian gauge, monopoles, and vortices in SU(3) lattice
  gauge theory}.
\newblock {\em Nucl. Phys. B} {\bf 2002}, {\em 639},~203--222,
  \href{http://xxx.lanl.gov/abs/hep-lat/0110196}{{\normalfont
  [hep-lat/0110196]}}.

\bibitem[Shiba and Suzuki(1994)]{Shiba:1994ab}
Shiba, H.; Suzuki, T.
\newblock {Monopoles and string tension in SU(2) QCD}.
\newblock {\em Phys. Lett. B} {\bf 1994}, {\em 333},~461--466,
  \href{http://xxx.lanl.gov/abs/hep-lat/9404015}{{\normalfont
  [hep-lat/9404015]}}.

\bibitem[Stack \em{et~al.}(1994)Stack, Neiman, and Wensley]{Stack:1994wm}
Stack, J.D.; Neiman, S.D.; Wensley, R.J.
\newblock {String tension from monopoles in SU(2) lattice gauge theory}.
\newblock {\em Phys. Rev. D} {\bf 1994}, {\em 50},~3399--3405,
  \href{http://xxx.lanl.gov/abs/hep-lat/9404014}{{\normalfont
  [hep-lat/9404014]}}.

\bibitem[Ambjorn and Greensite(1998)]{Ambjorn:1998qp}
Ambjorn, J.; Greensite, J.
\newblock {Center disorder in the 3-D Georgi-Glashow model}.
\newblock {\em JHEP} {\bf 1998}, {\em 05},~004,
  \href{http://xxx.lanl.gov/abs/hep-lat/9804022}{{\normalfont
  [hep-lat/9804022]}}.

\bibitem[Del~Debbio \em{et~al.}(1995)Del~Debbio, Di~Giacomo, and
  Paffuti]{DelDebbio:1994sx}
Del~Debbio, L.; Di~Giacomo, A.; Paffuti, G.
\newblock {Detecting dual superconductivity in the ground state of gauge
  theory}.
\newblock {\em Phys. Lett. B} {\bf 1995}, {\em 349},~513--518,
  \href{http://xxx.lanl.gov/abs/hep-lat/9403013}{{\normalfont
  [hep-lat/9403013]}}.

\bibitem[Di~Giacomo \em{et~al.}(2000)Di~Giacomo, Lucini, Montesi, and
  Paffuti]{DiGiacomo:1999yas}
Di~Giacomo, A.; Lucini, B.; Montesi, L.; Paffuti, G.
\newblock {Color confinement and dual superconductivity of the vacuum. 1.}
\newblock {\em Phys. Rev. D} {\bf 2000}, {\em 61},~034503,
  \href{http://xxx.lanl.gov/abs/hep-lat/9906024}{{\normalfont
  [hep-lat/9906024]}}.

\bibitem[Greensite and Lucini(2008)]{Greensite:2008ss}
Greensite, J.; Lucini, B.
\newblock {Is Confinement a Phase of Broken Dual Gauge Symmetry?}
\newblock {\em Phys. Rev. D} {\bf 2008}, {\em 78},~085004,
  \href{http://xxx.lanl.gov/abs/0806.2117}{{\normalfont
  [arXiv:hep-lat/0806.2117]}}.

\bibitem[Kraan and van Baal(1998{\natexlab{a}})]{Kraan:1998pm}
Kraan, T.C.; van Baal, P.
\newblock {Periodic instantons with nontrivial holonomy}.
\newblock {\em Nucl. Phys. B} {\bf 1998}, {\em 533},~627--659,
  \href{http://xxx.lanl.gov/abs/hep-th/9805168}{{\normalfont
  [hep-th/9805168]}}.

\bibitem[Kraan and van Baal(1998{\natexlab{b}})]{Kraan:1998kp}
Kraan, T.C.; van Baal, P.
\newblock {Exact T duality between calorons and Taub - NUT spaces}.
\newblock {\em Phys. Lett. B} {\bf 1998}, {\em 428},~268--276,
  \href{http://xxx.lanl.gov/abs/hep-th/9802049}{{\normalfont
  [hep-th/9802049]}}.

\bibitem[Lee and Lu(1998)]{Lee:1998bb}
Lee, K.M.; Lu, C.h.
\newblock {SU(2) calorons and magnetic monopoles}.
\newblock {\em Phys. Rev. D} {\bf 1998}, {\em 58},~025011,
  \href{http://xxx.lanl.gov/abs/hep-th/9802108}{{\normalfont
  [hep-th/9802108]}}.

\bibitem[Bogomolny(1976)]{Bogomolny:1975de}
Bogomolny, E.B.
\newblock {Stability of Classical Solutions}.
\newblock {\em Sov. J. Nucl. Phys.} {\bf 1976}, {\em 24},~449.

\bibitem[Prasad and Sommerfield(1975)]{Prasad:1975kr}
Prasad, M.K.; Sommerfield, C.M.
\newblock {An Exact Classical Solution for the 't Hooft Monopole and the
  Julia-Zee Dyon}.
\newblock {\em Phys. Rev. Lett.} {\bf 1975}, {\em 35},~760--762.

\bibitem[Hofmann(2011)]{Hofmann-book}
Hofmann, R.
\newblock {\em The Thermodynamics of Quantum Yang–Mills Theory}; WORLD
  SCIENTIFIC,  2011;
  \href{http://xxx.lanl.gov/abs/https://www.worldscientific.com/doi/pdf/10.1142/8015}{{\normalfont
  [https://www.worldscientific.com/doi/pdf/10.1142/8015]}}.

\bibitem['t~Hooft(1976)]{tHooft:1976snw}
't~Hooft, G.
\newblock {Computation of the Quantum Effects Due to a Four-Dimensional
  Pseudoparticle}.
\newblock {\em Phys. Rev. D} {\bf 1976}, {\em 14},~3432--3450.
\newblock [Erratum: Phys.Rev.D 18, 2199 (1978)].

\bibitem[Nahm(1983)]{Nahm:1983sv}
Nahm, W.
\newblock {SELFDUAL MONOPOLES AND CALORONS}.
\newblock  {12th International Colloquium on Group Theoretical Methods in
  Physics},  1983.

\bibitem[Garland and Murray(1988)]{Garland:1988bv}
Garland, H.; Murray, M.K.
\newblock {{Kac-Moody} Monopoles and Periodic Instantons}.
\newblock {\em Commun. Math. Phys.} {\bf 1988}, {\em 120},~335--351.

\bibitem[Nahm(1980)]{Nahm:1979yw}
Nahm, W.
\newblock {A Simple Formalism for the BPS Monopole}.
\newblock {\em Phys. Lett. B} {\bf 1980}, {\em 90},~413--414.

\bibitem[Diakonov and Petrov(2007)]{Diakonov:2007nv}
Diakonov, D.; Petrov, V.
\newblock {Confining ensemble of dyons}.
\newblock {\em Phys. Rev. D} {\bf 2007}, {\em 76},~056001,
  \href{http://xxx.lanl.gov/abs/0704.3181}{{\normalfont
  [arXiv:hep-th/0704.3181]}}.

\bibitem[Bruckmann \em{et~al.}(2009)Bruckmann, Dinter, Ilgenfritz,
  Muller-Preussker, and Wagner]{Bruckmann:2009nw}
Bruckmann, F.; Dinter, S.; Ilgenfritz, E.M.; Muller-Preussker, M.; Wagner, M.
\newblock {Cautionary remarks on the moduli space metric for multi-dyon
  simulations}.
\newblock {\em Phys. Rev. D} {\bf 2009}, {\em 79},~116007,
  \href{http://xxx.lanl.gov/abs/0903.3075}{{\normalfont
  [arXiv:hep-ph/0903.3075]}}.

\bibitem[Gerhold \em{et~al.}(2007)Gerhold, Ilgenfritz, and
  Muller-Preussker]{Gerhold:2006sk}
Gerhold, P.; Ilgenfritz, E.M.; Muller-Preussker, M.
\newblock {An SU(2) KvBLL caloron gas model and confinement}.
\newblock {\em Nucl. Phys. B} {\bf 2007}, {\em 760},~1--37,
  \href{http://xxx.lanl.gov/abs/hep-ph/0607315}{{\normalfont
  [hep-ph/0607315]}}.

\bibitem[Gupta \em{et~al.}(2008)Gupta, Huebner, and Kaczmarek]{Gupta:2007ax}
Gupta, S.; Huebner, K.; Kaczmarek, O.
\newblock {Renormalized Polyakov loops in many representations}.
\newblock {\em Phys. Rev. D} {\bf 2008}, {\em 77},~034503,
  \href{http://xxx.lanl.gov/abs/0711.2251}{{\normalfont
  [arXiv:hep-lat/0711.2251]}}.

\bibitem[Greensite and H\"ollwieser(2015)]{Greensite:2014gra}
Greensite, J.; H\"ollwieser, R.
\newblock {Double-winding Wilson loops and monopole confinement mechanisms}.
\newblock {\em Phys. Rev. D} {\bf 2015}, {\em 91},~054509,
  \href{http://xxx.lanl.gov/abs/1411.5091}{{\normalfont
  [arXiv:hep-lat/1411.5091]}}.

\bibitem[Diakonov(2009)]{Diakonov:2009jq}
Diakonov, D.
\newblock {Topology and confinement}.
\newblock {\em Nucl. Phys. B Proc. Suppl.} {\bf 2009}, {\em 195},~5--45,
  \href{http://xxx.lanl.gov/abs/0906.2456}{{\normalfont
  [arXiv:hep-ph/0906.2456]}}.

\bibitem[Greensite and Matsuyama(2018)]{Greensite:2018mhh}
Greensite, J.; Matsuyama, K.
\newblock {What symmetry is actually broken in the Higgs phase of a gauge-Higgs
  theory?}
\newblock {\em Phys. Rev. D} {\bf 2018}, {\em 98},~074504,
  \href{http://xxx.lanl.gov/abs/1805.00985}{{\normalfont
  [arXiv:hep-th/1805.00985]}}.

\bibitem[Greensite and Matsuyama(2020)]{Greensite:2020nhg}
Greensite, J.; Matsuyama, K.
\newblock {Higgs phase as a spin glass and the transition between varieties of
  confinement}.
\newblock {\em Phys. Rev. D} {\bf 2020}, {\em 101},~054508,
  \href{http://xxx.lanl.gov/abs/2001.03068}{{\normalfont
  [arXiv:hep-th/2001.03068]}}.

\end{thebibliography}

\end{document}